\newcommand{\beq}{\begin{equation}}
\newcommand{\eeq}{\end{equation}}
\newcommand{\bea}{\begin{eqnarray}}
\newcommand{\eea}{\end{eqnarray}}

\newcommand{\gsim}{\lower.7ex\hbox{$\;\stackrel{\textstyle>}{\sim}\;$}}
\newcommand{\lsim}{\lower.7ex\hbox{$\;\stackrel{\textstyle<}{\sim}\;$}}

\documentclass[prd,11pt,a4paper,onecolumn,floats,aps,nofootinbib,amssymb]{revtex4-1}

\usepackage{geometry,amsmath,amsfonts}
\usepackage{slashed}
\usepackage{graphicx}
\usepackage{epstopdf}
\usepackage{mathrsfs}
\usepackage{amssymb}
\usepackage{verbatim}
\usepackage{color}
\usepackage{multirow}
\usepackage{hyperref}
\usepackage[normalem]{ulem}
\usepackage{lineno}

\usepackage[caption=false]{subfig}

\def\stacksymbols #1#2#3#4{\def\theguybelow{#2}
    \def\vp{\lower#3pt}
    \def\sp{\baselineskip0pt\lineskip#4pt}
    \mathrel{\mathpalette\intermediary#1}}

\def\intermediary#1#2{\vp\vbox{\sp
     \everycr={}\tabskip0pt
     \halign{$\mathsurround0pt#1\hfil##\hfil$\crcr#2\crcr
              \theguybelow\crcr}}}

\def\ra{\rangle}
\def\la{\langle}

\def\be{\begin{equation}}
\def\ee{\end{equation}}
\def\bea{\begin{eqnarray}}
\def\eea{\end{eqnarray}}

\def\sp{\;\;\;,\;\;\;}

\def\la{\langle}
\def\ra{\rangle}

\newcommand{\ovl}[1]{\overline{#1}}

\def\lsim{\raise0.3ex\hbox{$\;<$\kern-0.75em\raise-1.1ex\hbox{$\sim\;$}}}
\def\gsim{\raise0.3ex\hbox{$\;>$\kern-0.75em\raise-1.1ex\hbox{$\sim\;$}}}

\def\inbar{\,\vrule height1.5ex width.4pt depth0pt}

\def\IC{\relax\hbox{$\inbar\kern-.3em{\rm C}$}}
\def\IQ{\relax\hbox{$\inbar\kern-.3em{\rm Q}$}}
\def\IR{\relax{\rm I\kern-.18em R}}
 \font\cmss=cmss10 \font\cmsss=cmss10 at 7pt
\def\IZ{\relax\ifmmode\mathchoice
 {\hbox{\cmss Z\kern-.4em Z}}{\hbox{\cmss Z\kern-.4em Z}}
 {\lower.9pt\hbox{\cmsss Z\kern-.4em Z}}
 {\lower1.2pt\hbox{\cmsss Z\kern-.4em Z}}\else{\cmss Z\kern-.4em Z}\fi}

\def\comment#1{}
\def\to{\rightarrow}

\def\u1x{U(1)_X}
\newcommand{\nc}{\newcommand}
\nc{\LL}{L}
\nc{\vv}{\tilde{v}}
\nc{\ccdot}{\!\cdot\!}
\nc{\gsm}{G_{SM}}
\nc{\vfive}{\mathbf{5}\oplus\mathbf{\overline{5}}}
\nc{\vten}{\mathbf{10}\oplus\mathbf{\overline{10}}}
\nc{\zhol}{Z^{\rm hol}}
\nc{\xfb}{\,{\rm fb}}

\begin{document}

%
%

\begin{flushright}
LPT-Orsay-17-09, CPHT-RR009.032017, SCIPP 17/03
\end{flushright}

\vspace*{1mm}

\title{\color{blue}The Waning of the WIMP?\\ A Review of Models, Searches, and Constraints}




\author{Giorgio Arcadi$^{a}$}
\email{arcadi@mpi-hd.mpg.de}
\author{Ma\'ira Dutra$^{b}$}
\email{maira.dutra@th.u-psud.fr}
\author{Pradipta Ghosh$^{b,c}$}
\email{pradipta.ghosh@th.u-psud.fr }
\author{Manfred Lindner $^{a}$}
\email{lindner@mpi-hd.mpg.de}
\author{Yann Mambrini$^{b}$}
\email{yann.mambrini@th.u-psud.fr}
\author{Mathias Pierre$^{b}$}
\email{mathias.pierre@th.u-psud.fr}
\author{Stefano Profumo$^{d,e}$}
\email{profumo@ucsc.edu}
\author{Farinaldo S. Queiroz$^{a}$}
\email{queiroz@mpi-hd.mpg.de}

\affiliation{}
\affiliation{$^a$Max Planck Institut f\"ur Kernphysik, Saupfercheckweg 1, D-69117 Heidelberg, Germany}
\affiliation{$^b$Laboratoire de Physique Th\'eorique, CNRS, Univ. Paris-Sud, Universit\'e  Paris-Saclay, 91405 Orsay, France}
\affiliation{$^c$Centre de Physique Th\'eorique, Ecole Polytechnique, CNRS, Universit\'e Paris-Saclay, 91128 Palaiseau Cedex, France}
\affiliation{$^d$Department of Physics, University of California, Santa Cruz, 1156 High St, Santa Cruz, CA 95060, United States of America}
\affiliation{$^e$Santa  Cruz  Institute  for  Particle  Physics,  Santa  Cruz,  1156  High  St,  Santa  Cruz,  CA 95060, United States of America}

\begin{abstract} 

Weakly Interacting Massive Particles (WIMPs) are among the best-motivated dark matter candidates. In light of no conclusive detection signal yet despite an extensive search program that combines, often in a complementary way, direct, indirect, and collider probes, we find it timely to give a broad overview of the WIMP paradigm. In particular, we review here the theoretical foundations of the WIMP paradigm, discuss status and prospects of various detection strategies, and explore future experimental challenges and opportunities. 

\end{abstract}

\maketitle

\tableofcontents

\section{Introduction}
A combination of cosmological observations including (between others) studies of the cosmic microwave background, distant supernovae, large samples of galaxy clusters, baryon acoustic oscillation measurements has firmly established a standard cosmological model where the Dark Matter (DM), a new yet-to-be discovered form of matter, accounts for about 85\% of the matter content of the Universe, and about 27\% of the global energy budget \cite{Ade:2015xua}. Cold Dark Matter (CDM) is a key ingredient to successfully explain the formation of large-scale structure, producing theoretical predictions in striking agreement with observations \cite{Blumenthal:1984bp,Bullock:1999he}.

\noindent
Little is, however, as of yet known about DM as a particle; any candidate for (most of) the DM must nevertheless be consistent with the following five observationally-motivated constraints: 

(i) The relic abundance of DM needs to account for the observed CDM abundance; 

(ii) the DM particle should be non-relativistic at matter-radiation equality to form structures in the early Universe in agreement with the observation. As a result, if the DM was produced as a thermal relic in the early universe, its mass cannot be arbitrarily light. Specifically, cosmological simulations rule out DM masses below a few keV \cite{Benson:2012su,Lovell:2013ola,Kennedy:2013uta}. 

(iii) The DM should be  electromagnetically effectively neutral, as a result of null searches for stable charged particles \cite{McDermott:2010pa,SanchezSalcedo:2010ev} as well as direct detection experiments, which we review below;

(iv) The DM particle must be cosmologically stable since its presence is ascertained today, implying that its lifetime is larger than the age of the Universe. Under certain assumptions, much stronger limits are applicable conservatively requiring a lifetime order of magnitude larger can be derived \cite{Audren:2014bca,Queiroz:2014yna,Baring:2015sza,Mambrini:2015sia,Giesen:2015ufa,Lu:2015pta,Slatyer:2016qyl,Jin:2017iwg}. 

(v) Cluster collisions, such as the Bullet Cluster \cite{Clowe:2006eq}, constrain the level of self-interactions that DM particles can have (see however refs.~\cite{Hochberg:2014dra,Hansen:2015yaa} for alternative scenarios).

Within the generous parameter space outlined by the observational requirements listed above, we will argue below that the paradigm of Weakly Interacting Massive Particles (WIMPs)~\cite{Steigman:1984ac} is one of the most compelling options for DM as a particle. As such, it has undergone very close and effective experimental scrutiny. In what follows we will outline the theoretical underpinning of the WIMP paradigm, discuss a few well-motivated ``simplified model'' realizations of schematic WIMP models, and we will attempt to give an up-to-date state of the art of WIMP searches and prospects for the future.


\section{The WIMP Paradigm}

The paradigm of thermal decoupling, based upon applications to cosmology of statistical mechanics and particle and
nuclear physics, is enormously successful at making
detailed predictions for observables in the early universe,
including the abundances of light elements and the cosmic
microwave background \cite{Peebles:1991ch}. It is somewhat natural to invoke a similar
paradigm to infer the abundance of DM as a thermal
relic from the early universe uniquely from the underlying
DM particle properties. 

Assuming there exist interactions between a cosmologically stable particle $\chi$ -the (generic) DM- with Standard Model (SM) particles, sizable enough that for a high enough temperature $T$ the DM is in thermal equilibrium with the primordial thermal bath, the cosmological evolution of the DM particle can be traced through the following Boltzmann equation:
\begin{equation}
\label{eq:bolbase}
\frac{dn_{\chi}}{dt} +3 H(T) n_\chi = -\langle \sigma v \rangle (n_{\chi}^2 -n_{\chi, eq}^2),
\end{equation}
describing the DM number density $n_\chi$ in turn defined as:
\begin{equation}
\label{eq:def1}
n_\chi (T)=\int \frac{d^3 p}{{\left(2 \pi\right)}^3}f_\chi (p,T),
\end{equation}
with $f_\chi$ being the DM distribution function. The quantity $\la \sigma v \ra$, dependent on temperature $T$, is the thermally averaged pair annihilation cross-section associated to the process $\chi \chi \rightarrow SM SM$ while $H(T)$ is the Hubble rate. $n_{\chi,eq}$ is the equilibrium number density obtained by eq.~(\ref{eq:def1}) by replacing $f_\chi$ with the equilibrium distribution function (by convenience one typically adopts the Maxwell-Boltzmann distribution):

\begin{equation}
n_{\chi, eq}= g_{\chi} \frac{m_{\chi}^2 T}{2\pi} K_2\left(\frac{m_\chi}{T}\right),
\label{Eq:numberdens}
\end{equation}
where $g_{\chi}$ is the number of internal degrees of freedom of the DM particle, $m_{\chi}$ its mass, and $K_2$ is the modified Bessel function of second type.


As the Universe expands, the scale factor increases and the temperature decreases. Assuming that $\chi$ continues to be in thermal equilibrium, eventually the temperature drops below the DM mass, and the annihilation rate for DM particles, which depends linearly on the number equilibrium density $n_{\chi,eq}$, enters the so-called ``Boltzmann-tail'', $n_{\chi, eq}\propto \exp(-m_\chi/T)\ (m_\chi\ll T)$; the annihilation rate then eventually fall below the Universe expansion rate, $H(T)$ (a power-law in temperature, $\propto T^2$ in the radiation-dominated universe), leading to the thermal freeze-out of this ``cold'' relic. Thereafter,  the DM comoving number density $Y_\chi=\frac{n_\chi}{s}$ is approximately constant\footnote{See ref.~\cite{DEramo:2017gpl} for an exception (``relentless'' DM) for modified expansion histories.}. 

The DM abundance is typically expressed as a fraction $\Omega_{\rm DM}=\rho_{\rm DM}/\rho_{\rm cr}$ of the critical density of the universe $\rho_{\rm cr}(T)=3 H(T)^2 M_{\rm PL}^2/8 \pi$ times $h^2$ where $h$ is the Hubble expansion rate today in units of 100 km/s/Mpc. $\rho_{\rm cr}(T_0) \simeq 10^{-5}~ \mathrm{GeV ~cm^{-3}}$ today.
The thermal relic density can be expressed, as function of the DM comoving density, as $\Omega_{\rm DM}=\frac{m_\chi s_0 Y_0}{\rho_c}$ where $Y_0 \equiv Y(T_0), s_0 \equiv s(T_0)$ with $T_0$ being the temperature of the Universe at present times and $s$ the entropy density of the Universe. $Y_0$ can be semi-analytically determined as~\cite{Gondolo:1990dk}:
\begin{equation}
Y_0 \simeq \sqrt{\frac{\pi}{45}}M_{\rm PL}{\left[\int_{T_0}^{T_f} g_{*}^{1/2} \langle \sigma v \rangle dT \right]}^{-1},
\end{equation}
where, $g_{*}^{1/2}\equiv g_{*}^{1/2}(T)$ is a function related to the relativistic degrees of freedom of the primordial thermal bath (we refer to ref.~\cite{Gondolo:1990dk} for its definition) while $T_f$ represents the freeze-out temperature which can be determined by solving the equation:
\begin{equation}
 \sqrt{\frac{\pi}{45}}M_{\rm PL} \frac{g_{*}^{1/2} m_\chi}{x^2}\langle \sigma v \rangle Y_{\chi, eq} \delta (\delta+2)=-\frac{d \log Y_{\chi, eq}}{dx},
\end{equation}
where $\delta=(Y_\chi-Y_{\chi, eq})/Y_{\chi, eq}$ is conventionally set to $1.5$.

By combining the expressions above the DM relic density can be numerically estimated as:
\begin{equation}
\Omega_{\rm DM}h^2 \approx 8.76 \times 10^{-11}\, {\mbox{GeV}}^{-2} {\left[\int_{T_0}^{T_f} g_{*}^{1/2} \langle \sigma v \rangle \frac{dT}{m_\chi} \right]}^{-1}.
\end{equation}

\begin{figure}[t]
\centering
\includegraphics[scale=0.6]{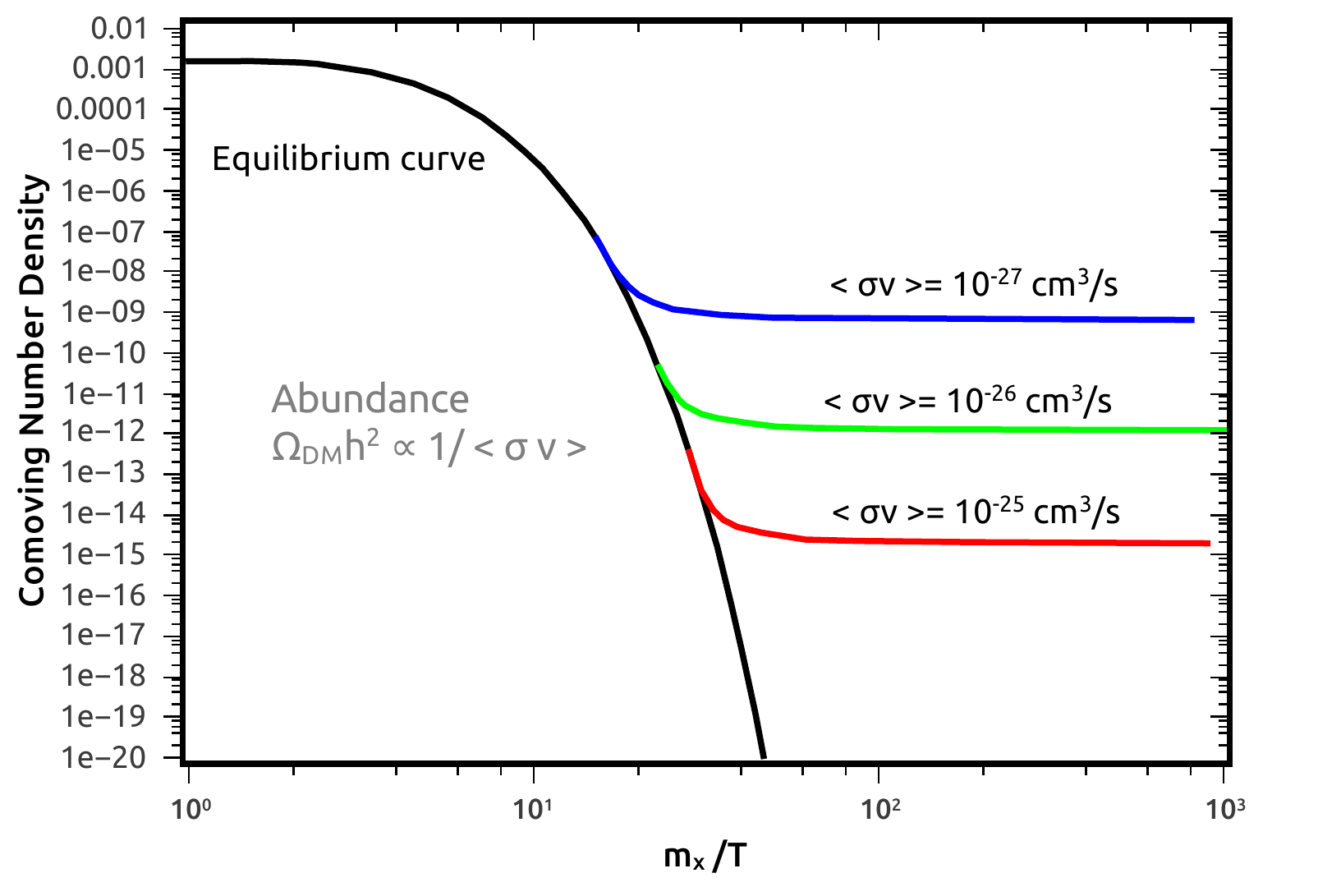}
\caption{Comoving number density evolution as a function of the ratio $m_{\chi}/T$ in the context of the thermal freeze-out. Notice that the size of the annihilation cross section determines the DM abundance since $\Omega_{DM} \propto 1/ \la \sigma v \ra$.}
\label{abundanceplot}
\end{figure}

The behavior of the solution of the Boltzmann equation is illustrated in fig.~(\ref{abundanceplot}). As expected, the
DM relic density is basically set by the inverse value of
the thermally averaged cross section (calculated at the
freeze-out temperature), with a logarithmic dependence on
$m_\chi$. It can be straightforwardly verified that the
experimental determination~\cite{Ade:2015xua} $\Omega_{\rm DM}h^2 \approx 0.12$ is matched by a value of the cross
section of the order of $10^{-9} {\mbox{GeV}}^{-2}$
corresponding to 
$\langle \sigma v \rangle \sim 10^{-26}\,{\mbox{cm}}^3 {\mbox{s}}^{-1}$. 

The WIMP ``miracle'' is the observation that for typical
weak-scale pair annihilation cross sections, say $\sigma\sim G_F^2  ~T^2$, with $G_F$ the Fermi constant, and $T\sim m_\chi/20$ the typical freeze-out temperature,  and for electroweak-scale  mass scales,
$m_\chi\sim E_{\rm EW}\sim 200$ GeV, the thermal relic density matches the observed cosmological density. 
It is important to realize that this coincidence is a
statement about cross sections (and, weakly, masses), and is
thus not unique to the weak scale and weak interactions;
however, what is indeed remarkable is the fact that independent theoretical reasons, such as naturalness and the hierarchy problem, indicate that it is plausible to expect new physics  at $E_{\rm EW}$; Moreover, weak interactions
are the only  gauge interactions in the Standard Model that
a DM particle might interact through.


%

The WIMP paradigm is thus an attractive solution of the DM
issues since the DM abundance is set to the observed value
by a new physics scale that is well motivated, and by
interactions mediated by one of the Standard Model gauge
interactions. As a result, concrete realizations of WIMP
models had been developed in different 
Beyond the Standard Model (BSM) frameworks, accessible to several different search strategies, as reviewed in 
the next sections.


Operationally, all the information about the particle physics framework connected to a specific DM particle candidate is contained in the thermally pair averaged cross section and in the freeze-out temperature that results from it. Its formal definition reads~\cite{Gondolo:1990dk}~\footnote{In scenarios where the DM is not the only new particle state other processes, like coannihilations, might be relevant for the DM relic density. A more general definition of $\langle \sigma v \rangle$, including also the processes, can be found e.g., in ref.~\cite{Edsjo:1997bg}.}:
\begin{equation}
\label{eq:sigmaG}
\langle \sigma v \rangle=\frac{1}{8 m_\chi^4 T {K_2\left(\frac{m_\chi}{T}\right)}^2}\int_{4 m_\chi^2}^{\infty} ds \sigma(s) \sqrt{s} \left(s-4 m_\chi^2\right)K_1\left(\frac{\sqrt{s}}{T}\right).
\end{equation}

\noindent
Since WIMPs freeze out in the non-relativistic regime, and thus $v\ll c$ (where $v$ is the WIMP relative velocity), a useful approximation consists of a velocity expansion (reviewed in the appendix) $\langle \sigma v \rangle \simeq a + b v^2$. The velocity expansion is however not valid in some relevant cases, like for example annihilations through the resonant exchange of an s-channel mediator~\cite{Griest:1990kh}. For this reason, all the numerical results presented in this work will rely on the full numerical determination of $\langle \sigma v \rangle$, as given in~(\ref{eq:sigmaG}) and on the solution of the DM Boltzmann equation, as provided by the numerical package micrOMEGAs~\cite{Belanger:2006is,Belanger:2008sj,Belanger:2013oya}. 

\section{Direct Detection}

The observation of a DM halo in our Galaxy motivates the search for DM scattering off of nuclei. Several experiments have played an important role in this direction \cite{Aprile:2010um,Tanaka:2011uf,Aprile:2011ts,Aprile:2011hi,Aalseth:2012if,Ahmed:2012vq,Aprile:2012nq,Agnese:2013rvf,Aprile:2013doa,Agnese:2013cvt,Adrian-Martinez:2013ayv,Aprile:2014eoa,Angloher:2014myn,Agnes:2014bvk,Aalseth:2014eft,Aalseth:2015mba,Aprile:2015ibr,Amole:2015pla,Angloher:2015ewa,Amole:2015lsj,Akerib:2015rjg,Agnes:2015ftt,Choi:2015ara,Akerib:2016lao,Akerib:2016vxi,Amole:2016pye,Aprile:2016swn,Hehn:2016nll,Armengaud:2016cvl,Adrian-Martinez:2016gti,Amole:2017dex}.  In this section
we focus on direct detection experiments looking for WIMPs scattering, but there are important searches stemming from
Neutrinos telescopes  by measuring the neutrino flux from
the Sun \cite{Bergstrom:1996kp,Gondolo:2004sc,Wikstrom:2009kw,Aartsen:2016exj}. 

Direct DM detection seeks to measure the nuclear recoil imparted by the scattering of a WIMP particle. 
The WIMP-nuclei differential scattering rate can be written as,
%
\begin{equation}
\frac{dR}{dE} (E,t) = \frac{N_T \rho_{\chi} }{m_{\chi}\, m_A} \int_{v_{\rm min}}^{v_{\rm esc}}  v f_E(\vec{v},t) \frac{d\sigma }{dE}(v,E) d^3\vec{v},
\label{scatteringrate}
\end{equation}
where $N_T$ is the number of target nuclei per kilogram of the detector, $m_{\chi}$ the DM mass, $\rho_{\chi}$ the local DM density ($\rho_{\chi} = 0.3\, \rm{GeV}/\rm{cm}^3$) \cite{Catena:2009mf,Weber:2009pt,Read:2014qva,Xia:2015agz,Iocco:2015xga}, $\vec{v}$ the velocity of the DM particle relative to the Earth, $f_E(\vec{v},t)$ the distribution of velocities of the WIMP in the frame of the Earth \footnote{The velocity distribution is understood as the probability of finding a WIMP  with velocity $v$ at a time $t$.}, $v_{\rm min}=\sqrt{m_N E/(2\mu^2)}$ is the minimum WIMP speed required to produce a detectable event at energy $E$, $v_{\rm esc}$ is the escape velocity i.e. the velocity for which the WIMP are no longer gravitationnally bounded to the Milky Way. $\mu = m_{\chi} m_N/ (m_{\chi} + m_N)$ is the WIMP-nucleus reduced mass ($m_N$ is the nucleus mass), $d\sigma/dE(v,E)$ the differential cross-section for the WIMP-nucleus scattering as follows,
\begin{equation}
\frac{d\sigma}{dE} = \frac{m_N}{2\mu^2 v^2} (\sigma_{SI} F^2(q) + \sigma_{SD} S(q)),
\end{equation}
where $F^2(q)$ and $S(q)$ are the spin-independent and spin-dependent form factors, as described e.g. in refs.~\cite{Jungman:1995df,Duda:2006uk,Bednyakov:2006ux,Schnee:2011ef}.

\begin{figure}[!t]
\includegraphics[scale=0.55]{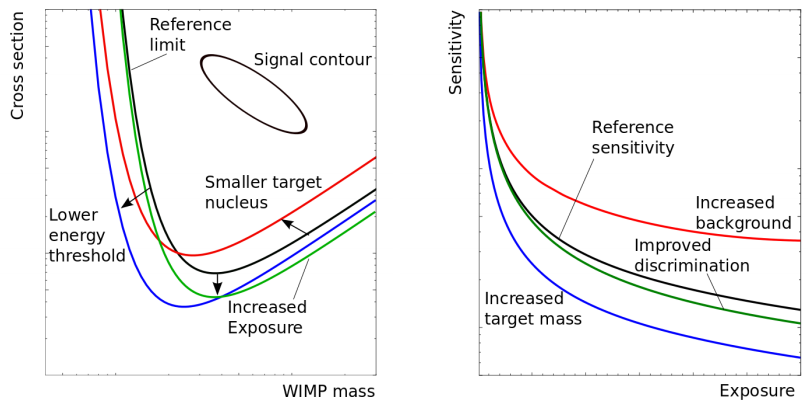}
\caption{{\it Left}: Illustrative impact of energy threshold, exposure and target nucleus. {\it Right}: Impact of background and exposure on the sensitivity. Taken from ref.~\cite{Undagoitia:2015gya}.}
\label{DDfigure}
\end{figure}

After measuring the scattering rate, the next and fundamental task is to discriminate signal from background. This is done by using the detector response to electron and nuclear scattering, 
which might vary from one experiment to another. For instance, in germanium detectors ionization yield is used to discriminate signal from background, whereas in experiments that use xenon, the ionization/scintillation ratio is the discriminating variable. What determines an experiment's sensitivity to a WIMP  signal is a combination of:
\begin{enumerate}
\item {\it Energy threshold}: drives the sensitivity to low WIMP masses, and consequently the sharpening of the direct detection limits on the scattering cross section at low masses as shown in fig.~(\ref{DDfigure});

\item {\it Control over the background and exposure:} determine the overall sensitivity of the experiment pushing the limits to lower scattering cross sections;

\item {\it Target:} has an impact on the experiment sensitivity to low and heavy WIMP masses, as well as on capability to probe spin-dependent scatterings.

\end{enumerate}

\noindent
All these facts are illustrated in fig.~(\ref{DDfigure}) where we plotted the impact of exposure, energy threshold and mass of the nucleus target on direct detection experiments sensitivity for WIMP-nucleon scatterings. 

{\it Left panel:} (i) Comparing the solid black and blue curves at low WIMP mass one can see that the energy threshold determines the smallest WIMP mass accessible to a given direct detection experiment. The blue line refers to a direct detection experiment with lower energy threshold; (ii) Notice that stronger bounds on the scattering cross section are possible with larger exposure as represented in the solid green line; (iii) The target nucleus can chance the WIMP mass where the strongest limit on the scattering cross section lies at and also the sensitivity to lower and larger WIMP masses. This is visible comparing the red and black lines. {\it Right panel:} (i) Comparing the black and blue curves we can see the importance of increasing exposure; (ii) Red and green curves exhibit the impact of background discriminations.

It is important to highlight that from going to the measured scattering rate in eq.~\eqref{scatteringrate} to the derivation of a limit on the WIMP-nucleon scattering cross section as a function of the WIMP mass, there are some assumptions that have to be made about the velocity distribution, nuclear form factor, type of WIMP-nucleon scattering, and local DM density that suffers from large uncertainties \cite{Pato:2010yq,Bidin:2014ola,Silverwood:2015hxa}. In particular, the common assumptions are that there is a smooth halo of DM particles in our galaxy well described by a Maxwellian velocity distribution \cite{Bozorgnia:2016ogo,Kelso:2016qqj,Kavanagh:2016xfi}, that the nucleus can be treated as a hard sphere as indicated by the Helm form factor \cite{Duda:2006uk}, and that the WIMP-nucleon scattering is elastic. Our results rely on the same set of assumptions throughout this manuscript (see refs.~\cite{McCabe:2010zh,Frandsen:2011gi,Fairbairn:2012zs,Davis:2014ama,OHare:2016pjy} for discussions on these topics). Interestingly, if the uncertainties present in the astrophysical input are under control and precise measurements on the scattering cross section can be realized, then one might even determine the nature of dark matter using direct detection experiments alone \cite{Queiroz:2016sxf}.

Anyhow, in summary, the present measurement of the scattering rate has not yet observed any excess over the background,
which after
some assumptions, translates into limits on the WIMP-nucleon
scattering cross section as a function of the DM mass. 
In this work we will be using the following limits and projections:

$\bullet$ {\it Current spin-independent limit:} 

\noindent
We adopt the latest result from LUX based on $3.35\times 10^4$~kg-day exposure. This recent limit represents a factor of four improvement over previous results. In particular, spin-independent cross sections above $2.2\times 10^{-46} cm^2$ are excluded at the 90\% C.L for a WIMP mass of $50$~GeV \cite{Akerib:2016vxi}. 

Similarly, PANDA-X collaboration with a very similar exposure found a limit which is basically identical to the one obtained by LUX. Thus, in the upcoming figures, whenever we quote LUX, it also reads PANDA-X \cite{Tan:2016zwf}.  

$\bullet$ {\it Current spin-dependent limit:} 

\noindent
We adopt the latest results from PANDA-X reported in ref.~\cite{Fu:2016ega} which overlaps with the XENON100 limits and LUX at higher WIMP masses at 90\% C.L. but
significantly improving them at lower masses. PANDA-X
result is based on $3.3 \times 10^4$~Kg of exposure, and in particular excludes  the WIMP-neutron spin-dependent scattering cross section of $4.1\times 10^{-41} cm^2$ for a
WIMP mass of $40$~GeV. In the figures these limits are
labeled as LUX to keep them uniform. Hopefully PANDA-X will
continue to run and improve their sensitivity and possibly
unearth a DM signal.

$\bullet$ {\it Projected spin-independent limit:}

\noindent
We adopt the projected XENON1T two years exposure spin-independent limits outlined in ref.~\cite{Aprile:2015uzo} and the LZ collaboration referred as baseline in ref.~\cite{Szydagis:2016few}.

$\bullet$ {\it Projected spin-dependent limit:}

\noindent
In the lack of published projections for spin-dependent 
scattering we simply rescaled the current spin-dependent limits taking into account the planned exposure. In practice, we derived the scaling factor between latest LUX spin-independent limit and the LZ projection, and then applied this same scaling factor to derive the LZ/XENON1T projection for spin-dependent scattering. In the light of no large background, the limits will roughly be improved simply by exposure, justifying our method.

\section{Indirect Detection}

\noindent
Indirect DM detection relies on the detection of the byproducts of WIMPs annihilations over the expected background at galactic or extragalactic scales, using Earth based telescopes such as H.E.S.S. and CTA, or satellites such as AMS and Fermi-LAT \cite{Abdo:2010ex,Ackermann:2011wa,Abramowski:2011hc,Abramowski:2012au,Ackermann:2012nb,Ackermann:2013uma,Ackermann:2013yva,Abramowski:2014tra,HESS:2015cda,Ackermann:2015tah,Ackermann:2015lka,Ackermann:2015zua,Abdalla:2016olq,Abdallah:2016ygi,Fermi-LAT:2016uux} \footnote{See refs.~\cite{Aleksic:2011jx,Hooper:2012sr,Grube:2012fv,Harding:2015bua,Ahnen:2016qkx,Queiroz:2016zwd,Profumo:2016idl,Doro:2017dqn,Campos:2017odj} for other competitive limits.}

In this regard, the search for gamma-rays and cosmic-rays and neutrinos offer an exciting possibility of DM detection. Here we will focus on gamma-rays. The gamma-ray flux from WIMP annihilation is proportional to:

\begin{itemize}
\item The number density squared of particles, i.e., $n_{\chi}^2=\rho^2/m_{\chi}^2$;
\item The WIMP annihilation cross section today, $\sigma$;
\item The mean WIMP velocity $v$;
\item Volume of the sky observed within a solid angle $\Omega$;
\item Number of gamma-rays produced per annihilation at a given energy, also known as the energy spectrum ($dN/dE$).
\end{itemize}

In summary, it is found to be:
%
\begin{equation}
\overbrace{\frac{d\Phi}{d\Omega dE}}^{\rm Diff. Flux} = {\color{blue} \frac{ \overbrace{ \sigma v }^{\rm Anni.\, Cross\, Section}}{8\pi m_{\chi}^2}} \times {\color{green} \underbrace{\frac{dN}{dE}}_{\rm Energy\, Spectrum}} \times {\color{red} \overbrace{\int_{\rm l.o.s} ds}^{\rm Line\, of\, Sight\, Integral}} \times {\color{magenta}  \underbrace{\rho^2 (\overrightarrow{r}(s,\Omega))}_{\rm DM\, Distribution}}.
\label{eq:flux}
\end{equation}

In eq.~\eqref{eq:flux} the DM density is integrated over the line of sight from the observer to the source.

The DM density is not tightly constrained, and several DM density profiles have been considered in the literature leading to either spike or core DM densities toward the center of galaxies \cite{Burkert:1995yz,Navarro:1995iw,Salucci:2000ps,Graham:2005xx,Salucci:2007tm,Navarro:2008kc}. In this work we adopt the  Navarro-Frenk-White (NFW) profile \cite{Navarro:1995iw} which reads,
%
\begin{equation}
\rho(r)  = \frac{r_s}{r} \frac{\rho_s}{[1+ r/r_s]^2},
\end{equation}
where $r_s=24.42$~kpc is the scale radius of the halo, as done by Fermi-LAT collaboration in ref.~\cite{Ackermann:2015zua}, and $\rho_s=0.184$ is a normalization constant to guarantee that the DM density at the location of the Sun is $0.3{\rm GeV/cm^3}$.

From eq.~(\ref{eq:flux}) it is clear that indirect detection probes complementary properties of the DM particles. It is sensitive to how the DM is distributed, to the annihilation cross section today, which might be different than the annihilation cross section relevant for the relic density, and to the WIMP mass.
Therefore, after measuring the flux in gamma-rays from a given source, we compare that with background expectations. If no excess is observed, we can choose a DM density profile and select an annihilation final state  needed for $dN/dE$, and then derive a limit on the ratio $\sigma v/m_\chi^2$ according to eq.~\eqref{eq:flux}. This is the basic idea behind experimental limits. Although, more sophisticated statistical methods have been conducted such as likelihood analysis.

An interesting aspect of indirect DM detection when it comes to probing WIMP models is the fact that if the annihilation cross section, $\sigma v$, is not velocity dependent, bounds on $\sigma v$ today are directly connected to the DM relic density. In particular, the observation of dwarf spheroidal galaxies in gamma-ray results in stringent limits on the plane {\it annihilation cross section vs WIMP mass}. If for a given channel the annihilation cross section of $10^{-26}cm^3s^{-1}$ is excluded for DM masses below $100$~GeV, it also means that one cannot reproduce the right relic density for WIMP masses below $100$~GeV \footnote{There are still some exceptions to this direct relation between non-velocity dependent annihilation cross section and relic density as discussed in detail in ref.~\cite{Griest:1990kh}.}. In other words, in this particular case, indirect detection limits will trace the relic density curve. This effect will be clearly visible in many instances.


\section{Collider Searches}

\begin{figure}[t]
\includegraphics[scale=0.8]{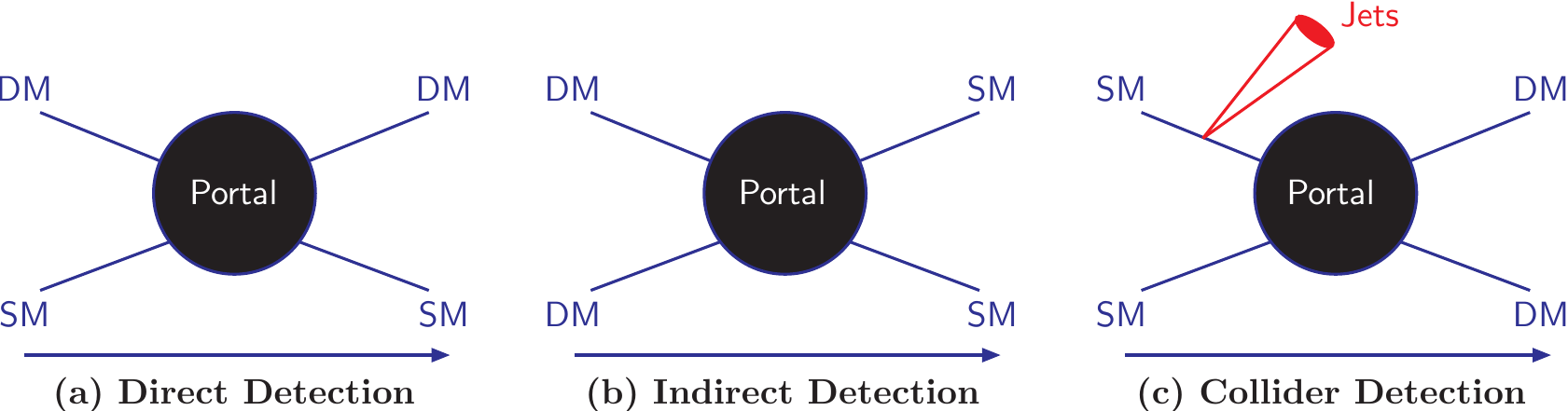}
\caption{Illustration of the DM interactions with SM particles. As of today we have no knowledge about how such interaction occurs. Thus, it is literally a black box.}
\label{monoX}
\end{figure}

LHC proton-proton collisions  might result in the  production of WIMPS in association with one or more QCD jets, photons. For other detectable Standard Model debris. Since WIMPs are electrically neutral and cosmologically stable particles, they manifest at colliders as missing energy. For this reason searches for DM are based on the observation of the visible counterpart of the event such as charged leptons, jets or a photon, generally referred to as mono-X searches, see fig.~(\ref{monoX}). By selecting events with large missing energy one can reduce the Standard Model background and potentially disentangle a DM signal. However, as mentioned above, what colliders identify is missing energy, and therefore they cannot uniquely ascertain the presence of DM in a signal event. They can simply confirm the presence of a neutral and stable particle, that might have even decayed outside the detector. 

Anyhow, colliders offer an exciting and complementary search strategy to identify WIMPs. Indeed, assuming that the production of WIMPs at colliders is uniquely connected to WIMP-nucleon scatterings at underground laboratories, one can use the non-observation signals with large missing to derive limits on the WIMP-nucleon scattering cross section \cite{Bai:2010hh,Goodman:2010ku,Goodman:2010yf,Rajaraman:2011wf,Busoni:2013lha,Alves:2014yha,Busoni:2014haa,DeSimone:2016fbz}.

We now review in some detail the specifics of given search channels for WIMPs at colliders.

\subsubsection{Mono-X Searches}
Mono-X searches stands for the search for WIMPs produced in association with one or more QCD jets or potentially other Standard Model particles, such as $\gamma$, $h$, $Z$ etc. The idea is to search for events with a jet with a high transverse momentum $p_T$ within an event with large missing transverse momentum. In particular, the most recent studies performed at the LHC include up to four jets and require the leading jet to have $p_T > 250$~GeV \cite{Ratti:2016pwi,Tolley:2016lbg}, while others do not limit the number of jets while selecting events with at least one jet with $p_T > 100$~GeV \cite{CMS:2016pod}. While being more inclusive, these recent searches have become more challenging due to the number of jets analyzed, requiring a substantial improvement on the background coming from $Z+jet$ and $W+jet$ channels.

There are important detector effects, such as fake jets, and QCD backgrounds that weaken the LHC sensitivity to WIMPs, and for these reasons mono-jet searches are subject to large systematics. Nevertheless, fortunately an enormous effort has been put forth in this direction with data driven background and optimized event selections, which combined with the increase in luminosity has led to an overall improvement on the LHC sensitivity to WIMPs.

That said, in the review, we will be using the latest results from CMS and ATLAS collaborations in the search for DM base on mono-X searches \cite{Aaboud:2016tnv,CMS:2016pod}.

Now we discuss the WIMP production at colliders we will address another collider limits relevant for DM purposes which has to do with the invisible width of the Higgs boson.
\subsubsection{Invisible Higgs Decays}
If WIMPs are lighter than $62.5$~GeV, the Higgs boson might invisibly decay into WIMP pairs.
In this case, one can use bounds from LHC on the invisible branching ratio of the Higgs, ${\rm Br (h\rightarrow inv)} \leq 0.25$ at 95\% C.L. \cite{Aad:2015pla,Khachatryan:2016whc}, to set constraints on WIMP models . Throughout the manuscript whenever applicable we compute the invisible decay rate of the Higgs into WIMPs and impose the upper limits above to obtain the limits displayed in the figures.

\subsubsection{Invisible Z Decays}

The decay width of the Z boson has been precisely measured and therefore stringent limits can be derived on any extra possible decay mode of the Z boson. In some of the models we discuss further, the DM particle does couple to the Z boson, thus when mass of the DM is smaller than half of the Z mass stringent limits are applicable. In particular, one can use only direct measurements of the invisible partial width using the single photon channel to obtain an average bound which is derived by computing the difference between the total and the observed partial widths assuming lepton universality. The current limit is
$\Gamma {\rm  (Z\rightarrow inv)} \leq 499 \pm 1.5$~MeV \cite{Olive:2016xmw}.

\section{Model Setup: Dark Portals}

In order to maximally profit of the information from the different kind of experimental searches we need an efficient interface between the experimental outcome and theoretical models. The processes responsible for the DM relic density and its eventual detection can be described by simple extensions of the SM in which a DM candidate interacts with the SM states (typically the interactions are limited to SM fermions) through a mediator state (dubbed portal). This idea is at the base of the so-called ``Simplified Models''~\cite{DiFranzo:2013vra,Berlin:2014tja,Abdallah:2014hon,Buckley:2014fba,Abdallah:2015ter,Duerr:2015wfa,Godbole:2015gma,Baek:2015lna,Carpenter:2016thc,Sandick:2016zut,Bell:2016ekl,Bell:2016uhg,Bauer:2016gys,Carpenter:2016thc,Sandick:2016zut,Sandick:2016zut,Khoze:2017ixx,ElHedri:2017nny} which are customarily adopted especially in the context of collider studies, see e.g., refs.~\cite{Jacques:2015zha,Xiang:2015lfa,Backovic:2015soa,Bell:2015rdw,Brennan:2016xjh,Goncalves:2016iyg,Englert:2016joy,Boveia:2016mrp,DeSimone:2016fbz,Liew:2016oon,Bauer:2017ota,Albert:2017onk,Kraml:2017atm}. 

We will adopt, in this review analogous setups, referring to them as ``Dark Portals''. We will then analyze a series of scenarios in which spin-0, spin-1/2 and spin-1 DM candidates (we will distinguish whenever relevant the cases of self- and not self-conjugated DM) interact with SM fields (mostly fermions) through spin-0 or spin-1 mediator fields~\footnote{We will briefly consider a single case of spin-1/2 mediator. We won't consider higher spin assignations. For these cases the interested reader can look for example at ref.~\cite{Kraml:2017atm}}. Compared to similar works we will, however adopt a broader perspective. In most of the cases considered we will indeed argue possible theoretical completions of the simplified models adopting specific choices of the parameters inspired by them. In the same fashion we will account, in our analysis, besides experimental constraints, the theoretical limitations of these frameworks (see e.g., refs.~\cite{Kahlhoefer:2015bea,Englert:2016joy,Bell:2016uhg,Goncalves:2016iyg,Bell:2016ekl} for more extensive discussions.)

Despite their simplicity, Dark Portals, with the exception of SM mediators, feature still several parameters; we will therefore have to rely on some specific assumptions in presenting our results. Throughout this work we will then follow these main guide lines for our analysis:
\begin{itemize}
\item we will always assume real couplings, hence no CP-violation; 
\item We will consider as free parameters the mass of the DM and of the portal mediator while the model couplings we will set to $O(1)$ values unless this option is precluded by theoretical considerations. There is no loss of generality in this choice. This choice maximizes the experimental sensitivity of these scenarios and allows at the same time to achieve the correct DM relic density in broad regions of the parameter space. Because of the already highlighted correlations between DM processes, although experimental limits would be more easily overcome by decreasing the model couplings, an analogous suppression of the DM pair annihilation would occur so that the DM would be in general overabundant unless rather fine tuned solutions, like s-channel resonances or coannihilations would be adopted. 
An exception is, of course, represented by the case of SM portals which are completely specified by just two parameters, i.e. the DM mass and couplings;
\item Our analysis will be mostly focused on the comparison between DM relic density and Direct Detection, especially in light of the incoming data releases of next generation multi-TON detectors. Collider and Indirect Detection limits will be reported only when their are competitive or feature a clear complementarity with DD limits.
\end{itemize}

Within these guidelines we will review a multitude of models which have been classified in five categories:
\begin{enumerate}
\item SM portals: Here the portal of DM interactions is represented by the Higgs or the $Z$-boson. As already pointed this scenarios are the most predictive since have only two free parameters. On the other hand this turns also in the most stringent limits so that these scenarios are strongly disfavored already by present limits;
\item BSM s-channel portals: Here the DM is coupled with SM fermions either by a spin-0 (real) or by a spin-1 new neutral state. These scenarios, besides DM portals, are the most sensitive to DM Direct Detection.
\item Portals evading Direct Detection: here we will instead consider the case, a pseudo-scalar mediator, in which constraints from Direct Detection are particularly weak so that the complementarity with other search strategies becomes crucial. We will also considered a more theoretically refined scenario in which the pseudoscalar mediator is part of a complex field. The presence of an additional scalar component will reintroduce DD bounds. A broad region of parameter space for thermal DM is nevertheless re-opened by considering a very light pseudoscalar, which can be interpreted as a pseudo-goldostone boson of a global symmetry carried by the original complex field;

\item Portals to secluded sectors: we assume here that the mediator field cannot be directly coupled with SM fermions. Even in these kind of constructions a portal can be originated by mass mixing with the SM Higgs, in the case of spin-0 mediators, and by kinetic mixing with the Z-boson in the case of a spin-1 field, so that the DM actually interacts through a double s-channel mediator, represented by a SM and a BSM state.
\item t-channel portals: in this alternative version of the Dark Portals the mediator field has not trivial quantum numbers with respect to the SM gauge group. In these kind of setups DM annihilation arise from t-channel interactions while s-channel interactions are responsible for Direct Detection. 
\end{enumerate}


\section{SM portals}
The first class of models which will be object of study are the SM Dark Portals~\footnote{An analogous study has been performed in ref.~\cite{Escudero:2016gzx}. Our results are in substantial agreement with the ones reported in this reference.}, i.e. models in which the DM interacts with the SM state through the Higgs or the Z-boson. In the case the DM is a pure SM singlet, gauge invariant renormalizable operator connecting the DM with the Z or the Higgs boson can be build only in the latter case and only for scalar and vectorial DM. In the other cases one should rely either on higher dimensional operators, or on the case that the coupling with the Higgs and/or the Z is originated by their mixing with new neutral mediators. The latter case can imply the presence of additional states relevant for the DM phenomenology and will be then discusses later on in the text. We will instead quote below some example of higher dimensional operator but we will not refer to any specific construction for our analysis. Alternatively one could assume that the DM has some small charge under $SU(2)$ or $U(1)_Y$, see e.g., refs.~\cite{Cohen:2011ec,Calibbi:2015nha,Yaguna:2015mva,Berlin:2015wwa,Angelescu:2016mhl,Kearney:2016rng}; we will not review these scenarios here.


\subsection{Higgs portal}

The most economical way to connect a SM singlet DM candidate with the SM Higgs doublet $H$ is through four field operators built to connect the Higgs bilinear $H^\dagger H$, which is a Lorentz and gauge invariant quantity, with a DM bilinear. Assuming CP conservation, the possible~\footnote{We limit, for simplicity to the lowest dimensional operators. Higher dimensional operators are discussed, for example, in~\cite{Greljo:2013wja}}  operators connecting the Higgs doublet with scalar, fermion and vector DM are given by \cite{Silveira:1985rk,McDonald:1993ex,Burgess:2000yq,Kanemura:2010sh,Andreas:2010dz,Djouadi:2011aa,Mambrini:2011ik,LopezHonorez:2012kv,Djouadi:2012zc}: 

\begin{equation}
\label{eq:HpLagrangian}
\xi \lambda^H_{\chi} \chi^{*} \chi H^\dagger H,
\hspace*{1cm} \xi \frac{\lambda^H_{\psi}}{\Lambda} \ovl \psi \psi H^{\dagger} H
~~~~{\rm and~~~~} \xi \lambda^H_V V^\mu V_\mu H^\dagger H,
\end{equation}
where, in the unitary gauge, $H ={\left(0\,\,\,\frac{v_h+h}{\sqrt{2}}\right)}^T$ with $h,\,v_h$ denoting the physical SM Higgs boson, Vacuum Expectation Value (VEV) and $\xi=1/2 (1)$ in case the DM is (not) its own antiparticle. From eq.~(\ref{eq:HpLagrangian}) note that stability of the DM is protected either by a discrete $\mathbb{Z}_2~({\rm for}~ \psi,\,V_\mu$ and when $\chi=\chi^*$) or by a $U(1)~({\rm for}~\chi\neq \chi^*)$ symmetry.

As already pointed in the case of scalar and vector DM it is possibly to rely on a dimension-4 renormalizable operator; on the contrary fermion DM requires at least a dimension-5 operator which depends on an unknown Ultra-Violet (UV) scale $\Lambda$.

After EW symmetry breaking (EWSB), trilinear couplings between the Higgs field $h$ and DM pairs are induced. In the case of fermionic DM it is possible to absorb the explicit $\Lambda$ dependence by a redefinition of the associated coupling, i.e., $\lambda_\psi^H \frac{v_h}{\Lambda}$ as $\lambda_\psi^H $, so that it does not appear explicitly in computations.

The models defined by Lagrangians of eq.~(\ref{eq:HpLagrangian}) have only two free parameters, the DM masses $m_{\chi,\psi,V}$ and couplings $\lambda_{\chi,\psi,V}^H$. The constraints on these models can be then straightforwardly summarized in bi-dimensional planes. 
%
\begin{figure}[t]
\subfloat{\includegraphics[width=6.5 cm]{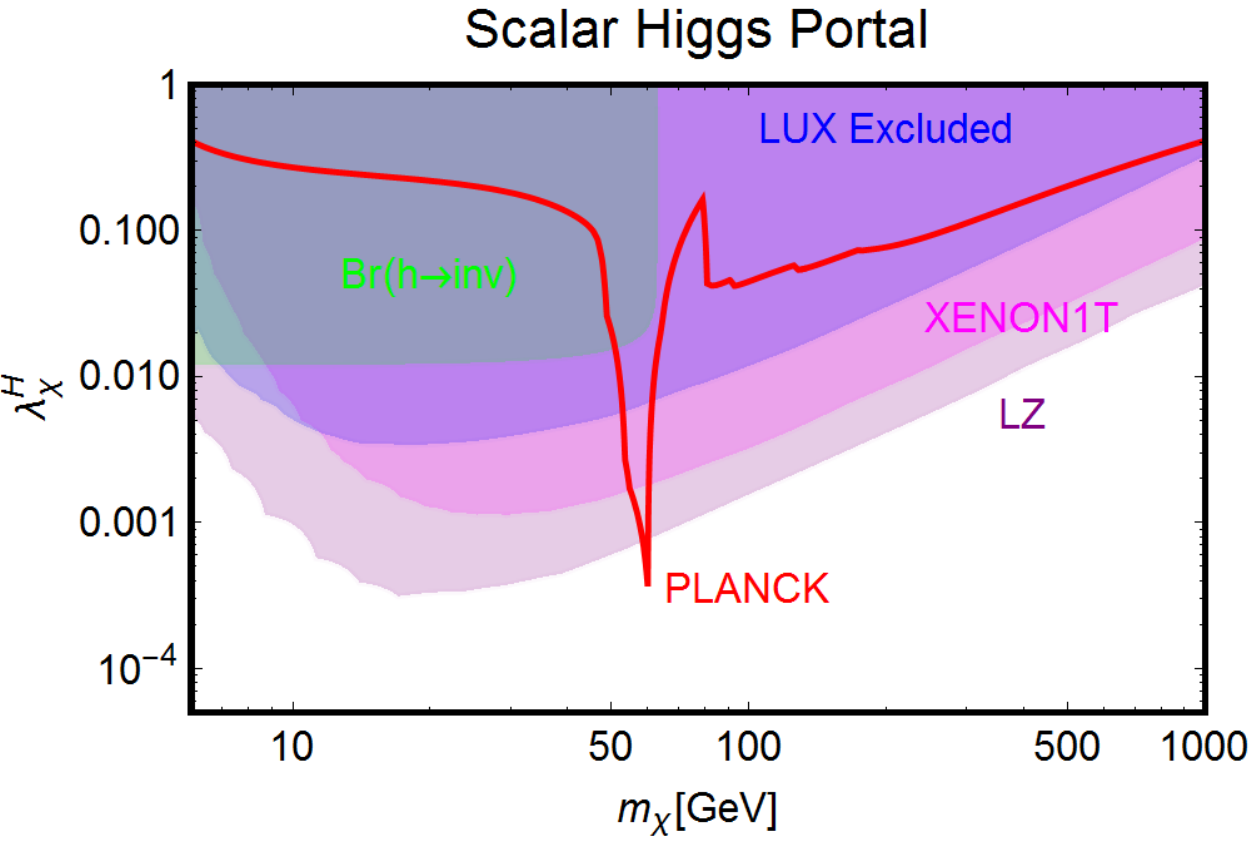}}
\subfloat{\includegraphics[width=6.5 cm]{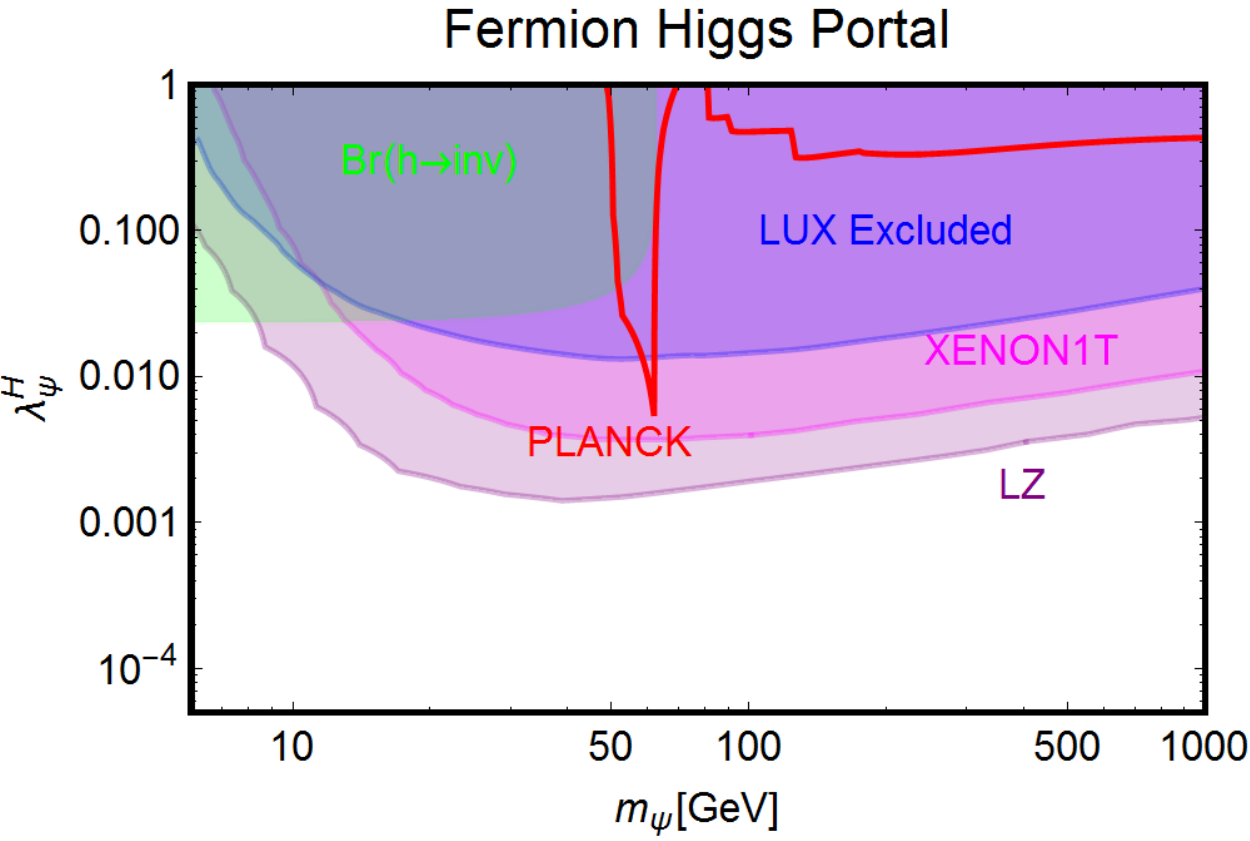}}\\
\subfloat{\includegraphics[width=6.5 cm]{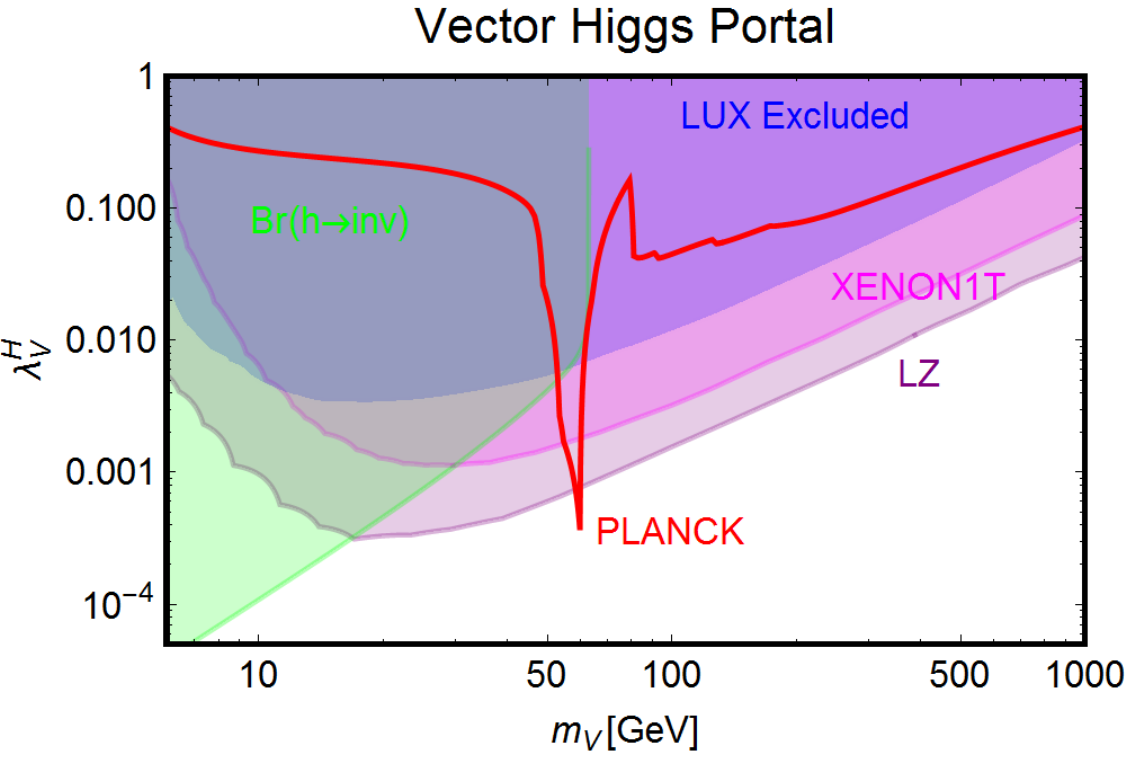}}
\caption{\footnotesize{The SM Higgs portal with scalar (upper left panel), fermionic (upper right panel) and vector (bottom) DM. In each plot, the red line represents the model points featuring the correct DM relic density. The blue region is excluded by the current LUX limits. The magenta coloured region would be excluded in case of absence of signals in XENON1T after two years of exposure time while the purple region is within reach of future LZ limits. Finally, the green region is excluded because of a experimentally disfavored invisible branching fraction of the SM Higgs boson.}}
\label{fig:ScalarHp}
\end{figure}

In figs.~(\ref{fig:ScalarHp}) we summarize our results for scalar,
fermion and vector DM, respectively. All the plots report basically
three set of constraints~\footnote{We will report in
the main text just the results of the analysis.
Analytical expressions of the relevant rates are
extensively reported in the appendix.}. The first one
(red contours) is represented by the achievement of the
correct DM relic density. The DM annihilates into SM
fermions and gauge bosons, through s-channel exchange
of the Higgs boson, and, for higher masses, also into
Higgs pairs through both s- and t-channel diagrams (in
this last case a DM particle is exchanged). Since the
coupling of the Higgs with SM fermions and gauge bosons
depends on the masses of the particles themselves, the
DM annihilation cross-section is suppressed, at the
exception of the pole region $m_\chi \sim m_h /2$,
until the $WW$, $ZZ$ and $\ovl t t$ final states are
kinematically accessible. Even in this last case, the
cosmologically allowed values for the couplings are 
in strong tension with the constraints from DM Direct
Detection (DD), which for all the considered spin
assignation of the DM, arise from Spin Independent (SI)
interactions of the DM with the SM quarks originated by
t-channel exchange of the Higgs boson. As can be easily
seen the entire parameter space corresponding to
thermal DM is already ruled out, at the exception,
possibly, of the pole region, for DM masses at least
below 1 TeV. Eventual surviving resonance regions will
be ruled-out in case of absence of signals at the
forthcoming XENON1T. As expected, the most constrained
scenario is the fermionic DM one because of the
further suppression of the p-wave suppression of its annihilation cross-section. 

Notice that, scalar and
vectorial DM, due to the s-wave annihilation cross
section, might also be probed through Indirect
Detection (ID). The corresponding limits are
nevertheless largely overcome by the ones from DD and
have been then omitted for simplicity. The limits from
DD experiments are complemented at low DM masses, i.e.,
$m_{\chi,\psi,V} < m_h/2$, by the one from invisible
decay width of the Higgs. Indeed this constraint would
exclude DM masses below the energy threshold of DD
experiments. All findings are in agreement with recent studies in the topic \cite{Cline:2013gha,Queiroz:2014pra,Buckley:2014fba,Kumar:2015wya,Abdallah:2015ter,Abercrombie:2015wmb,Chen:2015dea,Fedderke:2015txa,Anchordoqui:2015fra,Freitas:2015hsa,Dutra:2015vca,Duch:2015jta,Han:2016gyy,Bambhaniya:2016cpr}.

%

\subsection{Z-portal}

An interaction between the $Z$-boson and a SM singlet DM candidate is not gauge invariant for any dimension-4 operators. In the case of scalar and fermionic DM models, the simplest option is to consider a dimension 6 operator~\footnote{Similarly to the case of the Higgs portal we just quote, as an example, the lowest dimension operator. This is however not the only possible option.}~\cite{Cotta:2012nj,deSimone:2014pda,Kearney:2016rng}. In the case of scalar DM it is of the form: 
\bea
\mathcal{L}=  \lambda_\chi \frac{H^\dagger \overleftrightarrow{D^{\mu}} H}{\Lambda^2}
\chi^* \overleftrightarrow{\partial_{\mu}} \chi,
\eea
which give rise to a trilinear interaction between the $Z$ and a DM pair once the Higgs field in the Lagrangian is replaced by its VEV, so that $H \overleftrightarrow{D^\mu} H \rightarrow \frac{g v_h^2}{4 \cos \theta_W}$. $\Lambda$ is again the relevant cutoff scale of the effective theory. Similarly to the case of the fermionic Higgs portal we can absorb it in the definition of an dimensionless coupling as $\lambda^Z_\chi\equiv \lambda_\chi v^2_h/\Lambda^2$. 
In addition, after EWSB, an effective dimension-4 interaction like $(g^2/16\cos^2\theta_W)\lambda^{ZZ}_{\chi\chi}{|\chi|}^2Z^\mu Z_\mu$ can emerge from the dimension-6 SM gauge invariant operator $\lambda_{\chi\chi}(D^\mu H)^\dagger D_\mu H {|\chi|}^2/\Lambda^2$ such that $\lambda^{ZZ}_{\chi\chi}=\lambda_{\chi\chi}v^2_h/\Lambda^2$. For simplicity we maintain a rescaling with powers of $g$.

The interaction Lagrangian for the DM,
along with the relevant SM parts, can thus be written as: 
\bea
\mathcal{L}=&&i \frac{g}{4 c_W} \lambda_\chi^Z 
\chi^* \overleftrightarrow{\partial_{\mu}} \chi Z^\mu
+\frac{g}{4 c_W} \sum_f \overline{f} \gamma^\mu \left(V_f^Z-A_f^Z \gamma_5\right) f Z_\mu + \frac{g^2}{16 c^2_W} \lambda^{ZZ}_{\chi \chi} {|\chi|}^2 Z^\mu Z_\mu,
\eea
where $c_W=\cos\theta_W$ and $\theta_W$ is Weinberg angle \cite{Amsler:2008zzb}. Note that we have used a normalization 
of $g/4\cos\theta_W$ throughout in analogy to the SM
$\ovl f f Z$ couplings.

The interaction Lagrangian for fermion DM is built in a similar fashion as the scalar case. In the case of Dirac DM the starting operator is:
\begin{equation}
\mathcal{L}= \frac{H^\dagger \overleftrightarrow{D_\mu} H}{\Lambda^2}\left( \overline{\psi}\gamma^\mu \left(v_\psi^Z-a^Z_\psi \gamma^5\right) \psi\right),
\end{equation}
which, after the EWSB, together with
the apposite SM part leads to:
\begin{equation}
\label{eq:Zlagrangian}
\mathcal{L}=\frac{g}{4 \cos\theta_W} \overline{\psi}\gamma^\mu \left(V_\psi^Z-A^Z_\psi \gamma^5\right) \psi Z_\mu+
\frac{g}{4 \cos\theta_W}
\sum_f \overline{f}\gamma^\mu \left(V^Z_f-A^Z_f \gamma^5\right) f Z_\mu,
\end{equation}
with ${V}_\psi^Z={v}_\psi^Z \frac{v_h^2}{\Lambda^2}$ and ${A}_\psi^Z={a}_\psi^Z \frac{v_h^2}{\Lambda^2}$. In the case of Majorana DM $V_\psi^Z=0$ and we rescale the remaining DM coupling by a factor $1/2$. 

In the case of spin-1 DM we will consider two possible kind of interactions for, respectively, self- (abelian) and not self-conjugated (non abelian) DM. For the latter we can write the following Lorentz invariant interaction:
\bea
\label{eq:ZVlagrangian}
\mathcal{L}&&=\frac{g}{4 \cos\theta_W} \eta^Z_V [[VVZ]]
+ \frac{g}{4 \cos\theta_W} \sum_f \overline{f}\gamma^\mu \left(V^Z_f-A^Z_f \gamma^5\right) f Z_\mu,\nonumber \\
{\rm with~~}[[VVZ]] &&\equiv i \left[V_{\mu\nu} V^{\dagger\, \mu}Z^\nu-V^\dagger_{\mu\nu} V^{\mu}Z^\nu+\frac{1}{2}Z_{\mu\nu} \left(V^\mu V^{\dagger\,\nu}-V^\nu V^{\dagger\,\mu}\right)\right],
\eea
where $V_{\mu\nu},V^\dagger_{\mu\nu},Z_{\mu\nu}$ represent the respective field strengths. In eq.~(\ref{eq:ZVlagrangian}) the $[[VVZ]]$ coupling is normalized as $g/4\cos\theta_W$ while the model specific information are parametrized as $\eta^Z_V$.
%
In the case of self-conjugate spin-1 DM an interaction with the gauge boson be built through the Levi-Civita symbol as ref.~\cite{Mambrini:2009ad}:
\begin{equation}
\label{eq:abelianZ}
\mathcal{L}=\frac{g}{4 \cos\theta_W} \eta^{Z}_V \epsilon^{\mu \nu \rho \sigma} V_{\mu} Z_{\nu} {V}_{\rho \sigma}
+\frac{g}{4 \cos\theta_W}\sum_f\overline{f}\gamma^\mu \left(V^{Z}_f-A^{Z}_f \gamma^5\right) f Z_\mu.
\end{equation}

Similarly to the previous cases the coupling $\eta_V^Z$ in~(\ref{eq:abelianZ}) encodes a cut-off scale (see e.g.~\cite{Anastasopoulos:2006cz,Antoniadis:2009ze,Dudas:2009uq,Dudas:2013sia} for examples of construction of the effective theory). More contrived is instead a theoretical derivation of~(\ref{eq:ZVlagrangian}).

\begin{figure}[t]
\includegraphics[width=6.5 cm]{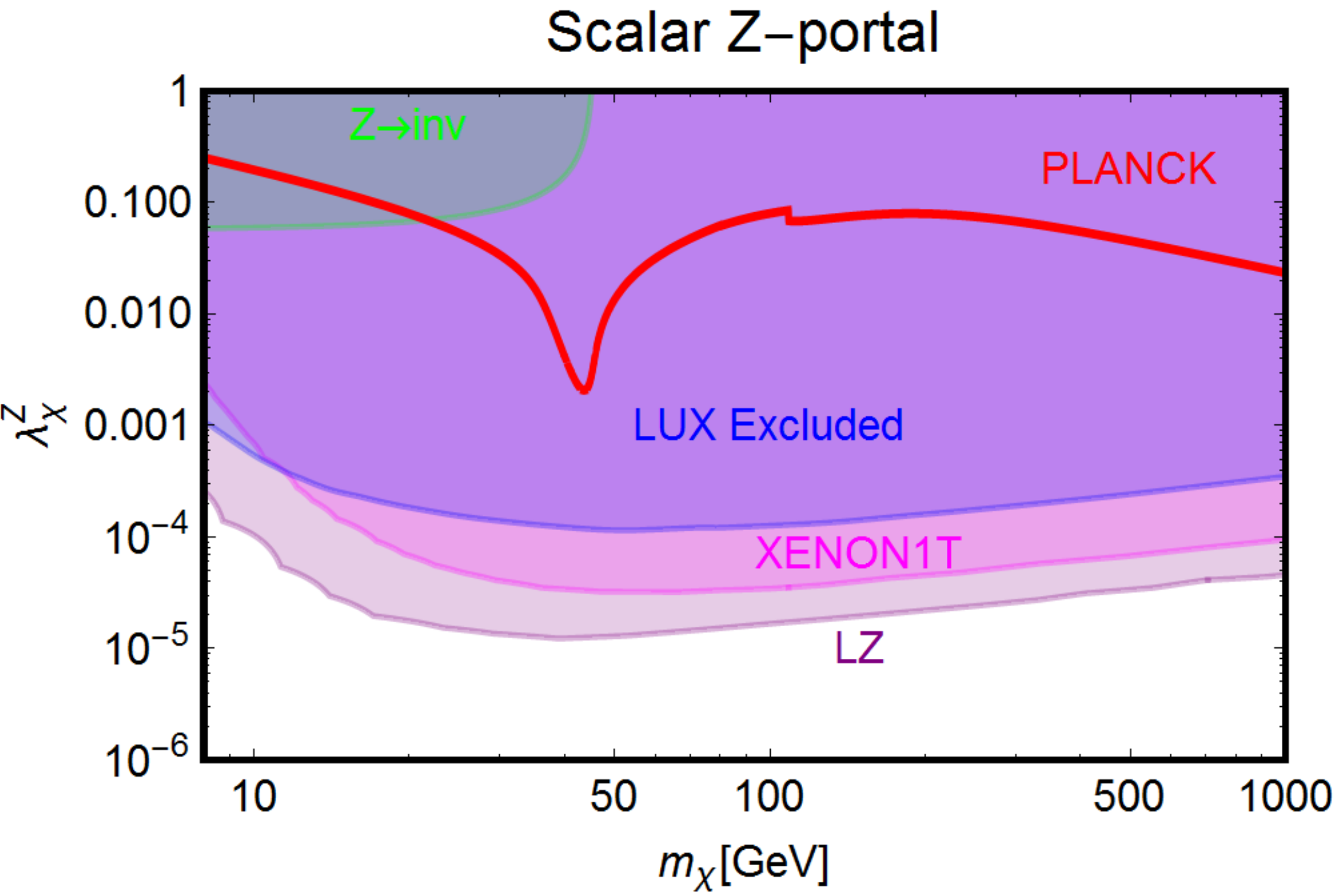}
\caption{\footnotesize{Combined constraints for Z-portal with scalar DM. Colour specifications are the same as
fig.~(\ref{fig:ScalarHp}), except the fact that now 
the green coloured region represents experimentally
excluded invisible decay width of $Z$-boson.}}
\label{fig:Zportal}
\end{figure}

\begin{figure}[t]
\includegraphics[width=6.5 cm]{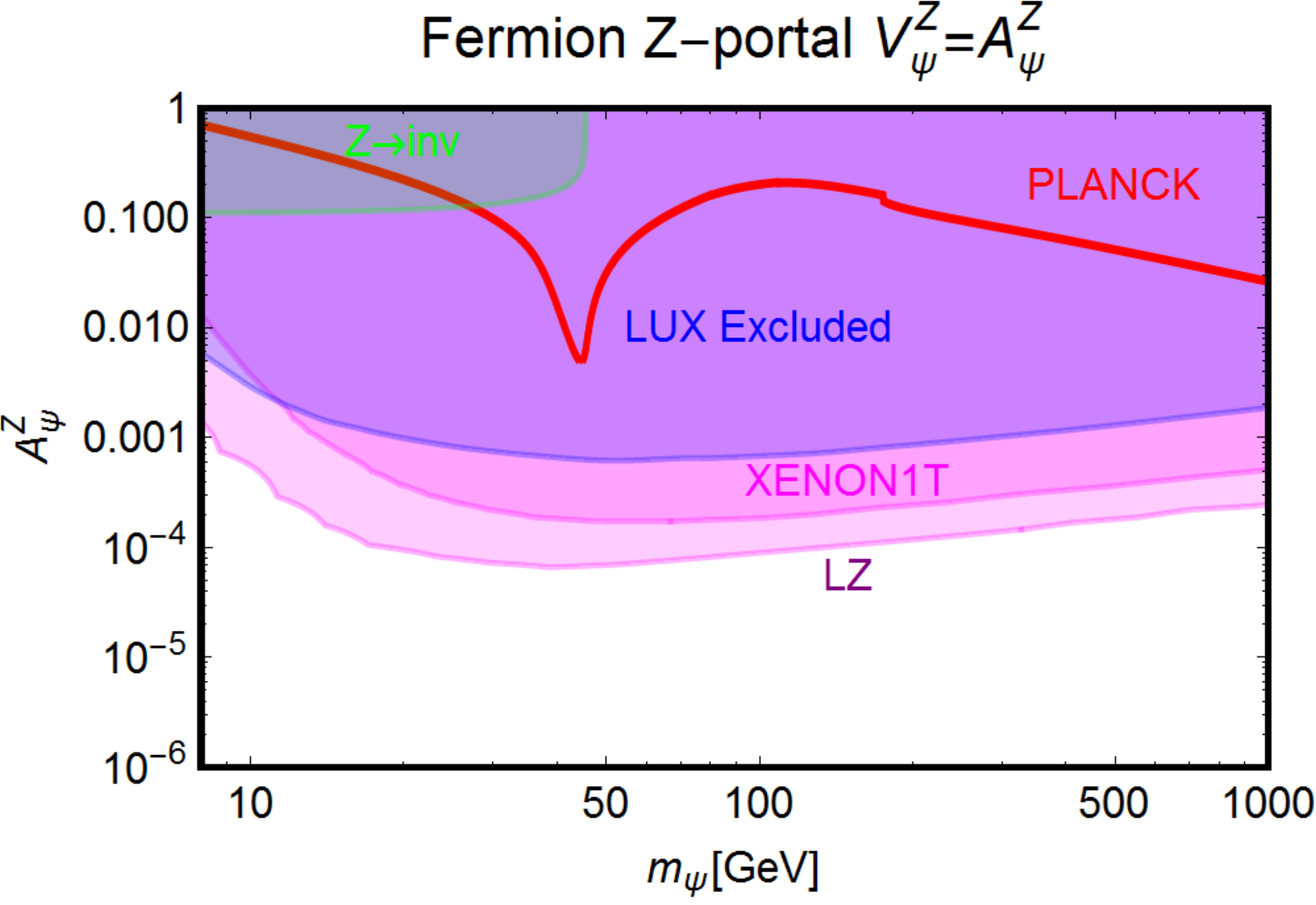}
\includegraphics[width=6.5 cm]{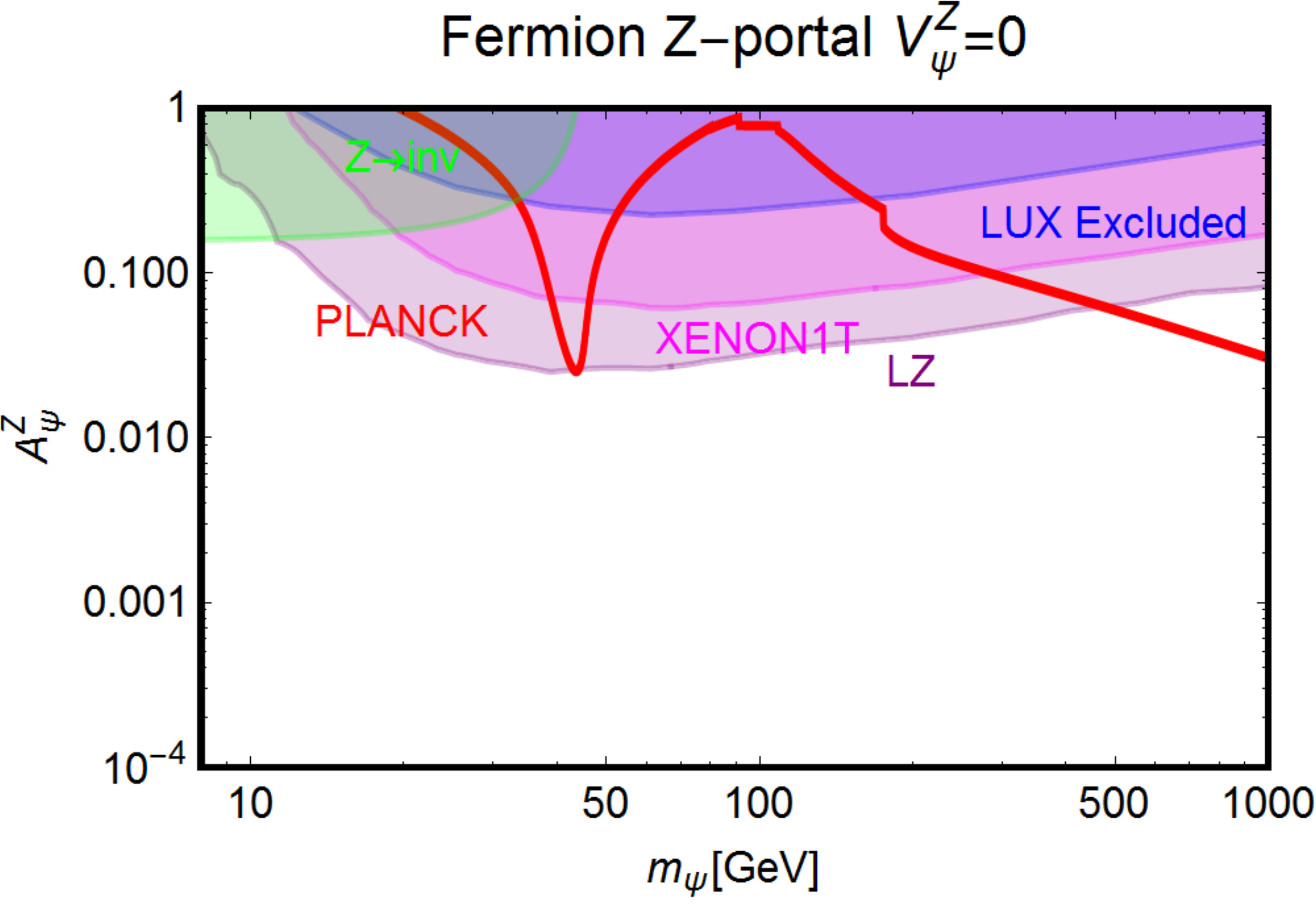}
\caption{\footnotesize{The same as fig.~(\ref{fig:Zportal}) but for Dirac fermion DM with both vectorial and axial couplings (left panel), set to the same value, and only axial couplings (right panel) with the $Z$-boson.}
}
\label{fig:FZportal}
\end{figure}

\begin{figure}[t]
\includegraphics[width=6.5 cm]{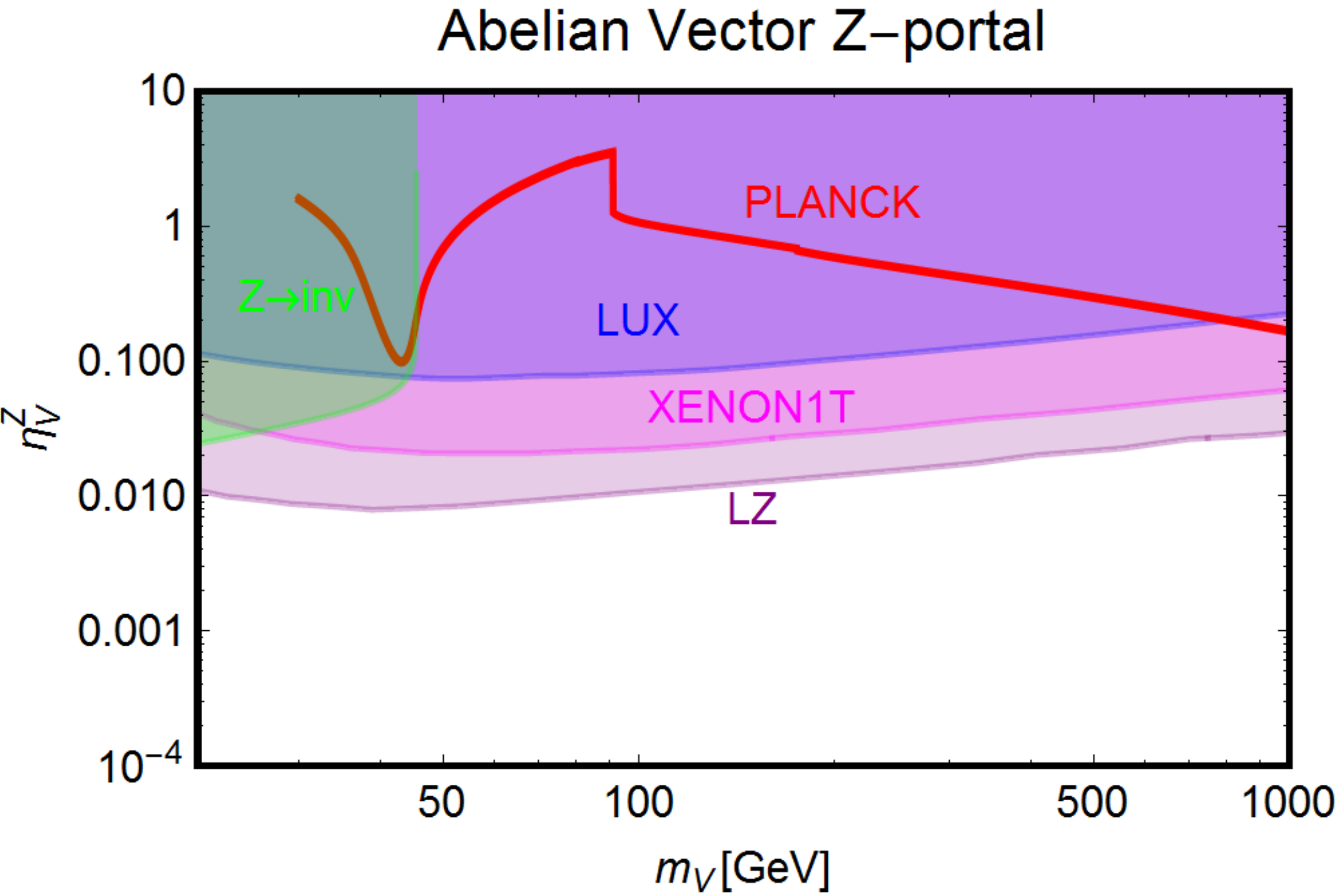}
\includegraphics[width=6.5 cm]{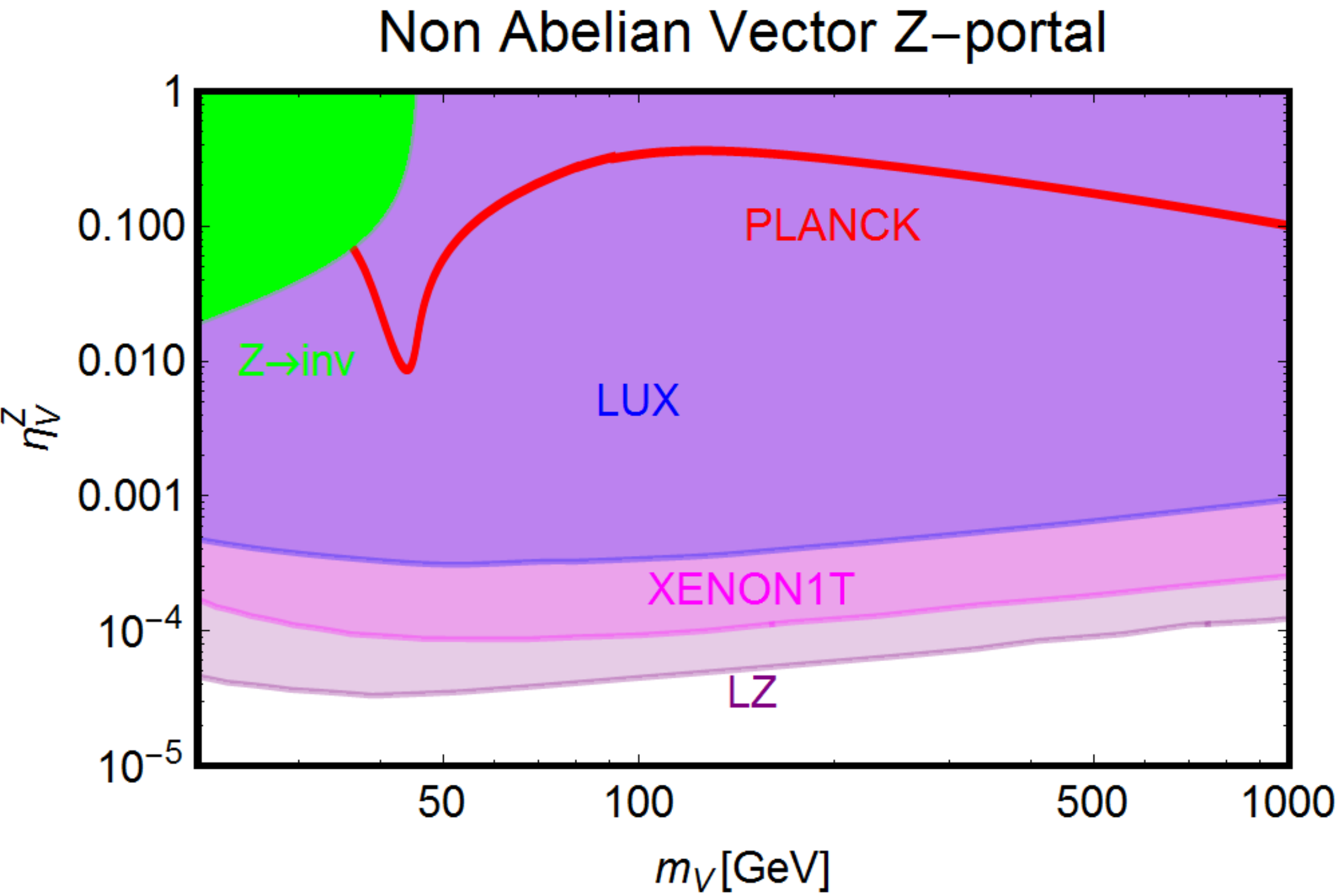}
\caption{\footnotesize{The same as fig.~(\ref{fig:Zportal}) but for Vector DM with (i) Abelian case (left-panel) and (ii)
Non-Abelian case (right-panel).}}
\label{fig:VZportal}
\end{figure}

Similarly to the Higgs portal, the Z-portal models are fully defined by two parameters so that one can repeat the same kind of analysis performed in the previous subsection. The results are summarized on figs.~(\ref{fig:Zportal})-(\ref{fig:VZportal})~\footnote{Similarly to the Higgs portal case we will report in the main text only the main results while discussing the computation more in detail in the appendix.}. 

%

As evident, in all but the Majorana $Z$-portal case, thermal DM is already excluded, even for masses above the TeV scale, by current constraints by LUX. These constraints are even stronger with respect to the case of the Higgs portal. This because, apart the lighter mediator, the scattering cross section on Xenon nuclei is enhanced by the isospin violation interactions of the $Z$ with light quarks. Low DM masses, possibly out of the reach DD experiments, are instead excluded by the limit on the invisible width of the $Z$. As already pointed the only exception to this picture is represented by the case of Majorana DM where the SI component of the DM scattering cross-section is largely suppressed due to the absence of a vectorial coupling of the DM with the $Z$. This scenario is nevertheless already (partially) within the reach of current searches for a Spin Dependent (SD) component of the scattering cross-section. The increased sensitivity by XENON1T will allow to exclude DM masses below 300 GeV, at the exception of the ``pole'' region.

\section{BSM s-channel portals}

The results presented in the previous case for the Higgs and $Z$-portal will be generalized and discussed in more details in the case of generic, BSM spin-0 and spin-1 mediators interacting with pairs of scalar, fermion or vector DM fields. Contrary to the case of Higgs and Z portal, interactions of the mediators with the gauge bosons are not mandatory. We will thus stick, in this section to the case, analogous to the so-called simplified models (citation), in which the DM is coupled only to SM fermions. The case of interactions with gauge bosons will be discussed separately later on in the text. Contrary to the aforementioned models, we will however assume interactions with both quarks and leptons. 

\subsection{Spin-0 portals}

\subsubsection{Scalar Dark Matter}

We will consider the following Lagrangian:
\begin{equation}
\label{eq:bsms0sdm1}
\mathcal{L}=\xi \mu_\chi^S {|\chi|}^2 S+ \xi \lambda_\chi^{S^2} {|\chi|}^2 S^2+ \frac{c_S}{\sqrt{2}} \frac{m_f}{v_h}\ovl f f S,
\end{equation}
where $S$ is a real scalar field and $\xi$ denotes the normalization factor, accounting, similarly to previous section, for the case DM coinciding (or not) with its own antiparticle. In the case of SM fermions we have assumed a Yukawa-like structure of the couplings while for the scalar $(\chi)$ we have parametrized all the information, including
possible normalization factors (e.g., factor of $1/2$ in the second term of eq.~(\ref{eq:bsms0sdm1})), in the respective couplings. Note that $\mu_\chi^S$ parameter has
the dimension of mass. Unless differently stated we will assume $\mu_\chi^S=\lambda_\chi^S m_S$ with $\lambda_\chi^S$ being a dimensionless coupling
and $m_S$ as the mass of $S$. We will also add self interaction term for the scalar field given by:
\begin{equation}
\label{eq:scalar_self}
\mathcal{L}_S=-\frac{1}{3!}m_S \lambda_S S^3.
\end{equation}

The assignation for the dimensional couplings, as well as the introduction of the Lagrangian term in eq.~(\ref{eq:scalar_self}), are inspired to scenarios in which the scalar field $S$ acquires a VEV. In this setup the Lagrangian of eq.~(\ref{eq:scalar_self}) originates from the quartic term in the scalar potential whose presence cannot be forbidden by any symmetry argument. In the same fashion a quartic interaction term $S^2 H^\dagger H$ with the SM Higgs doublet should be also included, responsible of a mixing of the $S$ and $h$ states. For simplicity we will assume here that the coupling of this last operator is negligible and postpone to a dedicated section the the discussion of the most general case.

\begin{figure}[t]
\includegraphics[width=4.9 cm]{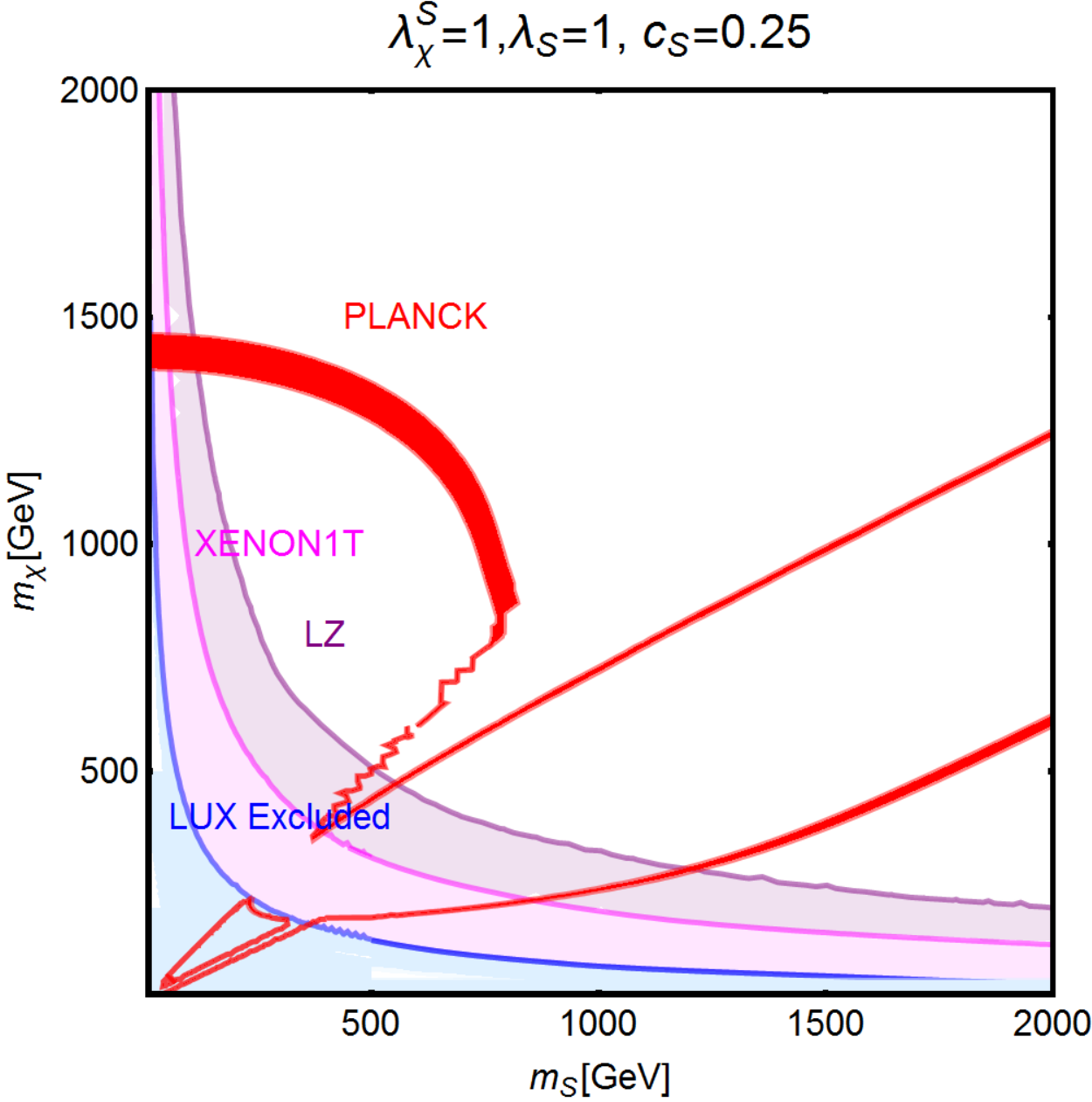}
\includegraphics[width=4.9 cm]{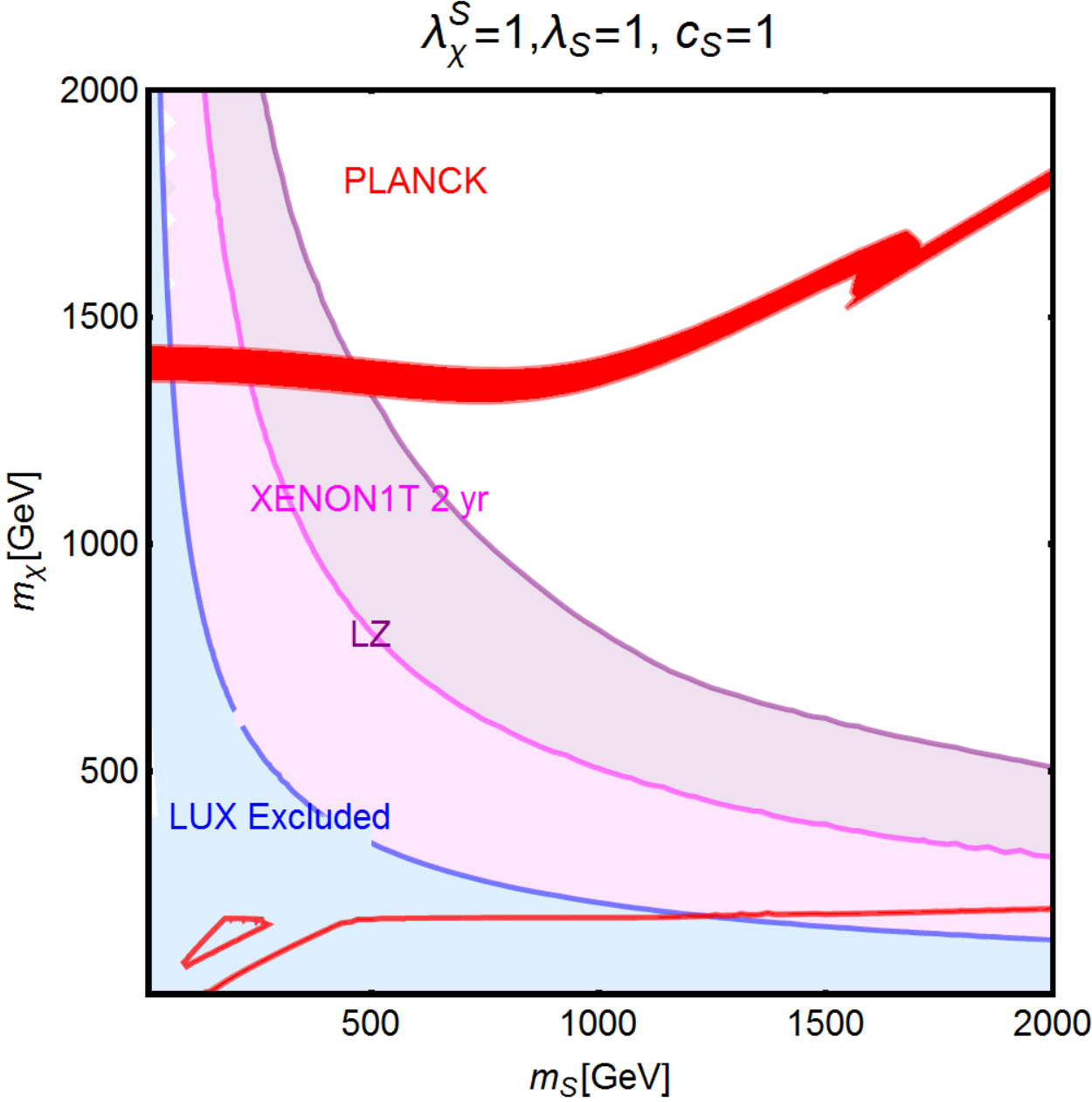}
\includegraphics[width=4.9 cm]{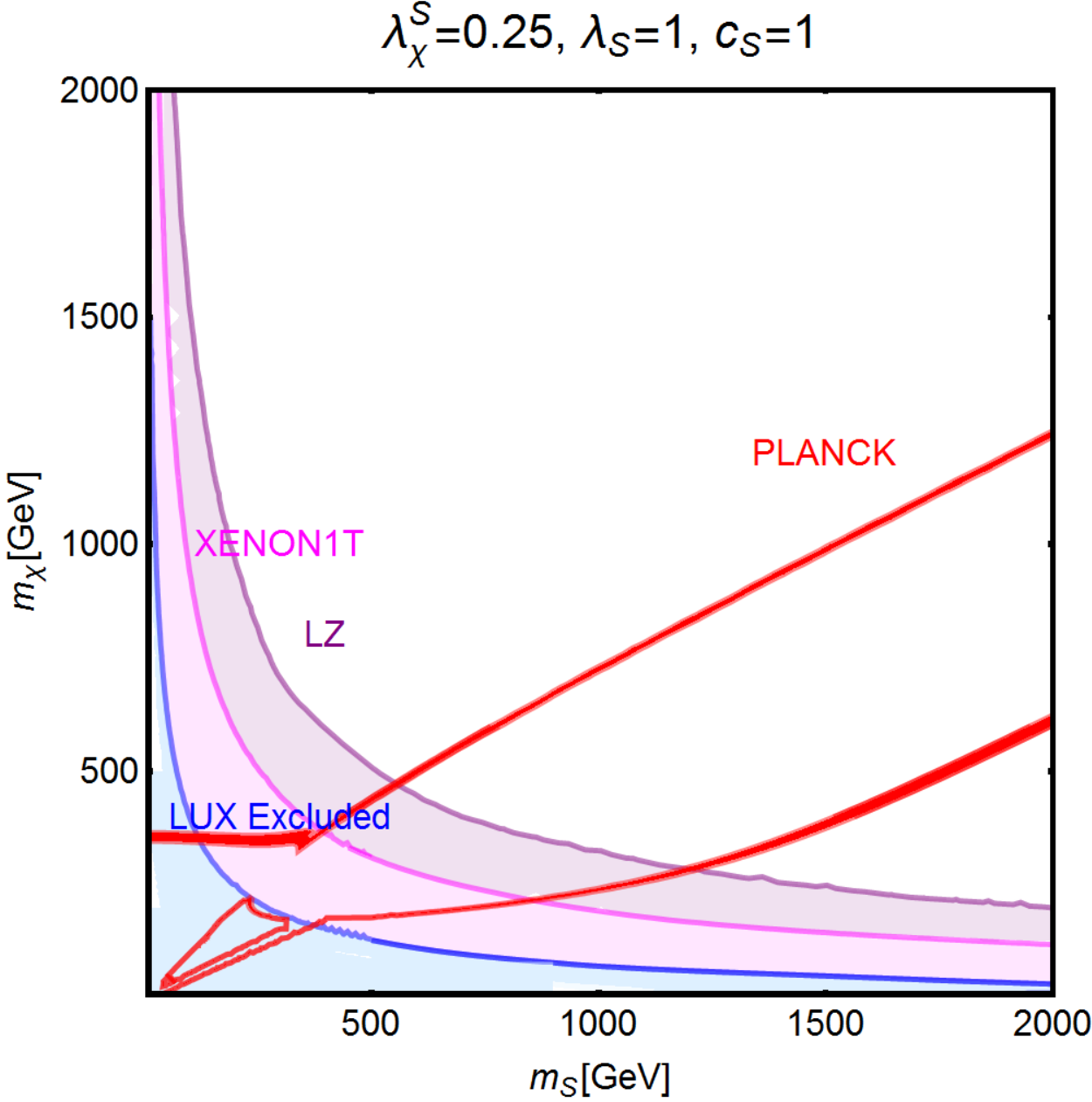}
\caption{\footnotesize{Combined constrains for a scalar DM with scalar mediator scenario in the bi-dimensional plane $(m_S,m_\chi)$ for three assignations of the relevant couplings, i.e., $(\lambda_\chi^S,\,\lambda_S,\,c_S)=(1,\,1,\,0.25),\, (1,\,1,\,1)\,{\rm and~} (0.25,\,1,\,1)$ (from left to right). Here the iso-contours of the correct DM relic density are represented by red coloured bands. The blue, magenta and purple coloured regions represent the current exclusion by LUX and the projected sensitivity of XENON1T (assuming 2 years of exposure time) and LZ, respectively.}}
\label{fig:Sportal}
\end{figure}

Contrary to the case of the SM portals, which have only the DM and its coupling as free parameters, we have expressed, as reported on fig.~(\ref{fig:Sportal}), our main results in the bi-dimensional plane $(m_\chi,m_S)$ for three coupling assignations $(\lambda_\chi^S,\lambda_S,c_S)=(1,1,0.25), (1,1,1), (0.25,1,1)$. fig.~(\ref{fig:Sportal}) hence shows the comparison between current DD limits, as well as the projected sensitivities by XENON1T and LZ, and the requirement of the correct DM relic density.

The results reported in fig.~(\ref{fig:Sportal}) can be explained as follows. t-channel exchange of the scalar mediator induces SI interactions of the DM, which are written, in the case of the proton as:
\begin{align}
& \sigma_{\chi N}^{\rm SI}=\frac{\mu_{\chi p}^2}{4 \pi}\frac{{(\lambda_\chi^S)}^2 c_S^2}{m_\chi^2 m_S^2}\frac{m_N^2}{v_h^2}{\left[f_p \frac{Z}{A}+f_n \left(1-\frac{Z}{A}\right)\right]}^2\nonumber\\
& \approx 1.8 \times 10^{-45}{\mbox{cm}^2}{(\lambda_\chi^S)}^2 c_S^2
{\left(\frac{400~\mbox{GeV}}{m_S}\right)}^2 {\left(\frac{400~\mbox{GeV}}{m_\chi}\right)}^2.
\end{align}

Here $A$, $Z$ represent, respectively, the atomic and proton number of the material constituting the detector,  $\mu_{\chi p}=m_\chi m_p/(m_\chi+m_p)$ denotes reduced mass
of the WIMP-proton system with $m_p$ representing the mass of the latter while $f_p$ and $f_n$ represent the effective couplings of the DM with protons and nucleons. In the case of a scalar mediator we have:
\begin{equation}
f_N=\sum_{q=u,d,s} f_q^N+\frac{6}{27}f^N_{\rm TG},\,\,\,\,N=p,n
\end{equation}
with $f_q^N$ being form factor whose physical meaning is associated to the contribution from up, down, strange and heavy quarks to the mass of the proton and the neutron. Notice that the factor ${\left[f_p \frac{Z}{A}+f_n \left(1-\frac{Z}{A}\right)\right]}$ is actually a rescaling factor which is introduced for a consistent comparison with experimental limits which customarily assume $f_p=f_n$~\cite{Feng:2013fyw}. It can be easily seen that in the case of the spin-0 mediator $f_p \simeq f_n$ so in the following numerical estimates we will automatically set $\left[f_p \frac{Z}{A}+f_n \left(1-\frac{Z}{A}\right)\right] \rightarrow f_p \sim 0.3$. As will be shown in the next section, for spin-1 mediators one expects in general $f_p \neq f_n$. This often translates into an enhancement of the cross-section and, hence, stronger limits on the model parameters. This can be already noticed by comparing the limits in the case of the Higgs and $Z$-portal. 

Current limits exclude then low values for both the mass of the DM and the one of the mediator. These limits will become, of course, progressively stronger, in case of absence of signals at XENON1T and/or LZ.

Concerning the DM relic density for $m_\chi < m_S,m_t$ the DM annihilation cross-section is suppressed by Yukawa structure of the couplings so that the correct relic density is obtained only around the resonance region $m_\chi \sim m_S/2$. 
This corresponds to the wide region between the two lines in fig.~\ref{fig:Sportal}. This region is of course proportional to $\lambda_\chi^S \times c_s $
and is then the same in the laft and right panel. The difference between these two figures is the disappearance of the region corresponding to the $SS$
final state, which annihilation cross section is proportional to $(\lambda_\chi^S)^4$ (see eq.~(\ref{Eq:chichiss})). This channel is then not sufficient to avoid an overdensity of the Universe for
$\lambda_\chi^S=0.25$. In the middle panel, the pole region is enlarge, to the point of covering almost all the parameter space, even joining the $SS$ final state at $m_S \simeq 1$ TeV.
For higher DM masses, instead, the correct relic density is achieved also far from s-channel resonances through either the $\ovl t t$ channel or the $SS$ channel, whether kinematically open. In such a case, the following analytical estimates, through the conventional velocity expansion, of the DM annihilation cross-section, can be obtained: 
\bea
\langle \sigma v \rangle (\chi \chi \rightarrow \ovl t t)\approx \frac{3}{16 \pi} {(\lambda_\chi^S)}^2 c_S^2 \frac{m_t^2}{v_h^2}\frac{1}{m_S^2} \approx 3.4 \times 10^{-25}{\mbox{cm}}^3 {\mbox{s}}^{-1}
{(\lambda_\chi^S)}^2 c_S^2 {\left(\frac{1~\mbox{TeV}}{m_S}\right)}^2
~{\rm for}~m_t < m_\chi < m_S,\nonumber\\
\eea
and:
\bea
 \langle \sigma v \rangle (\chi \chi \rightarrow \ovl t t) &&\approx \frac{3}{64 \pi} {(\lambda_\chi^S)}^2 c_S^2 \frac{m_t^2}{v_h^2}\frac{m_S^2}{m_\chi^4} \approx 1.2 \times 10^{-26}{\mbox{cm}}^3 {\mbox{s}}^{-1}
 {(\lambda_\chi^S)}^2 c_S^2{\left(\frac{2~\mbox{TeV}}{m_\chi}\right)}^4 {\left(\frac{m_S}{1.5~\mbox{TeV}}\right)}^2,\nonumber\\
\langle \sigma v \rangle (\chi \chi \rightarrow S S) &&\approx \frac{{(\lambda_\chi^S)}^4}{64 \pi m_\chi^2} \approx 5.8 \times 10^{-26}{\mbox{cm}}^3 {\mbox{s}}^{-1}{(\lambda_\chi^S)}^4{\left(\frac{1~\mbox{TeV}}{m_\chi}\right)}^2
~{\rm for}~m_t < m_S < m_\chi.
\label{Eq:chichiss}
 \eea

As evident both the $\ovl t t$ and $SS$ cross sections are s-wave dominated and then velocity independent. As a consequence residual annihilation would occur at present times to be probed by DM ID strategies. Similarly to the case of the Higgs portal, Direct Detection limit are much more competitive with respect to the ones from Indirect Detection, hence the latter have not been explicitly reported on fig.~(\ref{fig:Sportal}). We also notice that the dominant contribution of the annihilation cross-section into $SS$ depends only on the $\lambda_\chi^S$ coupling; as a consequence the scalar self coupling $\lambda_S$ does not play a relevant role for DM phenomenology.

\subsubsection{Fermionic Dark Matter}

The interaction of a fermionic DM and a scalar s-channel mediator can be described by the following phenomenological Lagrangian:
\begin{equation}
\mathcal{L}=\xi g_\psi \ovl \psi \psi S+\frac{c_S}{\sqrt{2}} \frac{m_f}{v_h}\ovl f f S+\mathcal{L}_S,
\end{equation}
where $\mathcal{L}_S$ has been introduced in the previous subsection. Contrary to the case of scalar DM, the operator $\ovl \psi \psi S$ is of dimension 4, so that $g_\psi$ is already an dimensionless parameter. Note that similar to eq.~(\ref{eq:bsms0sdm1}) we have parametrized $g_\psi$ to contain all the information of the $\ovl \psi\psi S$ vertex
including a normalization factor. One could think that an eventual VEV of the scalar mediator $S$ can be the origin of the DM mass, so that $g_\psi \sim m_\psi/v_S$, with $v_S$ being the VEV of $S$. We won't make this assumption in this work and regard $g_\psi$ as a generic dimensional-less constant.

\begin{figure}[t]
\includegraphics[width=4.9 cm]{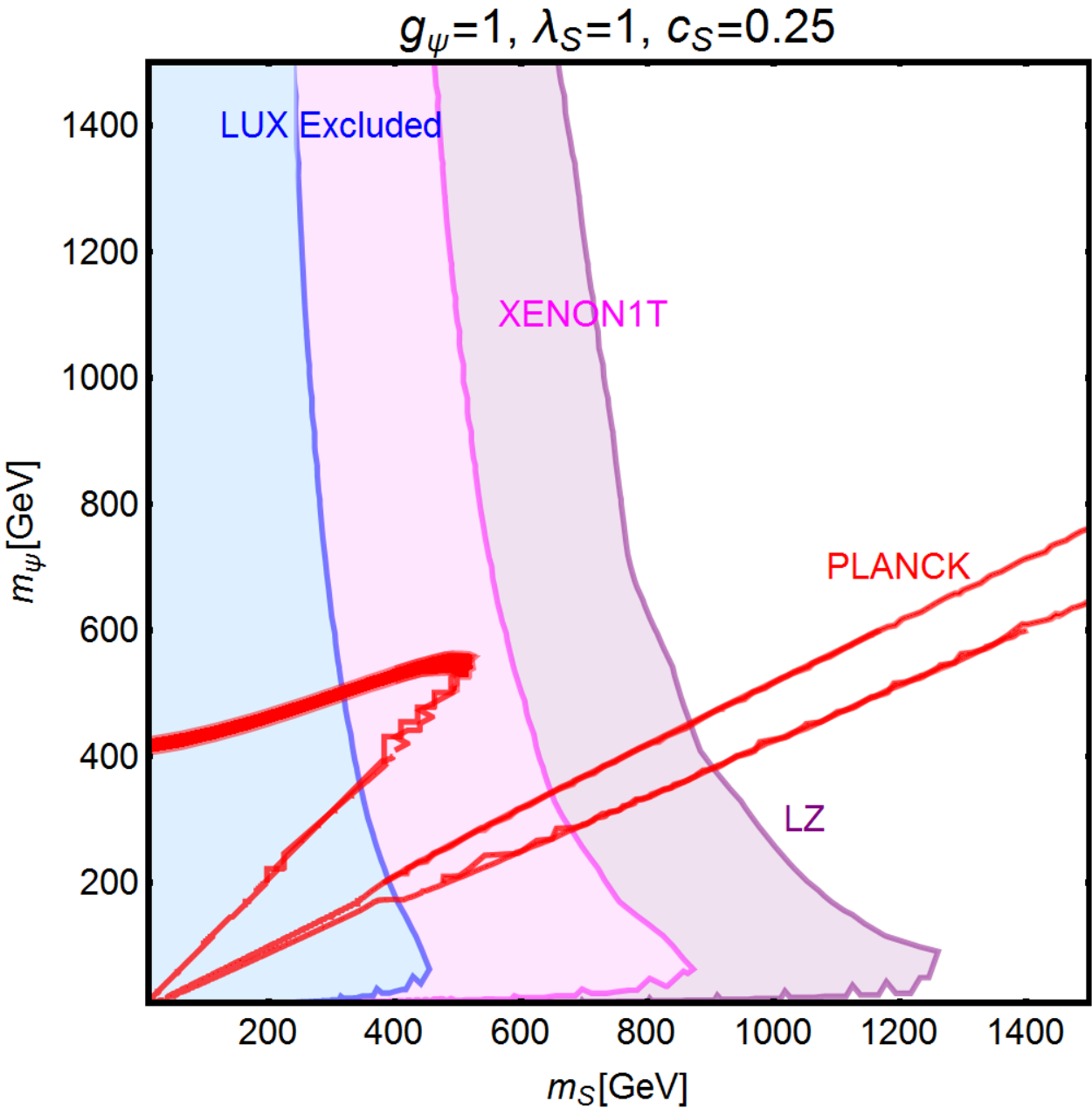}
\includegraphics[width=4.9 cm]{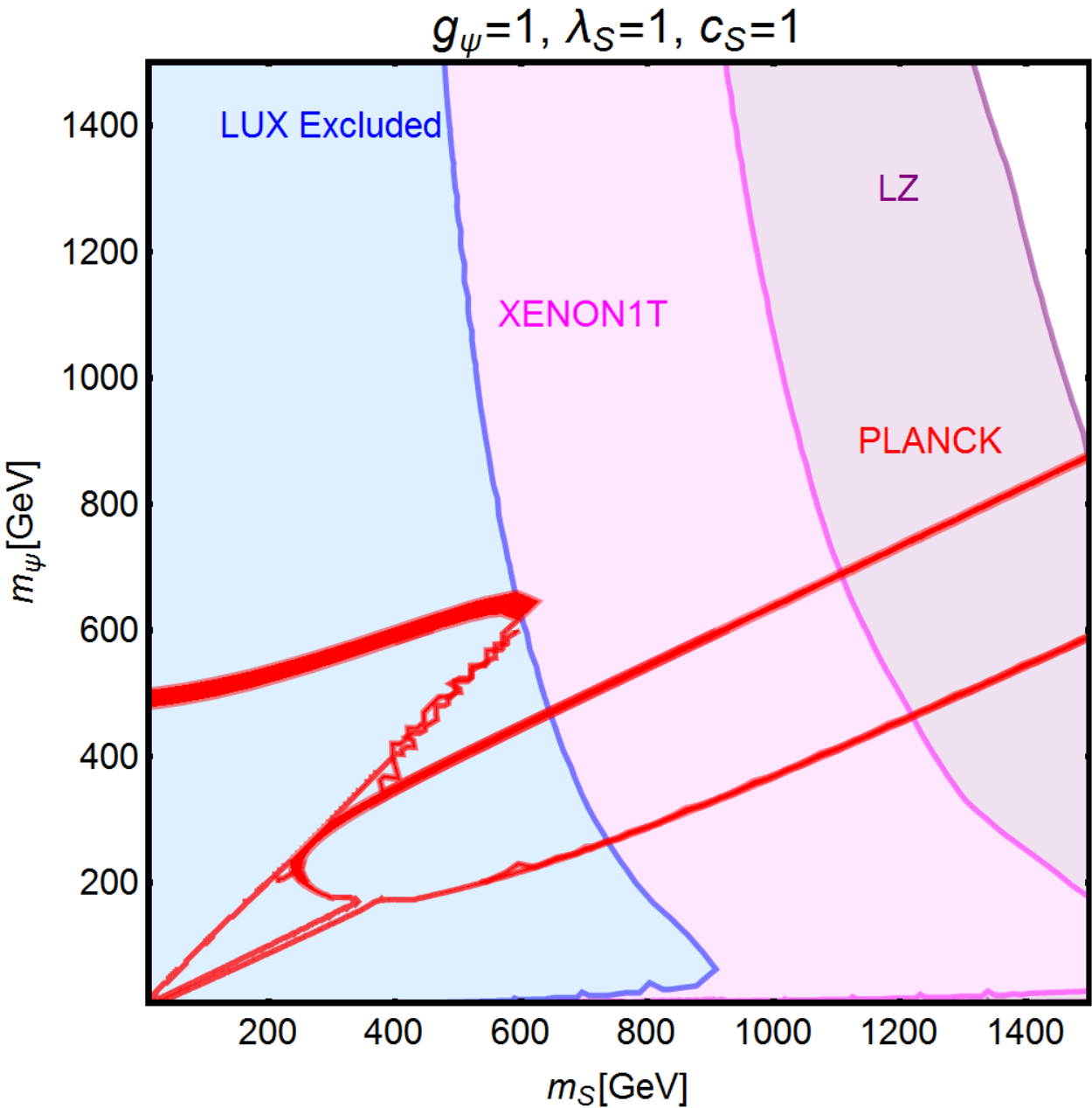}
\includegraphics[width=4.9 cm]{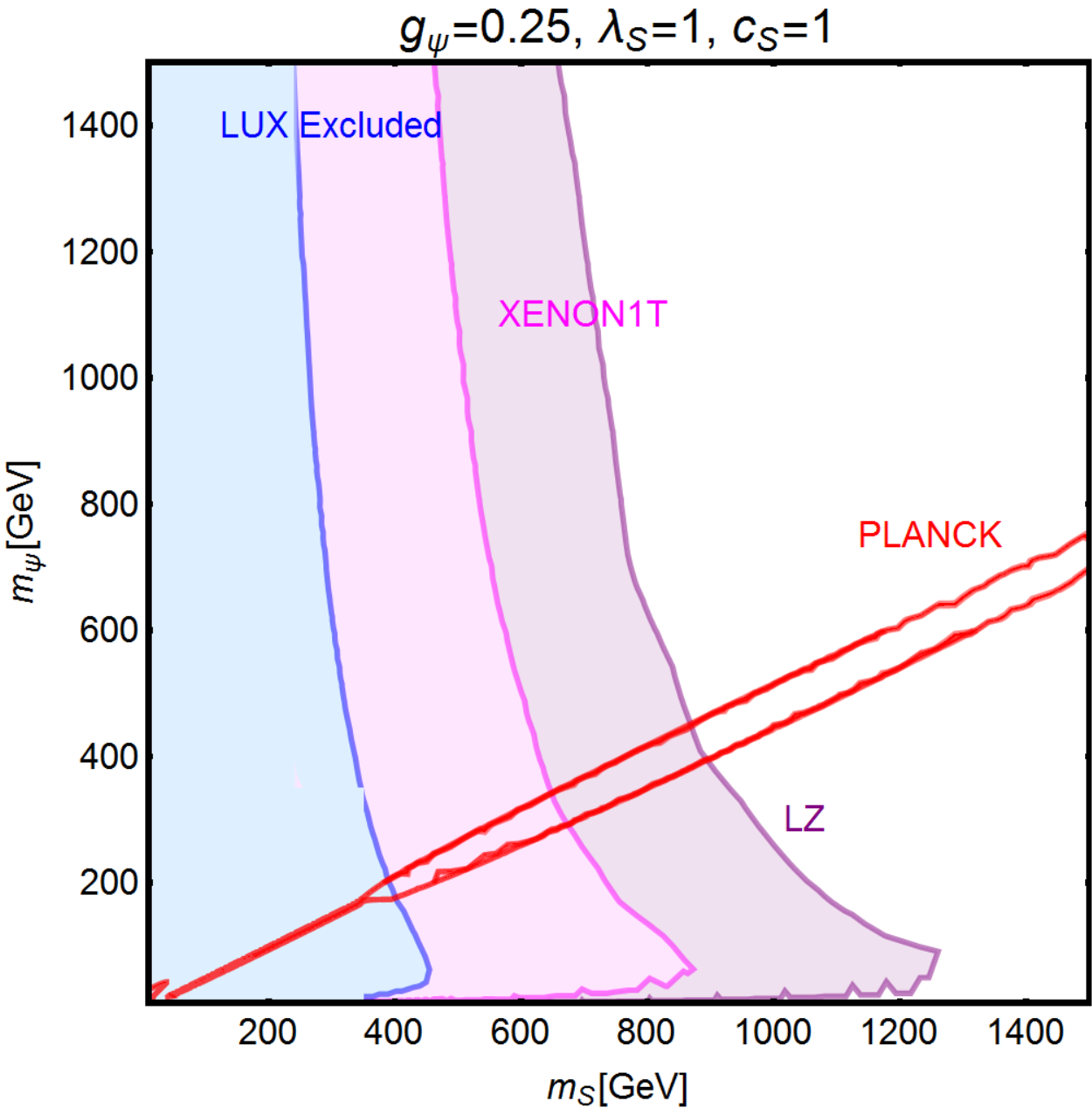}
\caption{\footnotesize{The same as fig.~(\ref{fig:Sportal}) but for a dirac fermion DM (i.e., replacing $\lambda^S_\chi$ by $g_\psi$).}}
\label{fig:Fportal}
\end{figure}

The main results of our analysis have been summarized in fig.~(\ref{fig:Fportal}). We have again considered the DM and scalar masses as free parameters and an analogous assignation of the couplings as in the previous subsection. 
%
The results reported in the figure can be described analytically as follows; the 
DM Direct Detection is again principally determined by SI interactions whose cross section is given by:
\begin{equation}
\sigma^{\rm SI}_{\psi p}=\frac{\mu_{\psi p}^2}{\pi}g_\psi^2 c_S^2 \frac{m_p^2}{v_h^2}f_N^2 \frac{1}{m_S^4}\approx \frac{1}{\pi}g_\psi^2 c_S^2 \frac{m_p^2}{v_h^2}f_N^2\frac{1}{m_S^4} \approx 2.9 \times 10^{-45}\, {\mbox{cm}}^{3} {\mbox{s}}^{-1} g_\psi^2 c_S^2 {\left(\frac{500~\mbox{GeV}}{m_S}\right)}^4,
\end{equation}
where $\mu_{\psi p}=m_\psi m_N/(m_\psi+m_N)$ denotes reduced mass
of the associated WIMP-proton system.
As evidenced by fig.~(\ref{fig:Fportal}), DM masses even above the TeV scale, are excluded by current DD limits for $m_S \lesssim 400-500\,\mbox{GeV}$. Values below the TeV scale for both the DM and mediator masses will be excluded in case of absence of signals from next generation of experiments.
The correct DM relic density can be achieved, without relying on s-channel resonances, only whether at least one between the $\ovl t t$ and $SS$ final states is kinematically accessible. In such a case the DM pair annihilation cross-section can be approximated as:
\bea
\langle \sigma v \rangle (\ovl \psi \psi \rightarrow \ovl t t)~ &&\approx \frac{3}{4\pi}g_\psi^2 c_S^2 \frac{m_t^2}{v_h^2}\frac{m_\psi^2}{m_S^4}v^2 \approx 1.5 \times 10^{-26} {\mbox{cm}^3} {\mbox{s}}^{-1} g_\psi^2 c_S^2 {\left(\frac{m_\psi}{300~\mbox{GeV}}\right)}^2 {\left(\frac{1~\mbox{TeV}}{m_S}\right)}^4\nonumber\\
&&~{\rm for}~m_t < m_\psi < m_S, 
\eea
\begin{align}
& \langle \sigma v \rangle(\ovl \psi \psi \rightarrow \ovl t t)\approx \frac{3}{64\pi}g_\psi^2 c_S^2 \frac{m_t^2}{v_h^2}\frac{1}{m_\psi^2}v^2 \approx 2.8 \times 10^{-26} {\mbox{cm}^3} {\mbox{s}}^{-1} g_\psi^2 c_S^2 {\left(\frac{600~\mbox{GeV}}{m_\psi}\right)}^2 ,\nonumber\\
& \langle \sigma v \rangle(\ovl \psi \psi \rightarrow S S)\approx \frac{3}{64\pi}g_\psi^4 \frac{1}{m_\psi^2}v^2 \approx 2.0 \times 10^{-26} {\mbox{cm}^3} {\mbox{s}}^{-1} g_\psi^4 {\left(\frac{1~\mbox{TeV}}{m_\psi}\right)}^2 ~{\rm for}~m_\psi > m_t,m_S.
\end{align}

Here $v^2 \sim 0.23$. 
We notice again that in the limit $m_\psi \gg m_S$ the scalar self-coupling $\lambda_S$ does not influence the DM relic density. The dependence on 
the couplings between the three figures (see fig.~\ref{fig:Fportal}) is the same than in the scalar case.
Contrary to the case of scalar DM, all the annihilation channels are velocity suppressed, hence cannot account for a sizable indirect signals.

\subsubsection{Vector Dark Matter}

For the description of the vector DM case we consider the following
Lagrangian:
\begin{equation}
\label{eq:bsmscap1}
\mathcal{L}=\frac{1}{2} m_V \eta_V^S V^\mu V_\mu S+ \frac{1}{8}\eta_V^{S^2} V^\mu V_\mu S S+ \frac{c_S}{\sqrt{2}}\frac{m_f}{v_h}S \ovl f f+\mathcal{L}_S,
\end{equation}
which is inspired by the construction proposed e.g., in refs.~\cite{Gross:2015cwa,Arcadi:2016kmk}. Note that 
all three terms of eq.~(\ref{eq:bsmscap1}) appear after the spontaneous symmetry breaking, once the portal field is expanded as $(S+v_S)/\sqrt{2}$ with $v_S$ as the concerned VEV. The quantity $m_V$ is expressed as $\eta^S_V v_S/2$.
A similar construction is also possible from a gauge invariant $D^\mu {\bf S} {(D_\mu {\bf S})}^*$ operator
for a compelx scalar field ${\bf S}$ with
$D_\mu=\partial_\mu-i \frac{1}{2} \eta^S_V V_\mu$. 
However, in this scenario the last term of eq.~(\ref{eq:bsmscap1}) would require new BSM charges
for the SM fermions. 

\begin{figure}[t]
\includegraphics[width=4.9 cm]{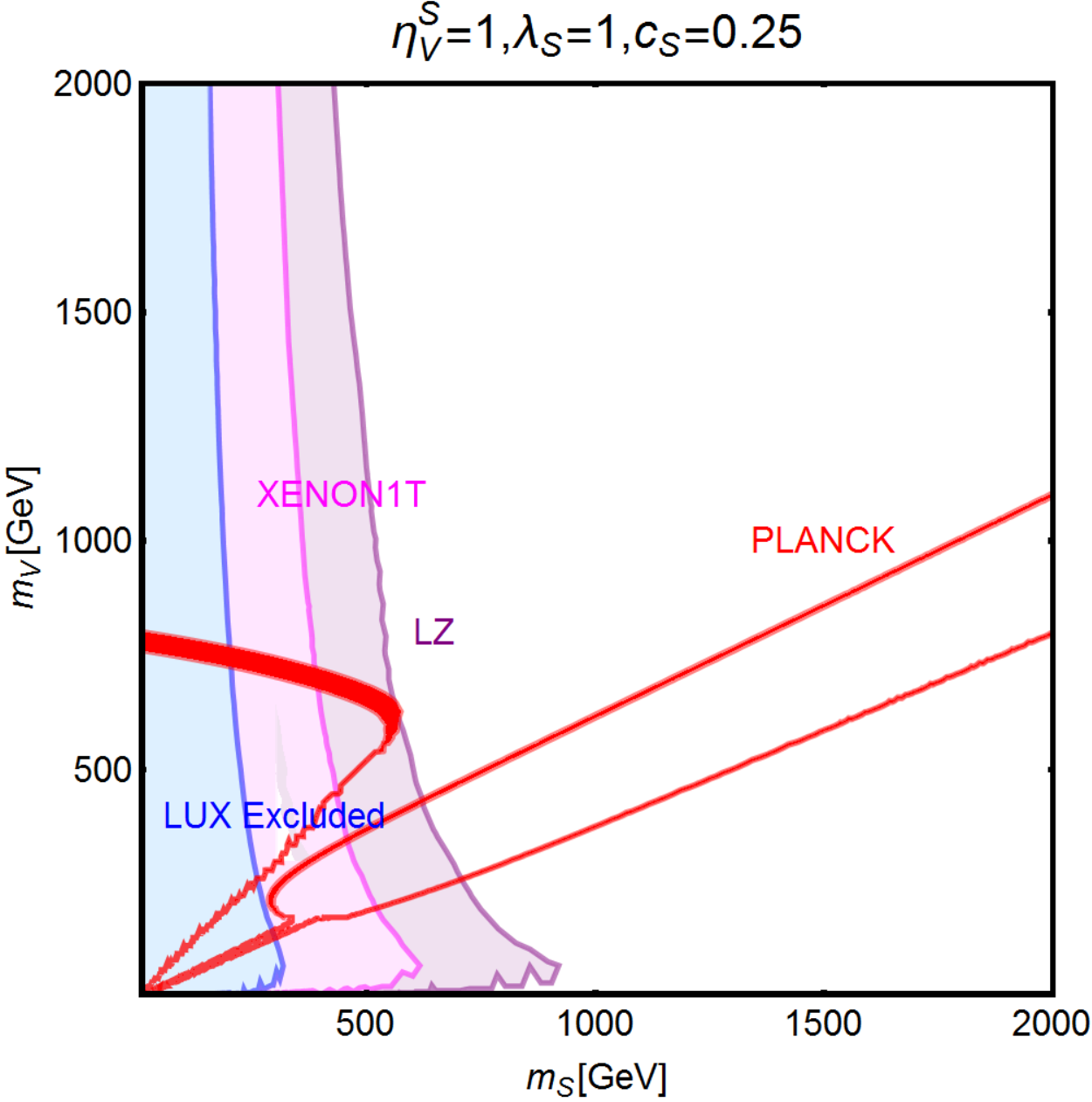}
\includegraphics[width=4.9 cm]{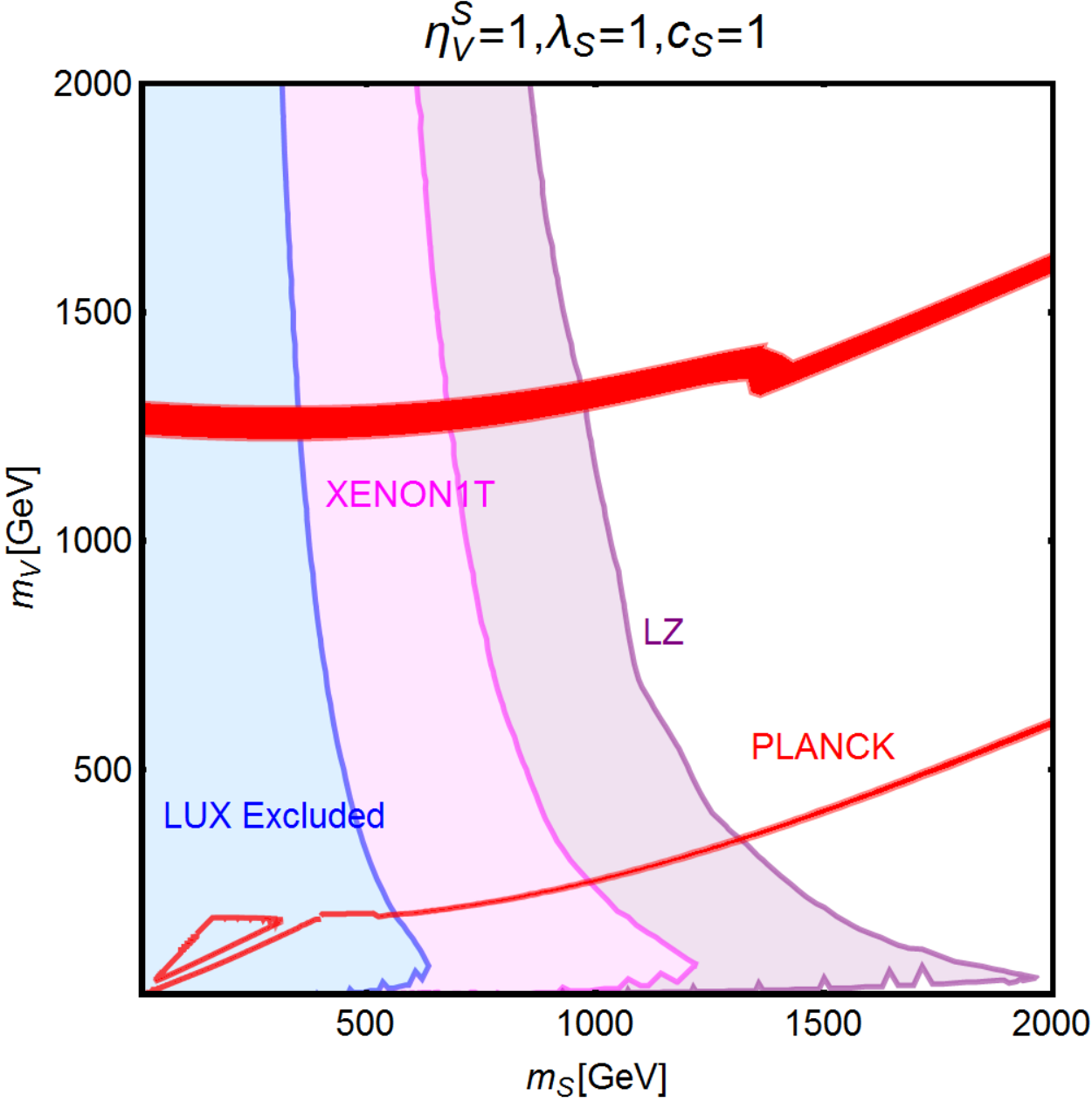}
\includegraphics[width=4.9 cm]{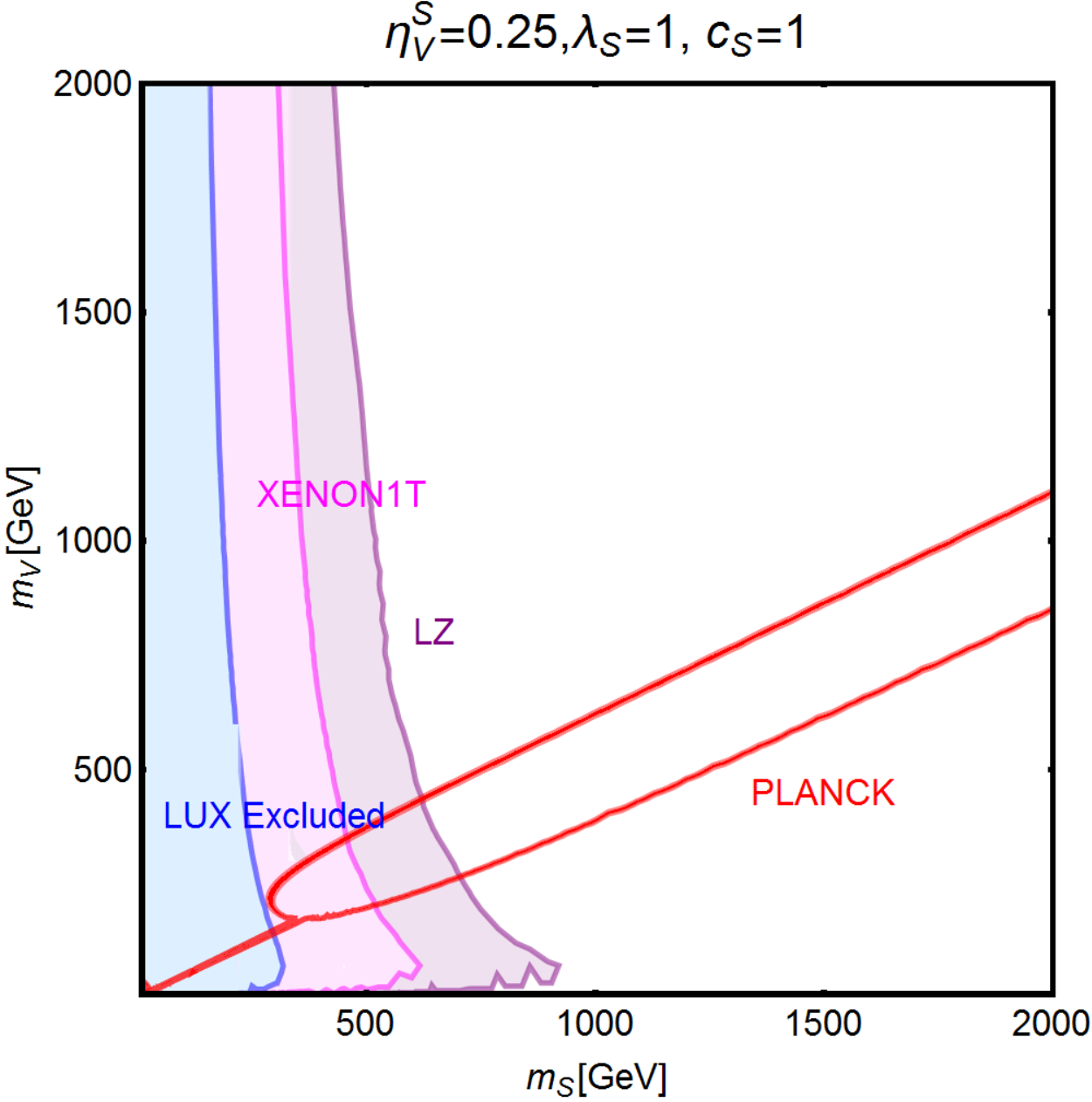}
\caption{\footnotesize{The same as fig.~(\ref{fig:Sportal}) for a vector DM with scalar mediator.}}
\label{fig:Vportal}
\end{figure}

This scenario has been analyzed with the same procedure as the scalar and fermionic DM cases. The results, reported in figs.~(\ref{fig:Vportal}) appear to be not very different from what obtained in the case of scalar DM, fig.~(\ref{fig:Sportal}). This can be explained by the fact that a vectorial DM can be viewed as three scalar degrees of freedom. 
The DM scattering rate on protons and its most relevant annihilation channels are described by the following analytical expressions:
%
\begin{equation}
\sigma_{Vp}^{\rm SI}=\frac{\mu_{Vp}^2}{4\pi} {(\eta_V^S)}^2 c_S^2 \frac{m_p^2}{v_h^2}f_p^2 \frac{1}{m_S^4} \approx 8.2 \times 10^{-45}{\mbox{cm}}^2 {(\eta_V^S)}^2 c_S^2 {\left(\frac{1~\mbox{TeV}}{m_S}\right)}^4.
\end{equation} 

The parameter $\mu_{V p}=m_V m_p/(m_V+m_p)$ as usual represents reduced mass
of the relevant WIMP-proton system and
%
\begin{equation}
\langle \sigma v \rangle (V V \rightarrow \ovl t t) \approx
\left \{
\begin{array}{cc} 
\frac{1}{4\pi} {(\eta_{V}^S)}^2 c_S^2 \frac{m_t^2}{v_h^2}\frac{m_V^2}{m_S^4} \approx 4.1 \times 10^{-26} {\mbox{cm}^3} {\mbox{s}}^{-1} {(\eta_V^S)}^2 c_S^2 {\left(\frac{m_V}{300~\mbox{GeV}}\right)}^2 {\left(\frac{1~\mbox{TeV}}{m_S}\right)}^4~{\rm if}~{m_S} < m_V, \\\\
\frac{1}{64\pi}{(\eta_V^S)}^2 c_S^2 \frac{m_t^2}{v_h^2}\frac{1}{m_V^2} \approx 2.8 \times 10^{-26} {\mbox{cm}^3} {\mbox{s}}^{-1} {(\eta_V^S)}^2 c_S^2 {\left(\frac{1~\mbox{TeV}}{m_V}\right)}^2~{\rm if}~ m_V > m_S.
\end{array}
\right.  
\end{equation}
\begin{equation}
\langle \sigma v \rangle(V V \rightarrow S S) \approx \frac{11}{2304\pi}{(\eta_V^S)}^4 \frac{1}{m_V^2} \approx 1.7 \times 10^{-26} {\mbox{cm}^3} {\mbox{s}}^{-1} {(\eta_V^S)}^4 {\left(\frac{1~\mbox{TeV}}{m_V}\right)}^2.
\end{equation}

\subsection{Spin-1 portals}

In this section we will analyze, in analogous fashion as the previous section, the case of a s-channel spin-1 mediator. Contrary to the scalar case, this kind of scenario offers a much richer collider phenomenology since one could assume gauge-like interactions (contrary to Yukawa-like interactions for spin-0 mediators) of the mediator with SM light quarks and leptons, leading to visible signals, implying stronger constraints though \cite{Profumo:2013sca,Alves:2013tqa,Alves:2015pea,Alves:2015mua,Allanach:2015gkd,Klasen:2016qux,Altmannshofer:2016jzy,Alves:2016cqf}. As a consequence we will consider a wider mass range, for both DM and the spin-1 mediator, in our analysis.

New spin-1 s-channel mediators can be straightforwardly associated to gauge bosons of extra $U(1)$ groups. Extra $U(1)$ symmetries are particularly common in extensions of the Standard Model and, in particular, in Grand Unified Theories (GUT). These $Z'$ particles can be coupled to SM fermions either indirectly, through kinetic mixing~\cite{Babu:1997st,Chun:2010ve,Mambrini:2011dw,Frandsen:2012rk} with the $Z$-boson, or directly in case the latter have non-trivial charges under the new symmetry group (see e.g., refs.~\cite{Langacker:2008yv,Han:2013mra} ). We will focus, in this section, on this last case while the case of kinetic mixing will be reviewed later in the text. Scalar or fermionic DM can be easily embedded in this kind of construction since they can be assumed to be new states charged under the new $U(1)$ but singlet with respect to the SM group. In such a case portal interactions simply arise from their covariant derivatives \footnote{Interestingly the $Z^{\prime}$ couplings with SM particles as presented here are similar to the ones in several existing electroweak extensions of the SM which can be embedded in these GUT models \cite{Mizukoshi:2010ky,Alvares:2012qv,Kelso:2013nwa,Dong:2014wsa}.}.

In the case of spin-1 DM we re-propose the two constructions, for self-conjugate (Abelian) and not self-conjugate (non-Abelian) DM already proposed in the Z-portal setup.


\subsubsection{Scalar Dark Matter}

Following the discussion above the interaction between a scalar DM and spin-1 ($Z'$) mediator, together with a piece connecting $Z'$
to the SM fermions, is described by the following Lagrangian:
\begin{equation}
\label{eq:new1}
\mathcal{L}=i g' \lambda_\chi^{Z'} \chi^* \overset{\leftrightarrow}{\partial_\mu}\chi Z'^\mu
+ g{'^2} \lambda_\chi^{Z'^{2}} {|\chi|}^2 Z'_\mu Z'^{\mu}+g' \sum_f \overline{f} \gamma^\mu \left(V_f^{Z'}-A_f^{Z'} \gamma_5\right) f Z'_\mu.
\end{equation}

Notice that trilinear interaction between DM pairs and the $Z'$, of the form reported above, is possible only in the case of a complex scalar DM.

\begin{figure}[t]
\includegraphics[width=4.9 cm]{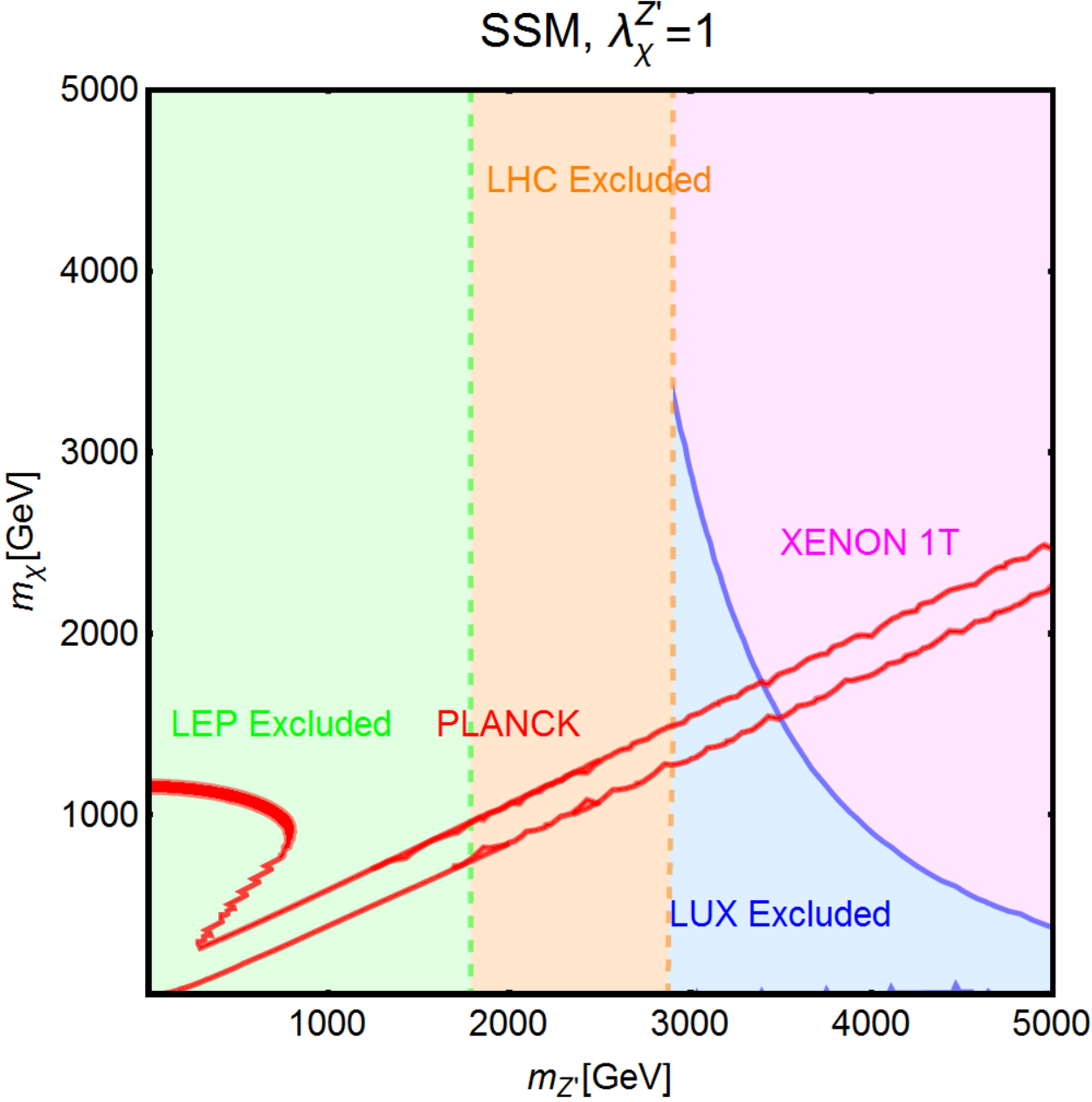}
\includegraphics[width=4.9 cm]{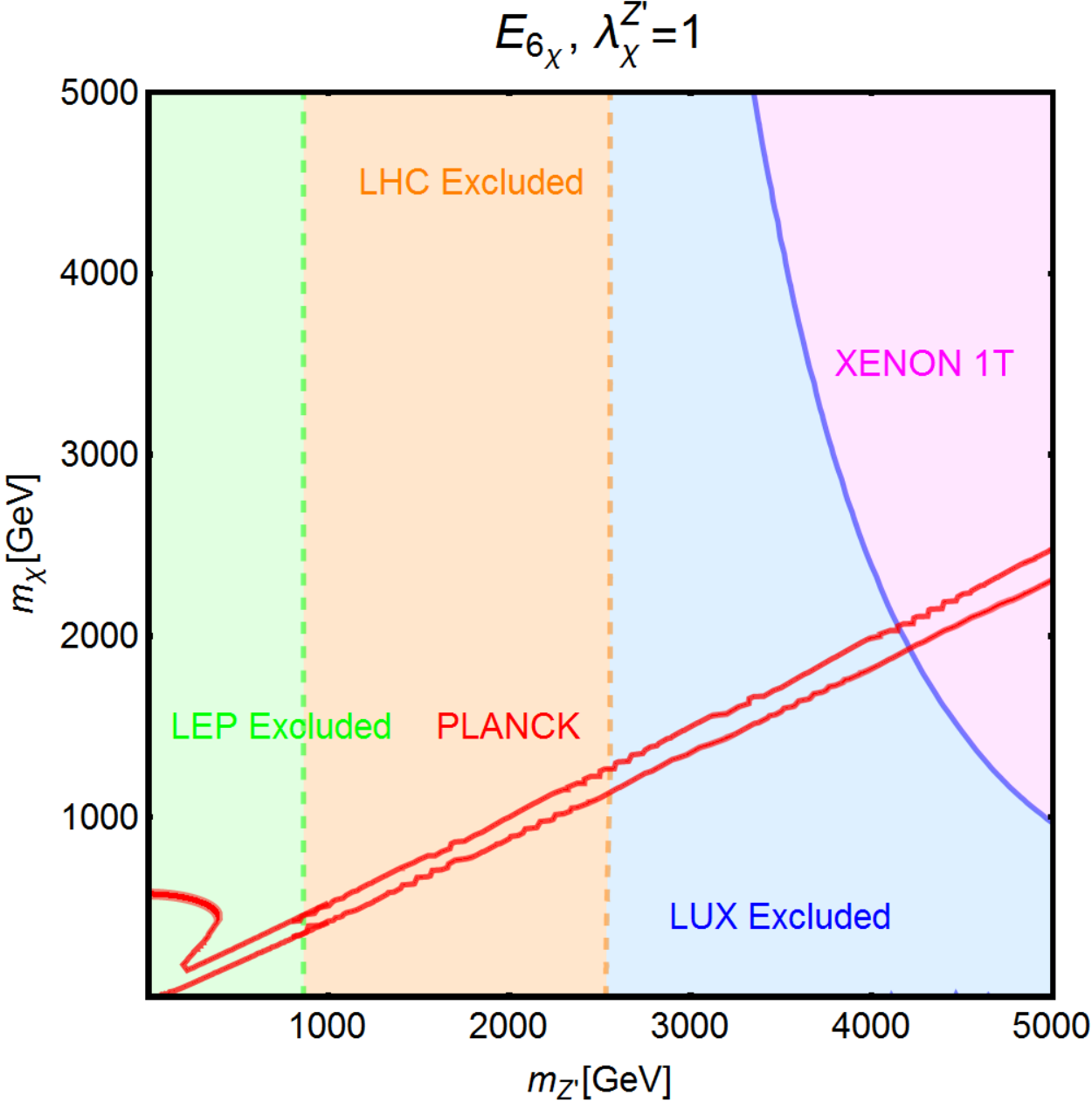}
\includegraphics[width=4.9 cm]{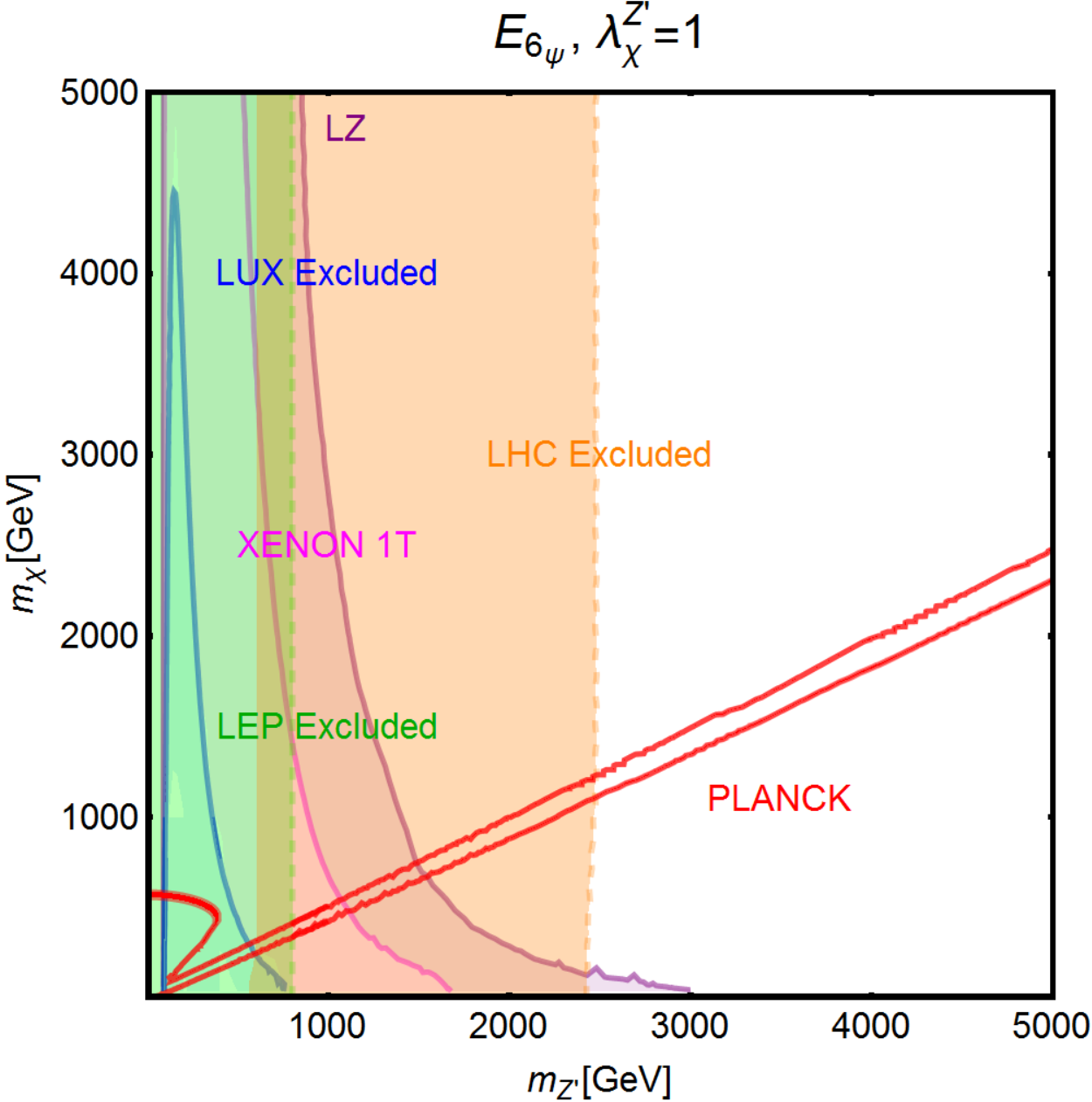}
\caption{\footnotesize{Summary of constraints for $Z'$
portal in the context of a scalar DM for the three different realizations i.e., SSM (left), $E_{6_\chi}$ (middle) and $E_{6_\psi}$ (left) (see table~(\ref{tab:Zpcouplings})). In these plots the red coloured contours represent the correct DM relic density. The blue coloured region is already excluded by LUX while the magenta, purple coloured regions are allowed by LUX but within the sensitivity of XENON1T (assuming two years of exposure) and LZ, respectively. Finally, the green and orange coloured regions represent the exclusions from dilepton searches by the LEP/Tevatron and LHC experiments.}}
\label{fig:SZprime}
\end{figure}

Similarly to the case of scalar mediator, our main parameters will be represented by the DM and $Z'$ masses. For what regards the couplings of the $Z'$ with the SM fermions we will consider some definite assignations, as dictated by the Sequential Standard Model (SSM), i.e., same couplings as the $Z$-boson, and some 
GUT-inspired realizations. According to this the coupling $g'$ will be set to $g \approx 0.65$ in the case of SSM and to $g_{\rm GUT}=\sqrt{{5}/{3}}\,g \tan \theta_W \approx 0.46$ for the GUT realizations. Finally, unless differently stated, we will set $\lambda_\chi^{Z'}=1$.
\begin{table}[t]
\centering
\begin{tabular}{|c|c|c|c|c|c|c|c|c|}
\hline
 & $V_u^{Z'}$ & $A_u^{Z'}$ & $V_d^{Z'}$ 
 & $A_d^{Z'}$ & $V_e^{Z'}$ & $A_e^{Z'}$ & $V_\nu^{Z'}$ & $A_\nu^{Z'}$ \\
 \hline
 SSM & $\frac{1}{4}-\frac{2}{3}\sin^2 \theta_W$ & $\frac{1}{4}$ & -$\frac{1}{4}+\frac{1}{3}\sin^2 \theta_W$
& $\frac{1}{4}$&-$\frac{1}{4}+\sin^2 \theta_W$ &
-$\frac{1}{4}$& $\frac{1}{4}$&$\frac{1}{4}$\\
\hline
$E_{6_\chi}$ & $0$ & -$\frac{1}{2\sqrt{10}}$ & -$\frac{1}{\sqrt{10}}$ & $\frac{1}{2\sqrt{10}}$ & $\frac{1}{\sqrt{10}}$
&$\frac{1}{2\sqrt{10}}$ &$\frac{3}{\sqrt{10}}$ &-$\frac{1}{2\sqrt{10}}$\\
\hline
$E_{6_\psi}$ & $0$ & -$\frac{1}{2\sqrt{6}}$ & $0$ & $\frac{1}{2\sqrt{6}}$ & $0$
&$\frac{1}{2\sqrt{6}}$ &$\frac{1}{4\sqrt{6}}$ &-$\frac{1}{2\sqrt{6}}$\\
\hline
\end{tabular}
\caption{\footnotesize{Table of couplings between the SM fermions and a Z' (see eq.~(\ref{eq:new1})) for the three different realizations of a Z' portal.}}
\label{tab:Zpcouplings}
\end{table}
%
The different assignations of the $V_f^{Z'}, A_f^{Z'}$ couplings
(see eq.~(\ref{eq:new1})) considered in our analysis for the three cases of SSM, $E_{6_\chi}$ and $E_{6_\psi}$ realizations
are exhibited in tab.~(\ref{tab:Zpcouplings}). For the same 
three realizations the effect of different constraints are summarized in fig.~(\ref{fig:SZprime}).

As first thing we notice, in the case of $SSM$ and $E_{6\chi}$ models, a much stronger impact of the limits from DM Direct searches with respect to the case of scalar mediator. The reason lies on the fact that SI interactions, with cross-section given by (as usual for the case of SI we will refer to scattering on protons):
%
\begin{equation}
\sigma_{\chi p}^{\rm SI}= \frac{\mu_{\chi p}^2}{\pi}\frac{g{'^4}}{m_{Z'}^4}\frac{{\left[ Z f_p+(A-Z) f_n \right]}^2}{A^2},\,\,\,f_p=2 V_u^{Z'}+V_d^{Z'},f_n=V_u^{Z'}+2 V_d^{Z'},
\end{equation}
are particularly efficient since, as evident from the fact that, for spin-1 mediators, the effective couplings of the DM with the proton and the neutron, $f_p$ and $f_n$, are just linear combination of couplings of the $Z'$ with up and down quarks~\footnote{The difference arises from the fact that in the case of spin-0 mediator the quantities $f_p$ and $f_n$ are originated by matrix element $\langle N| q q |N\rangle$ which is related to the mass of the nucleon. In the case of spin-1 mediator one instead evaluates the matrix element $\langle N|q \gamma^\mu q |N\rangle$ which is, instead, related to the electric charge of the nucleon.}. 
A further enhancement comes, in general, as already remarked, from the fact that $f_p \neq f_n$. As a consequence, an absence of signal from XENON1T would exclude values of the masses of the DM and of the $Z'$ even above 5 TeV. Sizable limits from DD, although weaker with respect to the previous two cases, are remarkably present also for the $E_{6\psi}$ realization, despite the assignations of the charges of the quarks under the new U(1) imply a null vectorial combination. Indeed non-null vectorial couplings, at the typical energy scale of DM scattering with nucleons, are radiatively generated by the axial couplings of the $Z'$, in particular with the top quark~\cite{Crivellin:2014qxa,DEramo:2014nmf,DEramo:2016gos}. An approximate expression, mostly valid for $m_{Z'}>m_Z$, for this RG induced couplings are given by:
\bea
 \widetilde{V}_u^{Z'}&&=\frac{\alpha_t}{2\pi}(3-8 s_W^2) A_u^{Z'} \log\left(\frac{m_{Z'}}{m_Z}\right)-(3-8 s_W^2) \left[\frac{\alpha_b}{2 \pi} A_d^{Z'}+\frac{\alpha_\tau}{6\pi}A_e^{Z'}\right] \log\left(\frac{m_{Z'}}{\mu_N}\right), \nonumber\\
 \widetilde{V}_d^{Z'}&&=-\frac{\alpha_t}{2\pi}(3-4 s_W^2) A_u^{Z'} \log\left(\frac{m_{Z'}}{m_Z}\right)+(3-4 s_W^2) \left[\frac{\alpha_b}{2 \pi} A_d^{Z'}+\frac{\alpha_\tau}{6\pi}A_e^{Z'}\right] \log\left(\frac{m_{Z'}}{\mu_N}\right).
\eea

Here $s_W\equiv \sin\theta_W$, $\alpha_{t,b,\tau}=y^2_{t,b,\tau}/4\pi$
with $y_f$ as the SM Yukawa couplings and $\mu_N$ is the characteristic energy scale of scattering interaction, here taken to be $1\,\mbox{GeV}$.

The DM annihilation cross section into SM fermions is instead velocity suppressed: 
\begin{equation}
\langle \sigma v \rangle (\chi \chi^{*} \rightarrow \ovl f f) \approx 
\left \{
\begin{array}{cc}
g{'^4}{(\lambda_\chi^{Z'})}^2 \frac{m_\chi^2}{3 \pi m_{Z'}^4} v^2 \sum\limits_f n_c^f  \left( {(V_f^{Z'})}^2 +{(A_f^{Z'})}^2 \right)~{\rm for}~ m_\chi < \frac{m_{Z'}}{2}, \\
g{'^4} {(\lambda_\chi^{Z'})}^2 \frac{1}{48 \pi m_\chi^2} v^2  \sum\limits_f n_c^f  \left( {(V_f^{Z'})}^2 +({A_f^{Z'})}^2 \right)~{\rm for}~ m_\chi > \frac{m_{Z'}}{2},
\end{array}
\right.
\end{equation}
where the sum runs over the final states kinematically accessible. 
The parameter $n^f_c$ represents colour factor for the final state fermions.
The correct relic density is thus achieved only around the pole region, namely $m_\chi \sim m_{Z'}/2$, unless the annihilation into $Z' Z'$ is kinematically accessible. This cross-section is s-wave dominated and can be simply approximated, for $m_\chi \gg m_{Z'}$:
\begin{equation}
\langle \sigma v \rangle (\chi \chi^{*} \rightarrow Z' Z') \approx \frac{g{'^4}{(\lambda_\chi^{Z'})}^4}{8 \pi m_\chi^2} \approx 3.7 \times 10^{-26} {\mbox{cm}^3}{\mbox{s}}^{-1} {\left(\frac{1.5~\,\mbox{TeV}}{m_\chi}\right)}^2,
\end{equation}
where, for definiteness, we have considered the SSM for the numerical estimates. As already pointed strong collider limits complement the ones from DM phenomenology. The panels of fig.~(\ref{fig:SZprime}) report two exclusion regions, green and orange. Both regions are related to limits associated to the couplings of the $Z'$ with SM leptons. The first ones, associated to the green regions, come from LEP and Tevatron~\cite{Alcaraz:2006mx,Aaltonen:2007al} and are based on possible modifications of the dilepton production cross-section. Since they do not necessarily rely on on-shell production of the $Z'$, once its couplings with the SM fermions are fixed, like in our case, they are straightforwardly translated into lower bounds on $m_{Z'}$. More specifically, these lower bounds are $1789, 853$ and $804$ GeV for, respectively, SSM, $E_{6_\chi}$ and $E_{6_\psi}$. These constraints are combined with the limits (orange regions in the plots) from LHC searches of dilepton resonances~\cite{ATLAS:2016cyf,Khachatryan:2016zqb,CMS:2016abv}. Contrary to the previous case, these limits are in principle sensitive to modification of the decay branching fraction of the $Z'$ as consequence, for example, of couplings with the DM~\cite{Arcadi:2013qia}. For the chosen assignation of the couplings the decay branching fraction into DM of the $Z'$ is small so that the limits substantially coincide with the ones reported by experimental collaborations. Other limits stemming from flavor and g-2 are weaken compared to the collider bounds \cite{Queiroz:2014zfa,Lindner:2016bgg}.

\subsubsection{Fermionic Dark Matter}

As already mentioned we will describe the interactions of a fermionic DM $\psi$ and a $Z'$, mediating its interactions with the SM fermions, through a Lagrangian of the form:
\begin{equation}
\mathcal{L}=g' \xi \overline{\psi} \gamma^\mu\left(V_\psi^{Z'}-A_\psi^{Z'} \gamma_5\right) \psi  Z'_\mu+g' \sum_f \overline{f} \gamma^\mu \left(V_f^{Z'}-A_f^{Z'} \gamma_5\right) f Z'_\mu,
\end{equation}
where $\xi=1 (1/2)$ for Dirac (Majorana) fermions. We remind that in the case of Majorana fermions $V_\psi^{Z'}=0$.

\begin{figure}[t]
\centering
\includegraphics[width=4.95 cm]{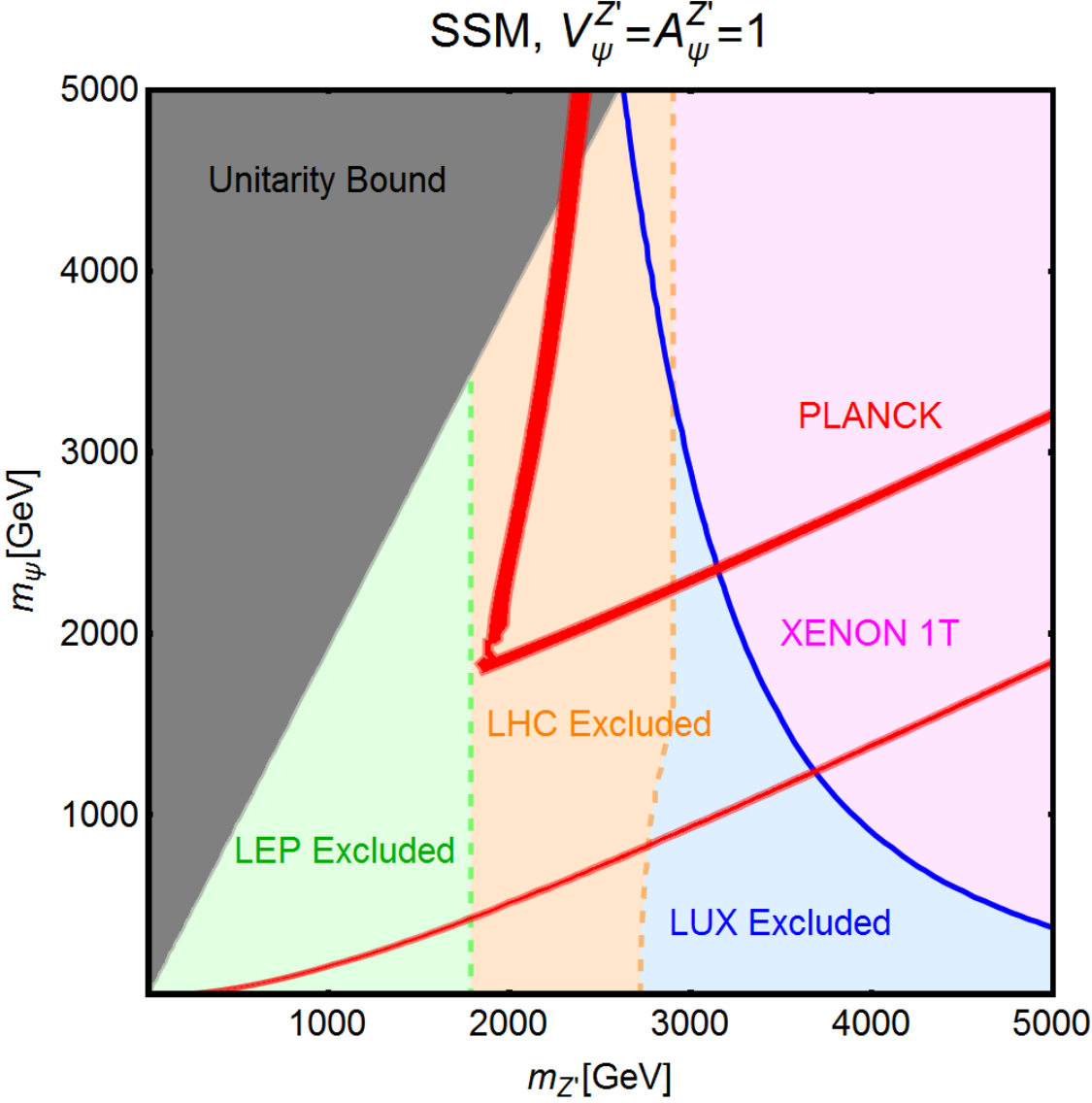}
\includegraphics[width=4.95 cm]{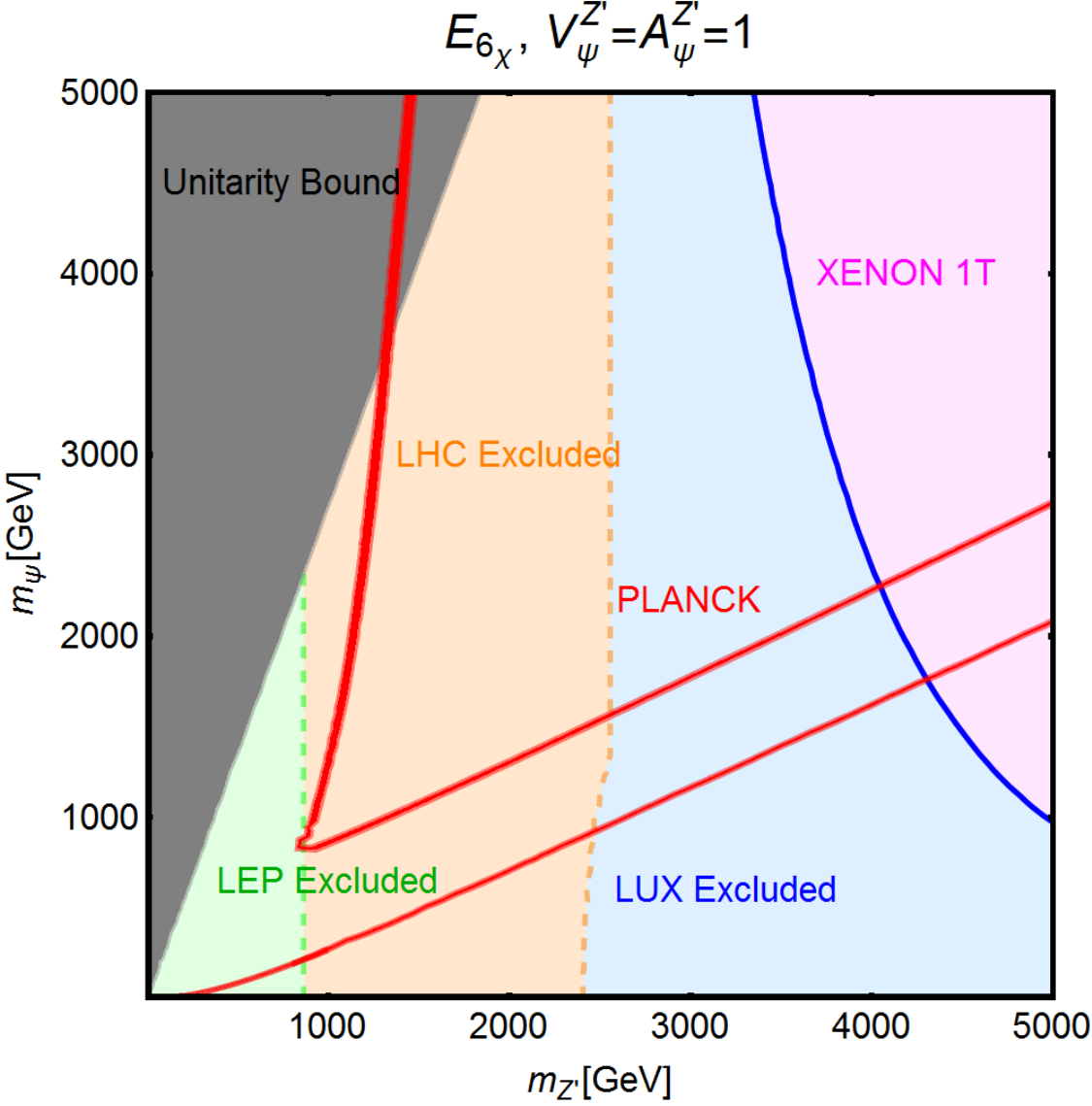}
\includegraphics[width=4.95 cm]{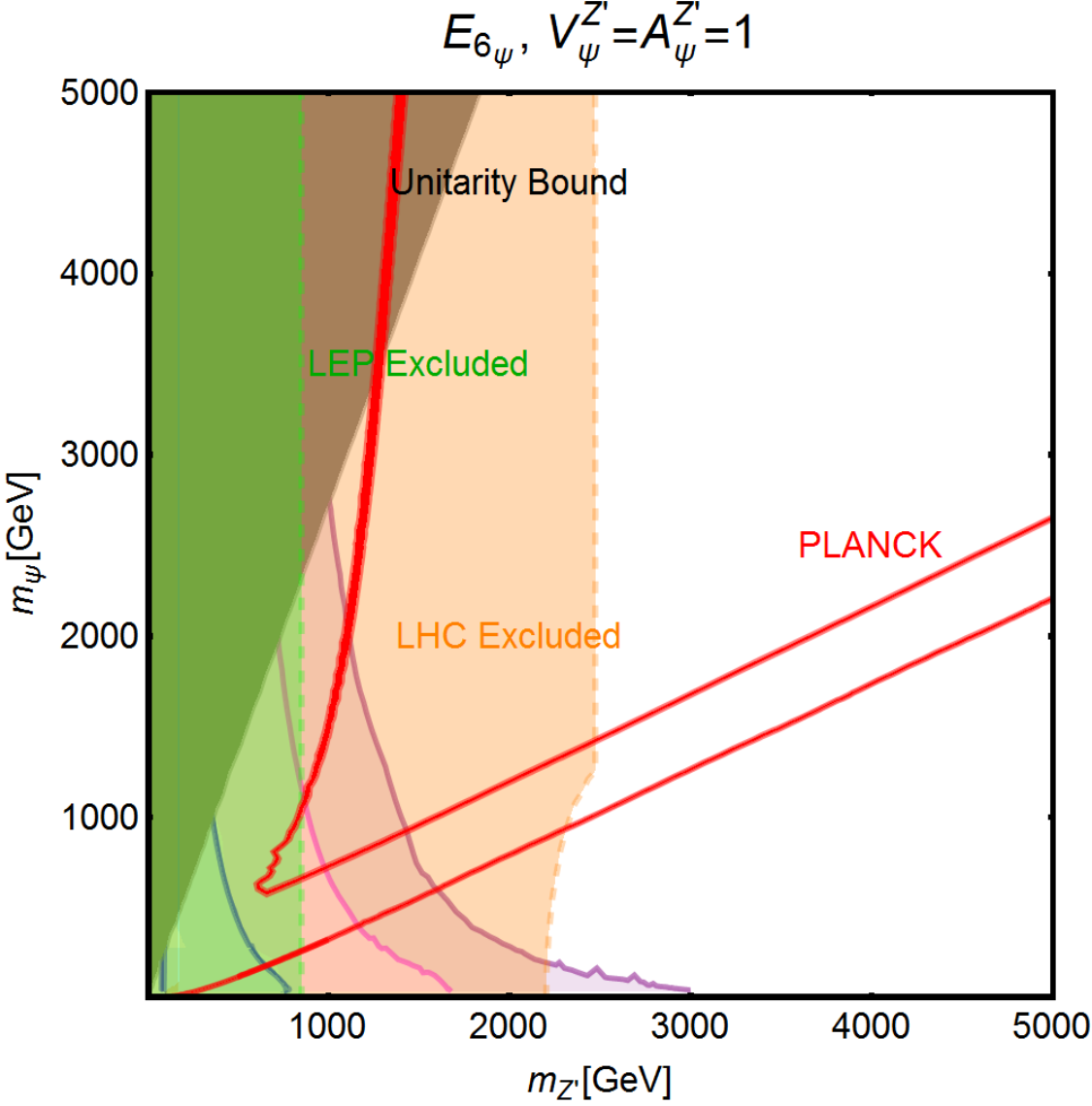}
\caption{\footnotesize{The same as fig.~(\ref{fig:SZprime}) but for a Dirac fermion DM where the gray coloured region 
shows exclusion from the unitarity bound.}}
\label{fig:FZprime}
\end{figure}

The combination of constraints is reported in fig.~(\ref{fig:FZprime}) for a Dirac fermion DM, following the style of previous figure.
%
%
\begin{figure}[t]
\includegraphics[width=4.9 cm]{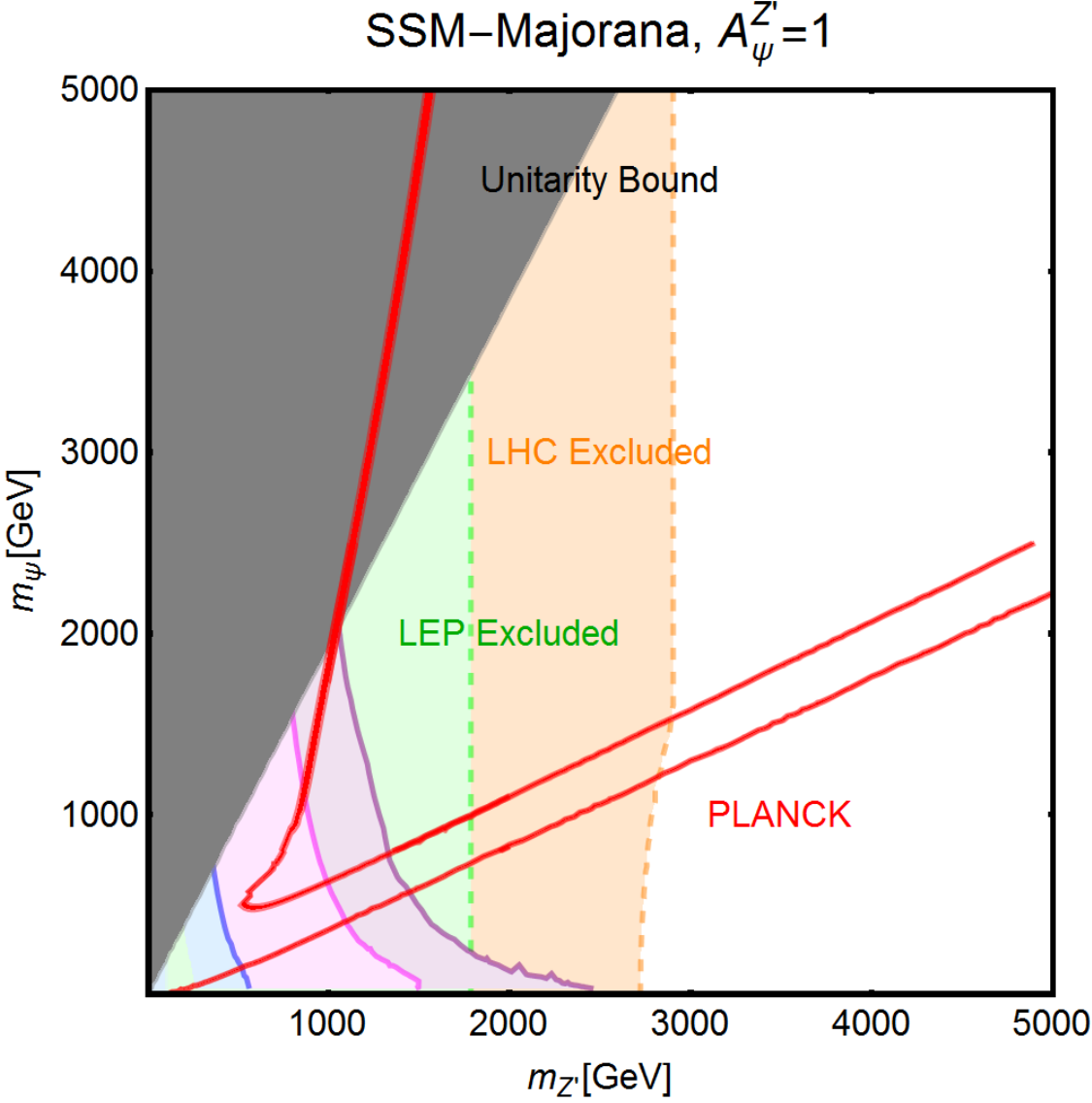}
\includegraphics[width=4.9 cm]{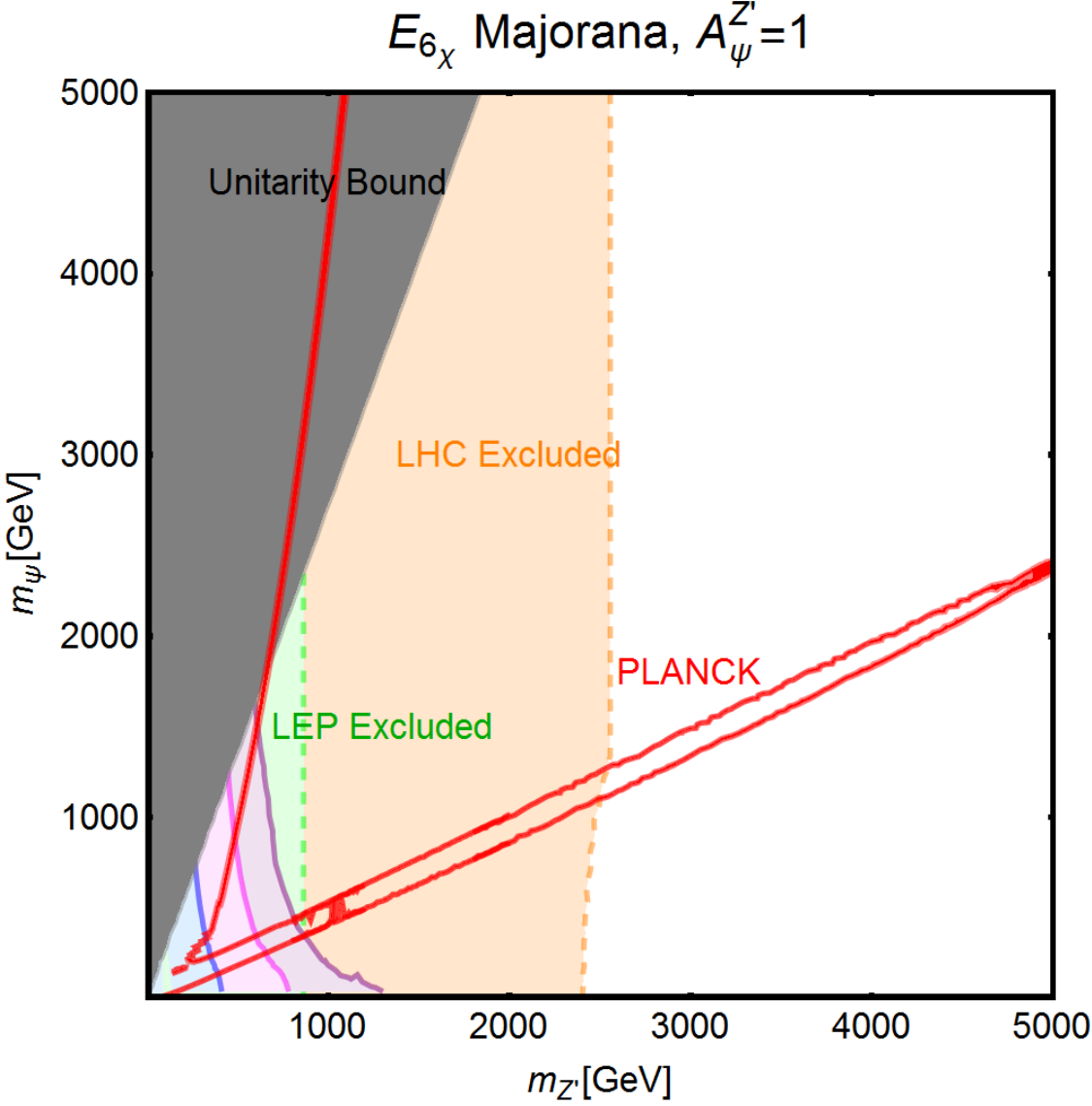}
\includegraphics[width=4.9 cm]{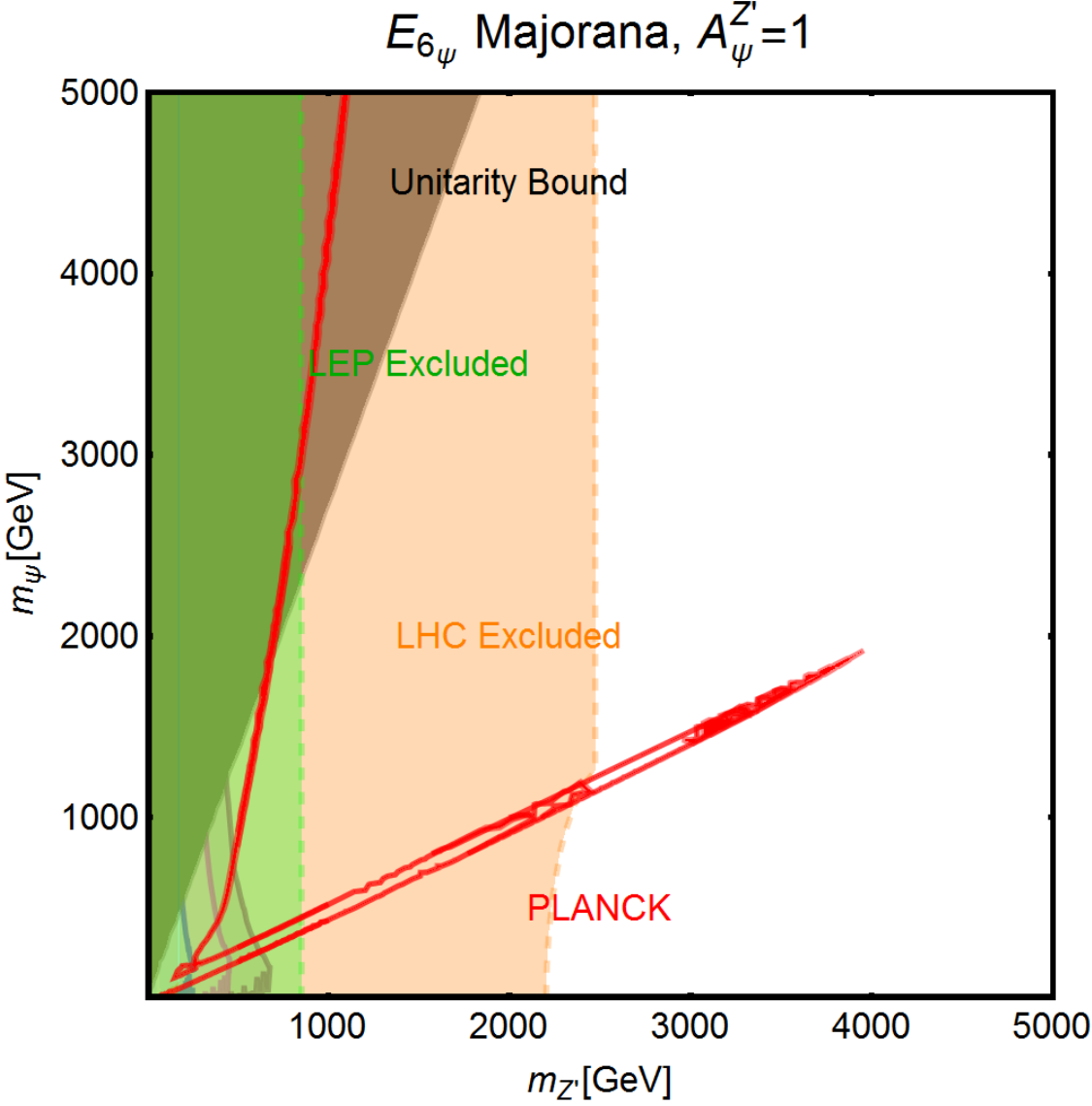}
\caption{\footnotesize{The same as fig.~(\ref{fig:FZprime}) but for a Majorana fermion DM.}}
\label{fig:FZprimemj}
\end{figure}
%
The SI cross section, from t-channel exchange of a $Z'$, in the case of Dirac fermions, exactly coincide with its corresponding for a complex scalar DM. As a consequence the excluded regions in the $(m_\psi,m_{Z'})$ plane are the same as shown in the previous subsection.

In the case of Majorana DM direct detection principally relies on SD interactions, to which Xenon based detectors are also sensitive. We have then reported, together with the most recent constraints~\cite{Akerib:2016lao,Fu:2016ega}, an estimation of the XENON1T and LZ sensitivities. As evident, even in the case of LZ, we have much weaker limits, not competitive with bounds from di-lepton searches.

On the contrary the regions corresponding to the correct DM relic density are sensitively different with respect to the case of scalar DM. Indeed the pair annihilation cross-section is not velocity suppressed and can be schematically expressed as: 
\begin{equation}
\langle \sigma v \rangle(\ovl \psi \psi \to \ovl f f) \approx
\left \{
\begin{array}{cc}
\frac{g'^4 m_\psi^2}{\pi m_{Z'}^4} \sum\limits_{f}  n_c^f \left({(V_f^{Z'})}^2+{(A_f^{Z'})}^2\right) \left({(V_\psi^{Z'})}^2+{(A_\psi^{Z'})}^2\right)~{\rm for}~ m_\psi < \frac{m_{Z'}}{2}, \\\\
\frac{g'^4}{16 \pi m_{\psi}^2} \sum\limits_f n_c^f \left({(V_f^{Z'})}^2+{(A_f^{Z'})}^2\right) \left(({V_\psi^{Z'})}^2+{(A_\psi^{Z'})}^2\right)~{\rm for}~ m_\psi > \frac{m_{Z'}}{2}.
\end{array}
\right .
\end{equation}

In addition the t-channel mediated annihilation process $\ovl \psi \psi \rightarrow Z' Z'$ is particularly efficient, being the corresponding cross-section given by:
\begin{equation}
\langle \sigma v \rangle (\ovl \psi \psi \rightarrow Z' Z') \approx \frac{g'^4}{\pi m_{Z'}^2} \left({(V_\psi^{Z'})}^2 {(A_\psi^{Z'})}^2+\frac{v^2}{3}{(A_\psi^{Z'})}^4 \frac{m_\psi^2}{m_{Z'}^2}\right).
\end{equation}

As evident a strong enhancement is originated by the velocity dependent term being proportional to $\frac{m_\psi^2}{m_{Z'}^2}$. As well know this kind of behavior lead to a violation of perturbative unitarity unless new degrees of freedom, like a dark Higgs~\cite{Kahlhoefer:2015bea}, are added to cure the pathological behavior of the theory. In absence of a UV completion, we have imposed, in our simplified framework, a unitarity constraint on the axial coupling $A_\psi^{Z'}$ of the form:
\begin{equation}
A^{Z'}_\psi \leq \frac{\pi m_{Z'}^2}{2 m_\psi^2}.
\end{equation}

\subsubsection{Vector Dark Matter}

As already mentioned we will discuss separately the cases of Abelian (real vector) and non-Abelian (complex vector) DM, in order to exploit different scenarios for what regards Direct Detection. Similarly to the case of $Z$-portal, we will consider the following two constructions for, respectively, non-Abelian and Abelian DM: 
\begin{equation}
\label{eq:NAVlagrangian}
\mathcal{L}=g' \eta^{Z'}_V [[VVZ']]+g'\sum_f\overline{f}\gamma^\mu \left(V^{Z'}_f-A^{Z'}_f \gamma^5\right) f Z'_\mu,
\end{equation}
\begin{equation}
\mathcal{L}=g' \eta^{Z'}_V \epsilon^{\mu \nu \rho \sigma} V_{\mu} Z'_{\nu} V_{\rho \sigma}+g'\sum_f\overline{f}\gamma^\mu \left(V^{Z'}_f-A^{Z'}_f \gamma^5\right) f Z'_\mu,
\end{equation}
where the second terms represent interactions among $Z'$ and the SM fermions.
As a convention we have normalized, in both cases, the DM coupling to the new gauge coupling $g'$. As already discussed in the case of Abelian DM we have considered a Chern-Simons type interaction~\cite{Mambrini:2009ad}. The interaction term of the complex vector DM can instead arise at the renormalizable level by considering the DM as the vector boson of an additional non abelian group, the minimal option would be SU(2), and the exactly mimicking the Electroweak group $SU(2) \times U(1)$ (this would require the presence of an additional $Z'$ which we assume to be heavy enough to have a negligible impact in the phenomenology). 

The parameter $\eta^{Z'}_V$ contains the 
model specific information for $[[VVZ']]$ interaction.

The most important difference among the two scenarios relies in the DM-nucleon scattering cross-section. In the non-Abelian case, interaction with the vectorial current $\ovl q \gamma^\mu q$ are possible, thus leading to the SI cross-section: 
\begin{equation}
\sigma^{\rm SI}_{Vp}=\frac{g'^4 {(\eta_V^{Z'})}^2 \mu_{Vp}^2}{\pi m_{Z'}^4}{\left(V_u^{Z'} \left(1+\frac{Z}{A}\right)+ V_d^{Z'} \left(2-\frac{Z}{A}\right)\right)}^2.
\end{equation} 

In the Abelian case, on the contrary, the only (momentum) unsuppressed interaction, is with the axial-vector quark current $\ovl q \gamma^\mu \gamma_5 q$, so that the interaction with nucleons is Spin Dependent with cross-section given by:
\begin{equation}
\sigma^{\rm SD}_{Vn}=\frac{3 g'^4 {(\eta_V^{Z'})}^2 \mu_{Vn}^2}{\pi m_{Z'}^4}\frac{{\left(A_u^{Z'} \left(\Delta_u^p S_p^A+\Delta_d^p S_n^A\right)+ A_d^{Z'} \left(\left(\Delta_d^p+\Delta_s^p\right)S_p^A+\left(\Delta_u^p+\Delta_s^p\right)S_n^A\right)\right)}^2}{(S_p^A+S_n^A)^2}.
\end{equation}
Here $\Delta^p_{i}$ denotes spin content of the `i'-th quark flavour inside proton 
and $S^A_p,\,S^A_n$ represent proton and neutron contribution to
the spin of nucleus, respectively. For Xenon based detectors $S^A_p \ll S^A_n$ so that the reference cross section is the one of DM on neutrons.

\begin{figure}[t]
\includegraphics[width=6.0 cm]{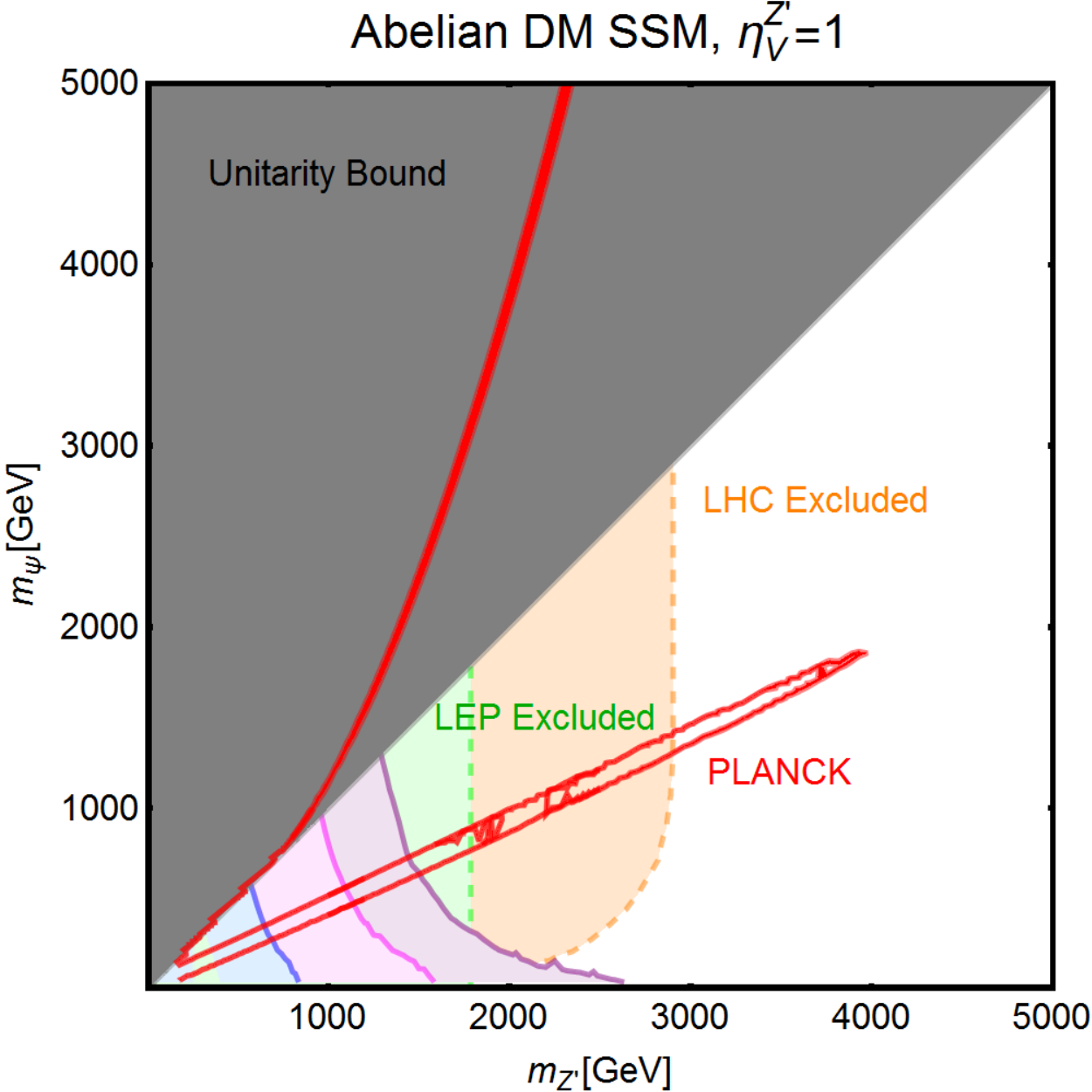}
\includegraphics[width=6.0 cm]{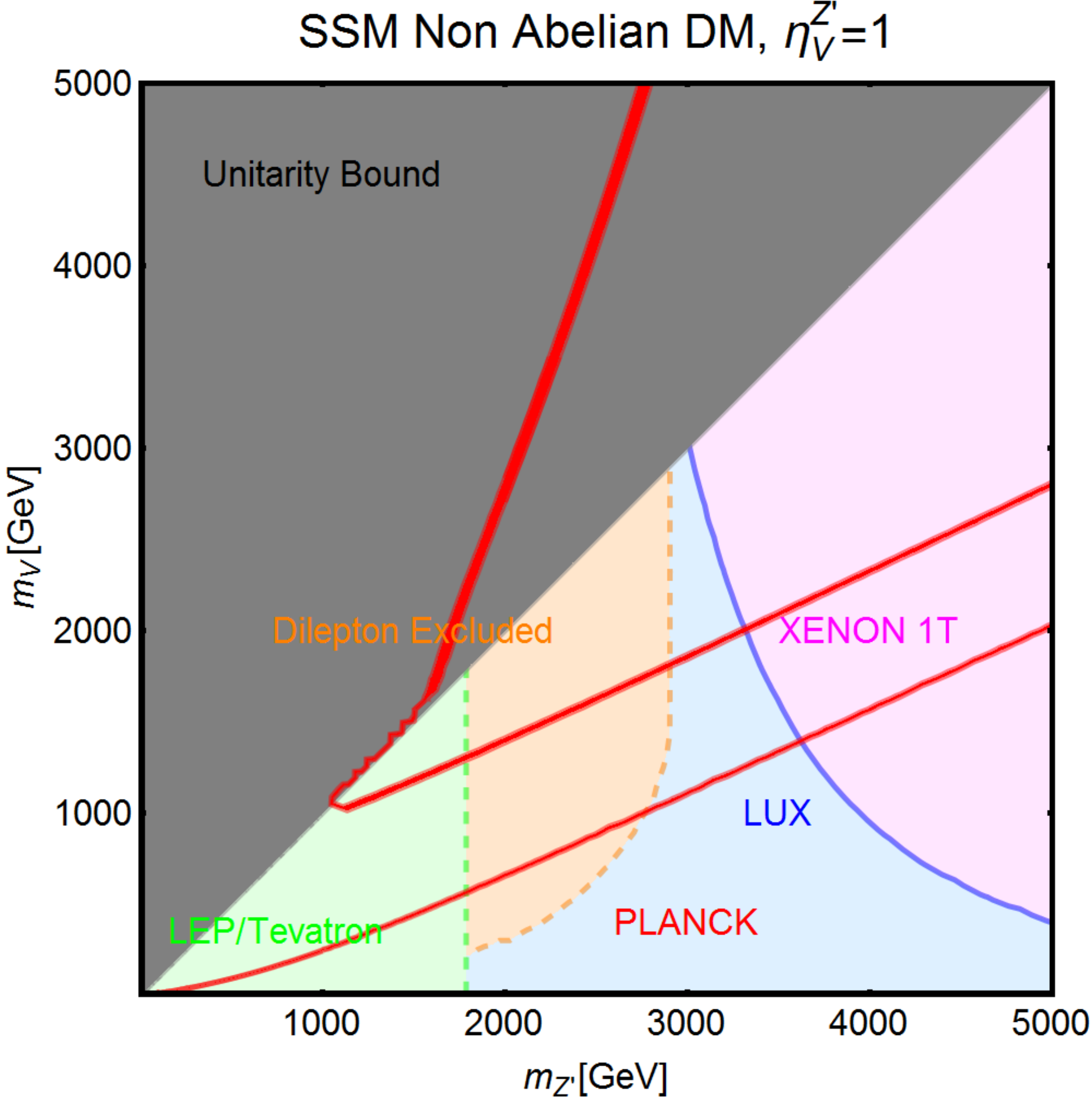}
\caption{\footnotesize{Combined constraints for Abelian (left panel) and non-Abelian (right-panel) vectorial DM interacting with a $Z'$ mediator. In both cases with have chosen SSM couplings of the $Z'$ with the SM fermions (see table ~(\ref{tab:Zpcouplings})). Colour scheme is the same as
fig.~(\ref{fig:FZprime}).}}
\label{fig:VZprime}
\end{figure}
%

The Abelian and non Abelian DM have very different properties also for what regards annihilations. In the first case we have that the annihilation cross-section into SM fermions is strongly suppressed, being, in fact, given by:
\bea
\langle \sigma v \rangle && (VV \rightarrow \ovl f f) \approx
\nonumber \\
&&\frac{g'^4 {(\eta_V^{Z'})}^2 m_f^2}{9 \pi m_{Z'}^4}v^2 \sum_f n_c^f {(A_f^{Z'})}^2+\frac{10}{81 \pi}g'^4 {(\eta_V^{Z'})}^2 \frac{m_V^2}{(m_{Z'}^2-4 m_V^2)^2}v^4 \sum_f n_c^f \left({(V_f^{Z'})}^2+{(A_f^{Z'})}^2\right).
~~~~~
\eea

Its first non-zero contribution, the p-wave, is suppressed by the final state mass. Ad exception of the case in with the mass of the ${Z'}$ is not too far from the one of the top quark, the DM annihilation cross section is actually dominated by the $d$-wave contribution and then results suppressed as $v^4$. On the contrary, the DM features a very efficient annihilation into $Z'$ pairs, whether kinematically allowed:
\begin{equation}
\langle \sigma v \rangle (VV \rightarrow Z' Z') \approx \frac{1}{18 \pi} g'^4 {(\eta_V^{Z'})}^4 \frac{m_V^2}{m_{Z'}^4}.
\end{equation} 

Its contribution to the DM relic density is nevertheless limited by the unitarity constraint~\cite{Griest:1989wd,Beacom:2006tt}.
In the case of non-Abelian vector DM the annihilation cross-section into SM fermions is only p-wave suppressed:
\begin{equation}
\langle \sigma v \rangle (VV \rightarrow \ovl f f) \approx \frac{2 g'^4 {(\eta_V^{Z'})}^2 }{\pi}v^2 \frac{m_V^2}{(m_{Z'}^2-4 m_V^2)^2}\sum_f n_c^f \left({(V_f^{Z'})}^2+{(A_f^{Z'})}^2\right),
\end{equation} 
while the annihilation cross-section into $Z' Z'$, instead, features a similar behaviour with respect to the Abelian case:
\begin{equation}
\langle \sigma v \rangle (VV \rightarrow Z' Z') \approx \frac{1}{4 \pi} g'^4 {(\eta_V^{Z'})}^4 \frac{m_V^2}{m_{Z'}^4}.
\end{equation}

The different limits on the two scenarios are shown, as customary, in the plane $(m_{Z'},m_V)$, in fig.~(\ref{fig:VZprime}). In the case of non-Abelian DM the weaker limits from DD do not coincide with a larger viable region for thermal DM since the contemporary suppression of the DM annihilation cross-section into fermions (the annihilation into $Z'Z'$ is strongly limited by unitarity) allow for the correct relic density only above the limit from LHC dilepton searches ad exception of a tiny region in correspondence of s-channel resonance. Much worse is the situation in the case of non-Abelian DM. Indeed, already current limits from Direct Detection overcome the ones from LHC and exclude thermal DM for both $m_{Z'}$ and $m_V$ below 5 TeV.




\section{Dark Portals (partially) evading Direct Detection}

\subsection{Pseudoscalar portal}

We will investigate in this subsection the phenomenology of a pseudoscalar s-channel portal. Under the assumption, performed along this paper, of CP-invariant interactions, only fermionic DM can be considered in this case. For simplicity we will limit to describe Dirac fermionic DM since the Majorana case features no substantial differences. We will then consider the following Lagrangian:
\begin{equation}
\mathcal{L}=i \lambda_\psi^a \ovl \psi \gamma_5 \psi a+i \sum_f \frac{c_a}{\sqrt{2}} \frac{m_f}{v_h}\ovl f \gamma_5 f a,
\end{equation}
where we have assumed, similarly to the case of scalar mediator, Yukawa-like couplings of the mediator with SM fermions, ensuring a $SU(2)$ invariant construction.

DM relic density is determined by annihilation into SM fermions pairs and, whether kinematically accessible, $aa$ pairs. At the leading order in the velocity expansion the corresponding cross section can be analytically approximated as follows:
%
\begin{equation}
\langle \sigma v \rangle{(\ovl\psi\psi \to \ovl ff)} \approx \sum_f \frac{n_c^f c_a^2 {(\lambda_\psi^a)}^2}{2 \pi}\frac{m_f^2}{v_h^2}\times \left \{
\begin{array}{cc}
\frac{m_\psi^2}{m_a^4} ~{\rm for}~ m_\psi < m_a, \\
\frac{1}{16 m_\psi^2} ~{\rm for}~ m_\psi > m_a,
\end{array}
\right.
\end{equation}
this cross-section is s-wave dominated (hence capable of Indirect DM signals). Given the Yukawa structure of the couplings with SM fermions this cross-section is sizable, away from s-channel resonances, only for $m_\psi > m_t$. In such a case a numerical estimate is given by:
\begin{equation}
\langle \sigma v \rangle(\ovl\psi\psi \to \ovl t t)\approx\left \{
\begin{array}{cc}
2.5 \times 10^{-25} {\mbox{cm}}^3 {\mbox{s}}^{-1} {(\lambda_\psi^a)}^2 c_a^2 {\left(\frac{m_\psi}{300\,\mbox{GeV}}\right)}^2 {\left(\frac{1\,\mbox{TeV}}{m_a}\right)}^4 ~{\rm for}~ m_\psi < m_a, \\
1.9 \times 10^{-24} {\mbox{cm}}^3 {\mbox{s}}^{-1} {(\lambda_\psi^a)}^2 c_a^2 {\left(\frac{m_\psi}{300\,\mbox{GeV}}\right)}^2 {\left(\frac{1\,\mbox{TeV}}{m_a}\right)}^4 ~{\rm for}~ m_\psi > m_a.
\end{array}
\right.
\end{equation}  

The annihilation cross-section into $aa$ pairs is, instead, p-wave suppressed:
\begin{equation}
\langle \sigma v \rangle{(\ovl \psi \psi \to aa)} \approx \frac{{(\lambda_\psi^a)}^2}{192 \pi m_\psi^2}v^2 \approx 2.3 \times 10^{-26} {\mbox{cm}}^3 {\mbox{s}}^{-1}{(\lambda_\psi^a)}^2 {\left(\frac{500\,\mbox{GeV}}{m_\psi}\right)}^2.
\end{equation}

The most peculiar feature of the pseudoscalar portal scenario is, nevertheless, the weakness of the interactions possibly responsible of DD~\footnote{For this reason it is also dubbed as coy DM~\cite{Boehm:2014hva}.}. Indeed, tree-level interactions between the DM and SM quarks (and gluons), mediated by a pseudoscalar, are momentum suppressed. They can be described by the following scattering rate
for a target nucleus of mass $m_T$~\cite{Arina:2014yna}:
\bea
\label{eq:coyDD}
 \frac{d \sigma_T}{dE_R}&&=\frac{{(\lambda_\psi^a)}^2 c_a^2}{128 \pi}\frac{q^4}{m_a^4}\frac{m_T^2}{m_\chi m_N}\frac{1}{v_E^2}\sum_{N,N'=p,n} g_N g_{N'} F_{\Sigma^{''}}^{NN'}(q^2), \nonumber\\
 g_N &&= \sum_{q=u,d,s} \frac{m_N}{v}\left[1-\frac{\overline{m}}{m_q}\right]\Delta_q^{N}, ~~{\rm with~~} \overline{m}={\left(1/m_u+1/m_d+1/m_s\right)}^{-1},
\eea
where $v_E$ represents the DM speed in the Earth frame, $E_R$ is the nuclear
recoil energy with recoil velocity $q$, and $F^{NN'}_{\Sigma^{''}}$ are
(squared) form factors whose (approximate) analytical expressions are found, for example, in ref.~\cite{Fitzpatrick:2012ix}. The cross-section of eq.~(\ref{eq:coyDD}) does not correspond neither to SI nor to SD (although the latter is a good approximation) interactions; for this reason we have expressed it directly in terms of a differential cross-section on a nucleus with mass $m_T$.

As can be easily argued the $q^4$ dependence implies a very suppressed scattering rate so that a potentially detectable signal is produced only for $m_a \sim O(10\,\mbox{MeV})$~\cite{Arina:2014yna}. We won't consider here these low values of the mass since they are subject to bounds from low energy observables and rare flavor processes (see e.g., ref.~\cite{Dolan:2014ska} for an extensive analysis). We also remark that Xenon based detectors, like LUX, XENON and LZ would be in any case not suited for probing the interaction cross-section of eq.~(\ref{eq:coyDD}) since it originates from a coupling of the DM mostly with unpaired protons, not present in Xenon, having instead, isotopes with an odd number of neutrons.

In the setup considered the most relevant interactions originate at the loop level, from a box diagram in which two pseudoscalars, one DM and one quark state are exchanged~\cite{Ipek:2014gua}, and are of SI-type. The corresponding cross-section is given by: 
\bea
 \sigma_{\psi p}^{\rm SI}&&=\frac{\mu_{\psi p}^2}{\pi}{\left|\sum_q \alpha_q f^p_q\right|}^2
~~{\rm with~~} \alpha_q=\frac{m_q^2 {(\lambda_\psi^a)}^2 c_a^2}{128 \pi^2 m_a^2 \left(m_\psi^2-m_q^2\right)}\frac{m_\psi m_q}{v^2}\left[F\left(\frac{m_\psi^2}{m_a^2}\right)-F\left(\frac{m_q^2}{m_a^2}\right)\right],\nonumber\\
 F(x)&&=\frac{2}{3x}\left[4+f_{+}(x)+f_{-}(x)\right],\nonumber\\
f_{\pm}(x)&&=\frac{1}{x}\left(1 \pm \frac{3}{\sqrt{1-4x}}\right){\left(\frac{1+\pm \sqrt{1-4x}}{2}\right)}^3 \log\left(\frac{1 \pm \sqrt{1-4x}}{2}\right).
\eea
with the coefficients $f_q^p$ being the same as the ones defined in the case of real scalar portals ($f^p_{c,b,t}=\frac{2}{27}f_{TG}^p$).

\begin{figure}[t]
\includegraphics[width=4.9 cm]{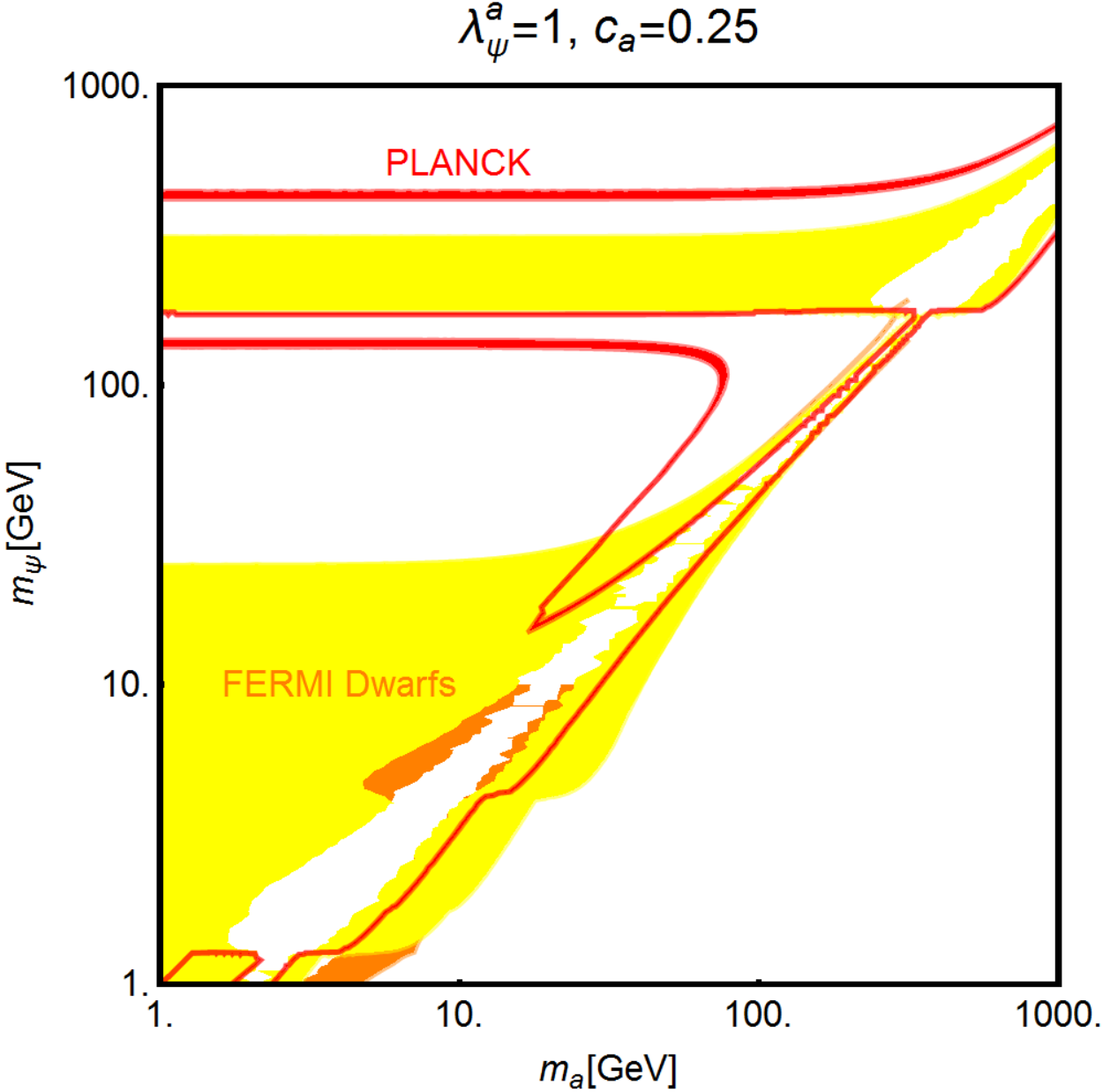}
\includegraphics[width=4.9 cm]{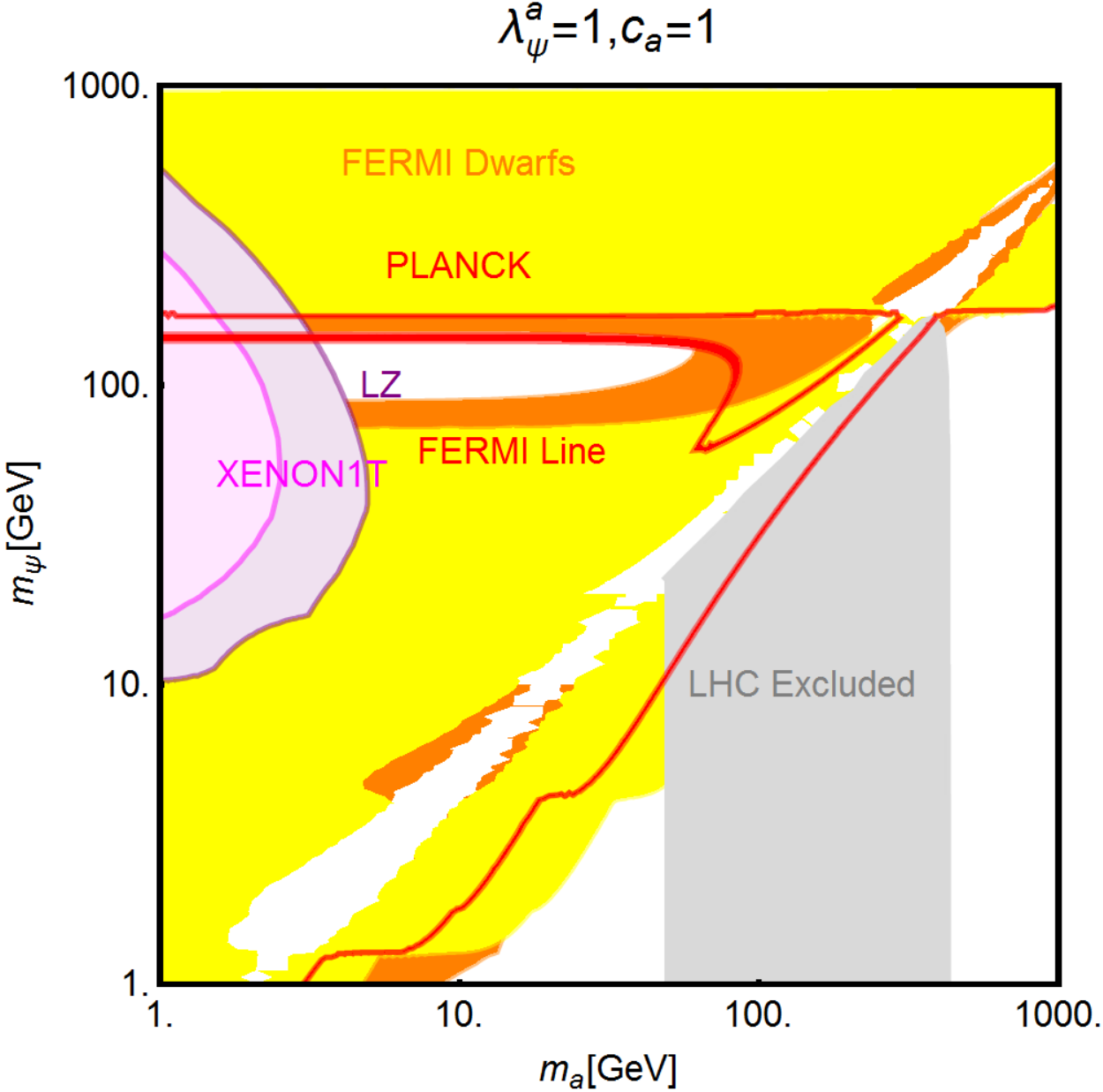}
\includegraphics[width=4.9 cm]{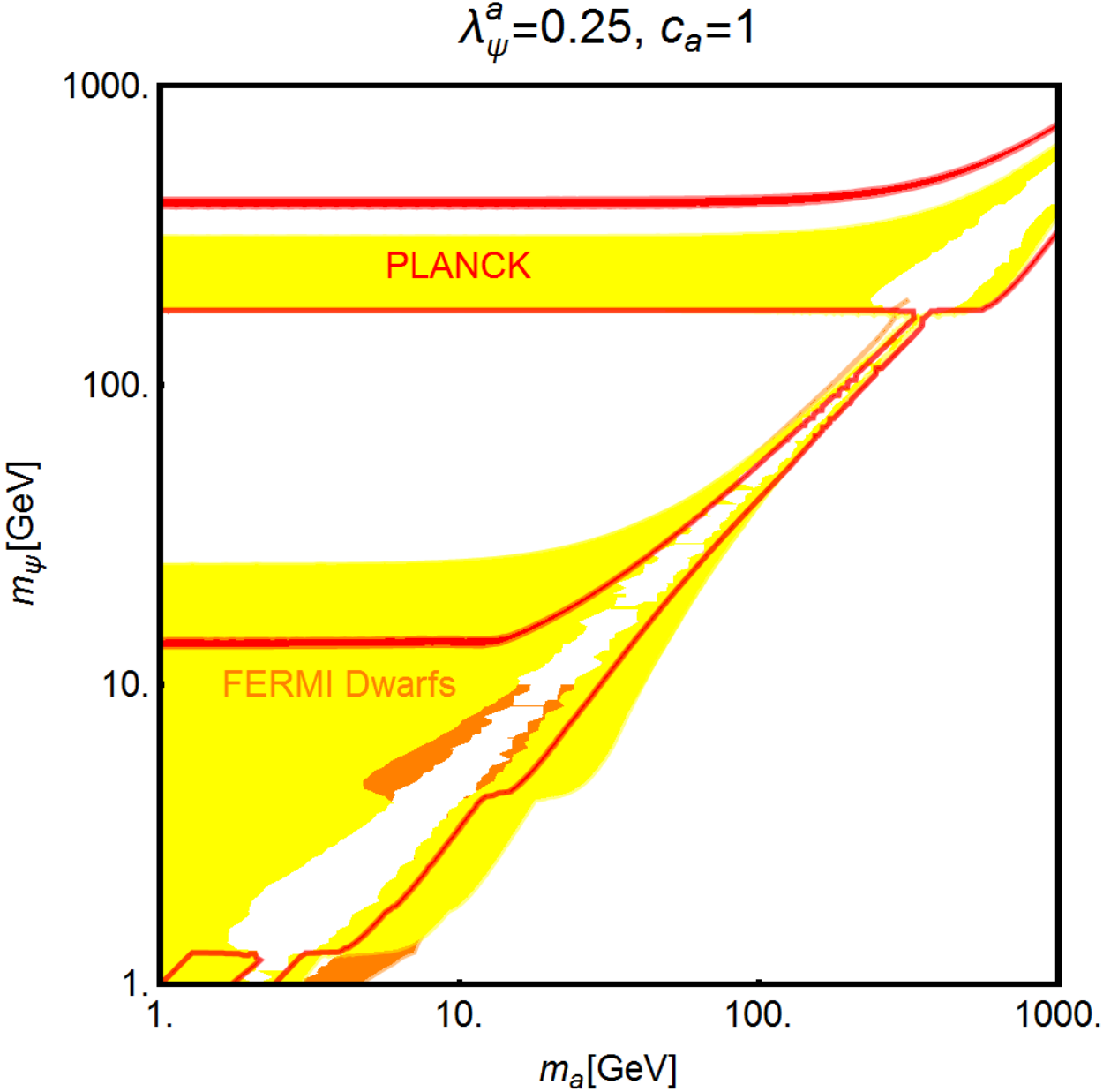}
\caption{\footnotesize{Summary of constraints from the phenomenology
of a fermionic DM in the case of s-channel pseudoscalar mediator. These are reported in the bi-dimensional $(m_a,\,m_\psi)$ plane for three assignations of $(\lambda_\psi^a,\,c_a)$, namely, from left to right, $(1,\,0.25)$, $(1,\,1)$ and $(0.25,\,1)$. The red coloured lines are the iso-contours of the correct DM relic density. In the yellow coloured regions the DM annihilation cross-section into SM fermions, computed at present time, exceeds the limit by FERMI from searches of signals in Dwarf Galaxies~\cite{Fermi-LAT:2016uux}. In the orange coloured regions the loop-induced annihilation cross-section into photon lines exceeds limits set by searches of gamma-ray lines~\cite{Ackermann:2015lka}. For $(\lambda_\psi^a,\,c_a)=(1,\,1)$ an exclusion region (gray coloured) from the LHC searches of monojet events~\cite{CMS:2016pod}, as well as projected excluded regions by XENON1T (magenta coloured) and LZ (purple coloured) are also reported.}}
\label{fig:FPSsimplified}
\end{figure}

As evidenced in fig.~(\ref{fig:FPSsimplified}), this cross-section is very suppressed so that no constraints come from present experiments. On the contrary, for $O(1)$ values of both the $\lambda_\psi^a$ and $c_a$ couplings, next generation detectors can partially probe the parameter space corresponding to thermal DM.

More stringent constraints come from DM indirect Detection. Contrary to the scalar mediator case the DM annihilation cross-section into SM fermions is s-wave dominated. This processes lead to potential signals in the gamma-ray continuum which can be probed by the FERMI satellite~\cite{Fermi-LAT:2016uux}. In addition, the DM can pair annihilate into two photons, through an effective coupling between the pseudoscalar mediator and photons, generated by triangle loops of SM fermions~\cite{Fermi-LAT:2016uux}. This process, responsible of the generation of gamma-ray lines, is strongly constrained by the negative results in present searches~\cite{Ackermann:2015lka}.

The summary of our analysis is presented, in the usual fashion, in fig.~(\ref{fig:FPSsimplified}). The figure features three panels corresponding to the assignations $(\lambda_\psi^a,\,c_a)=(1,\,0.25), (1,\,1), (0.25,\,1)$.
As already anticipated the most stringent constraints comes from Indirect DM searches in DSph galaxies. Indeed the absence of signals excludes thermal DM for mass below approximately 50 GeV. The most disfavored scenario turns to be the one corresponding to the assignation $\lambda_\psi^a=c_a=1$. Indeed a complementary constraint comes from searches of gamma-ray lines so that the excluded ranges of DM masses reaches order of 100 GeV. This specific assignation of the couplings has been also investigated at LHC through searches of events with monojet and missing energy. Due to the absence of any signal, the region of the parameter space corresponding to $100 \lesssim m_a \lesssim 500\,\mbox{GeV}, m_a > 2 m_\psi$ is currently excluded.

\subsection{Scalar + Light Pseudoscalar portal}
In this subsection we will consider the case in which the pseudoscalar mediator is actually a component of a complex field $\Phi\rightarrow (S+ia)/\sqrt{2}$ described by the following lagrangian: 
\bea
\mathcal{L}_\Phi&&=\partial_\mu \Phi \partial^\mu \Phi^{*}+\mu_\Phi^2 |\Phi|^2-\lambda |\Phi|^4+\frac{\epsilon_\Phi^2}{2}\left(\Phi^2 + \mbox{h.c.}\right).
\eea


We further that, after EWSB, the scalar component of $\Phi$ gets non-zero VEV $(v_\Phi)$, generating a mass term of its scalar component, $m_S=\sqrt{2\lambda}v_\Phi$ while leaving the pseudoscalar component massless. A mass for this second field is originated by an explicit mass term $m_a=\sqrt{2} \epsilon_\Phi$. In this kind of setup it is rather natural to identify the pseudoscalar component with a pseudo Goldstone boson associated to a $U(1)$ global symmetry carried by the complex scalar $\Phi$ and then assume that $m_a \ll m_S$~\cite{Mambrini:2015nza,Arcadi:2016dbl,Arcadi:2016acg}. We further assume that after EWSB breaking it is possible to write interaction terms of the fields $S$ and $a$ both with SM fermions and a fermionic DM candidate:  
\bea
\label{eq:bsmscapsca}
 -\mathcal{L}&&=\frac{m_S^2}{2}S^2+\frac{m_a^2}{2} a^2+\sqrt{\frac{\lambda}{2}}m_S Sa^2+\sqrt{\frac{\lambda}{2}}m_S S^3+\frac{\lambda}{4}{\left(S^2+a^2\right)}^2 \nonumber\\
&& +m_\psi\ovl \psi \psi+g_\psi \left(S \ovl \psi \psi+i a \ovl \psi \gamma^5 \psi\right) +\sum_f c_S \frac{m_f}{v_h}\left(S \ovl f f+i a \ovl f \gamma^5 f\right).
\eea
We have again assumed Yukawa-like
interactions among $S,\,a$ and the SM fermions where
the concerned couplings, including a normalization factor 
of $1/\sqrt{2}$, are parametrized as $c_S$.

While the generation of the interaction term with SM fermions would be theoretically challenging (the simplest option would be represented by the mixing with the SM Higgs, as considered in the next section), the coupling of $\psi$ could be elegantly derived by considering the DM chiral with respect to the $U(1)$ global symmetry carried by the field $\Phi$, i.e. $\psi\equiv \psi_L+\psi_R$, and the assume an interaction of the form $(g_\psi \ovl \psi_L \psi_R \Phi +\mbox{h.c.})$. This interaction would also originate the DM mass so that $g_\psi \sim m_\psi/v_\Phi\sim \sqrt{2\lambda}m_\psi/m_S$~\cite{Arcadi:2016dbl,Arcadi:2016acg}. We won't explicitly consider this scenario here and rather regard $m_\psi$ and $g_\psi$ as independent parameters. 

The main feature of this double s-channel portal scenario is the lower amount of correlation between DM Direct Detection and relic density, due to the presence of a light mediator state (see also ref.~\cite{Duerr:2016tmh} for a similar idea).
Indeed, DM direct detection relies only on the coupling of the DM with the scalar component of the $\Phi$ field. DM relic density is, instead determined by different annihilation channels, including $f \ovl f$, $aa$, $Sa$ and $SS$ involving interactions of both $S$ and $a$ fields. The expressions for the annihilation cross-sections into fermion and $S$ pairs can be can be straightforwardly derived from the cases of the scalar and pseudoscalar portals and won't be then rediscussed in detail. The annihilation cross-section into $a$ pairs, despite the presence of an additional contribution from s-channel exchange of $S$ is only moderately altered with respect to the case of the pseudoscalar portal and we just then refer to its detailed expression presented in the appendix. The most prominent feature, relevant for DM relic density, is the presence of $Sa$ as final state for annihilation processes, when kinematically allowed, annihilation channel. To this corresponds, in fact, an efficient s-wave annihilation cross-section which can be analytically approximated as:
\begin{equation}
\langle \sigma v \rangle {(\ovl \psi \psi\to Sa)}=\left \{
\begin{array}{cc}
\frac{g_\psi^2 \lambda m_S^4}{512 \pi m_\psi^6}\approx 1.6 \times 10^{-25} {\mbox{cm}}^3 {\mbox{s}}^{-1} g_\psi^2 \lambda {\left(\frac{m_S}{1\,\mbox{TeV}}\right)}^4 {\left(\frac{600\,\mbox{GeV}}{m_\psi}\right)}^6 & m_\psi < m_S, \\
\frac{g_\psi^4}{16 \pi m_\psi^2} \approx 2.3 \times 10^{-25} {\mbox{cm}}^3 {\mbox{s}}^{-1} g_\psi^4 {\left(\frac{1\,\mbox{TeV}}{m_\psi}\right)}^2 & m_\psi > m_S.
\end{array}
\right.
\end{equation}

\begin{figure}[t]
\includegraphics[width=4.9 cm]{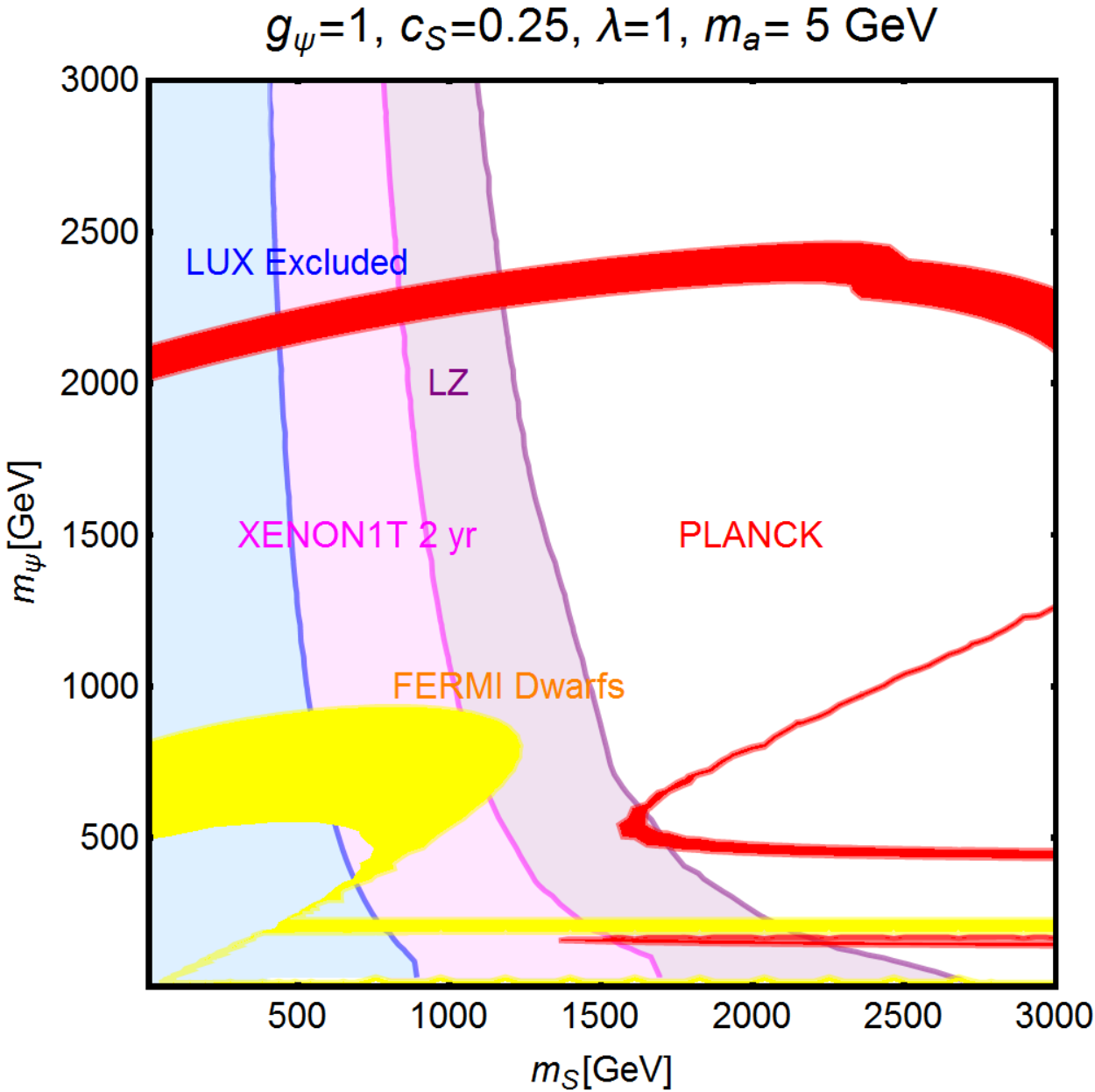}
\includegraphics[width=4.9cm]{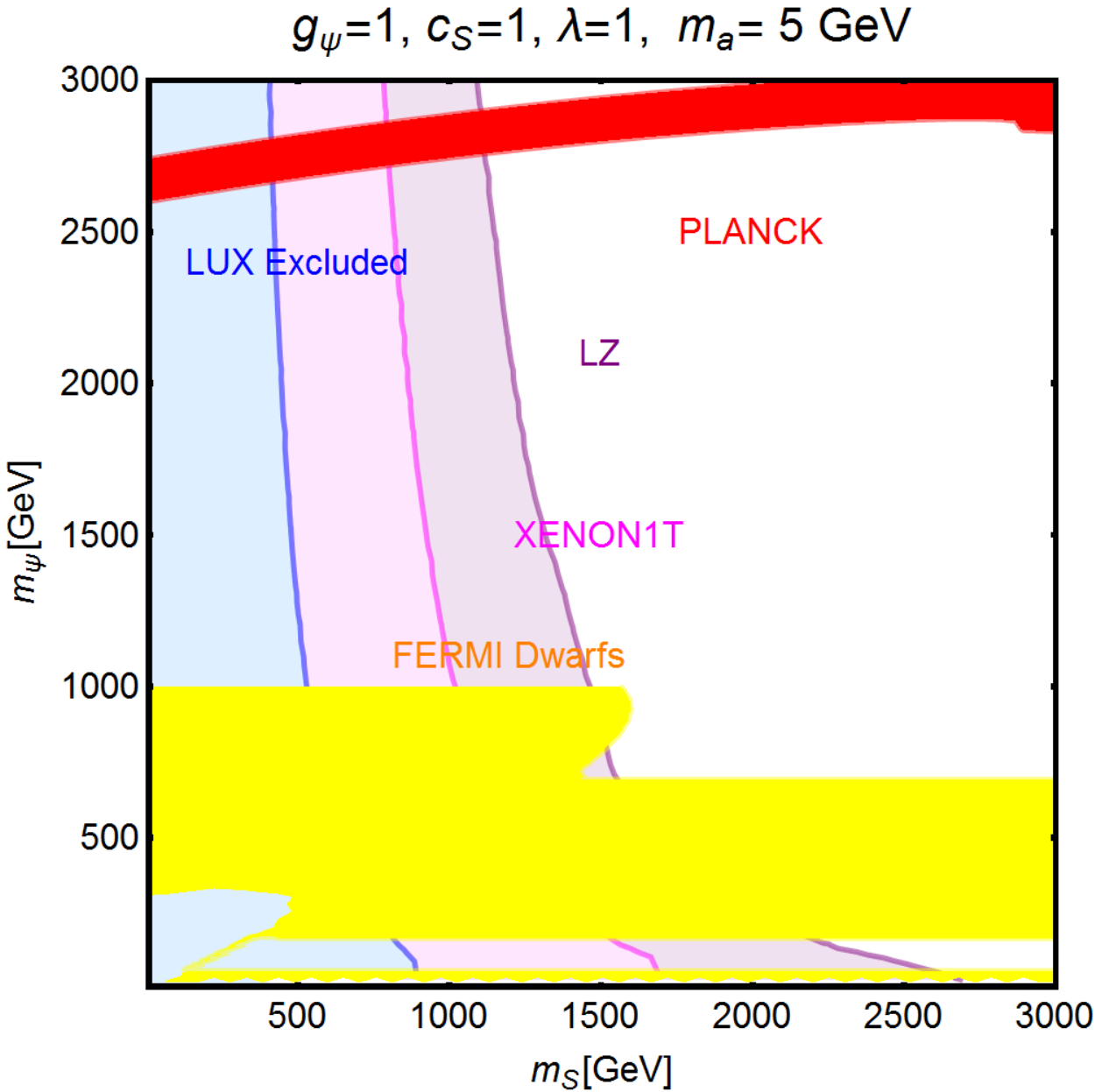}
\includegraphics[width=4.9 cm]{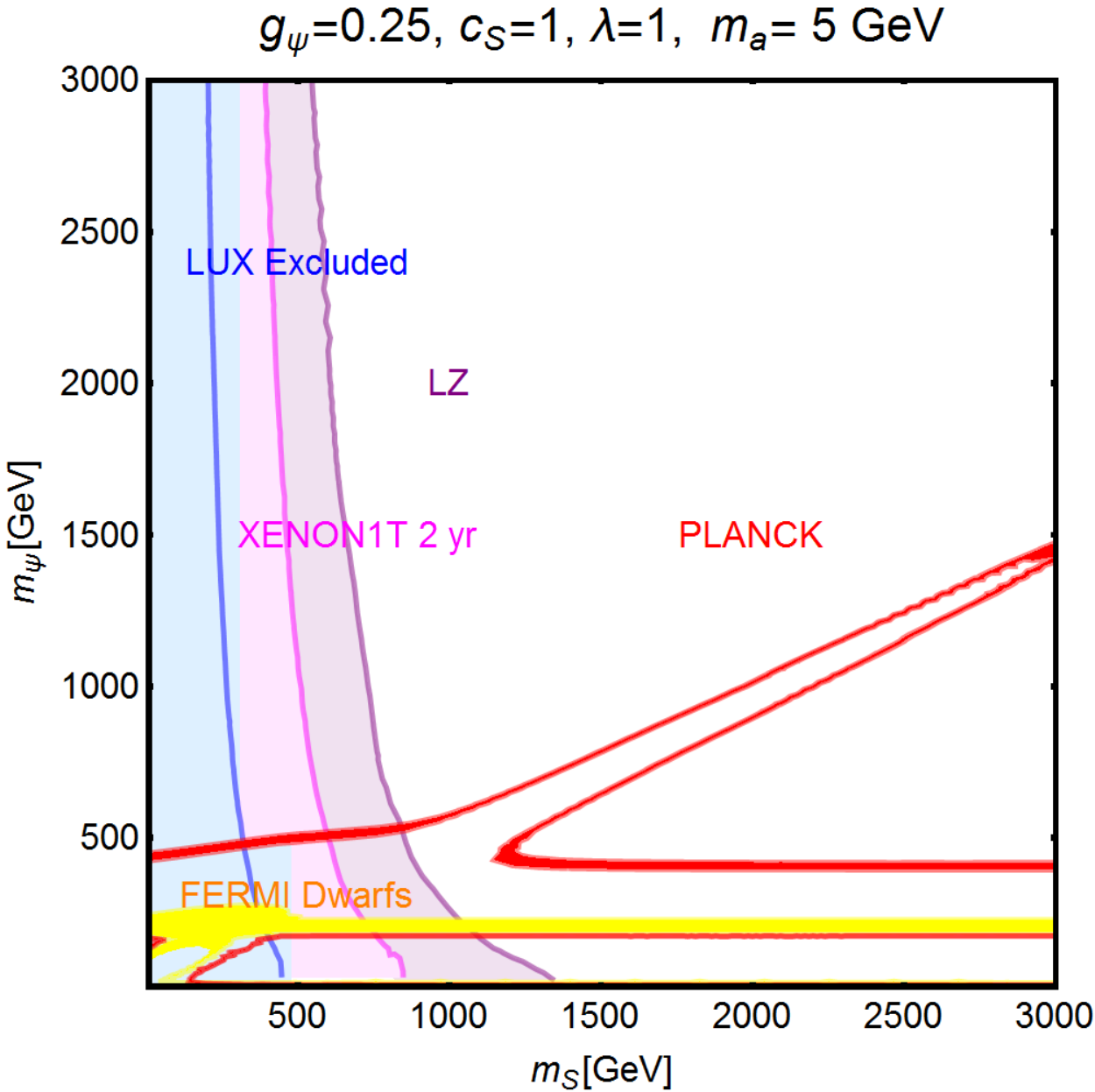}
\caption{\footnotesize{Summary of constraints for the Scalar + pseudoscalar portal with a fermionic DM in the bi-dimensional plane $(m_\psi,m_S)$ with $m_a,\,\lambda$ set to $5$ GeV, $1$, respectively. The three panels of the plot refer to the three assignations $(g_\psi,\,c_S)=(1,\,0.25)$ (left), $(1,\,1)$ (middle), $(0.25,\,1)$ (right). In each plot the red coloured lines are the contours of the correct DM relic density. The blue, magenta and purple coloured regions represent the current exclusion by LUX and projected exclusions by XENON1T and LZ, respectively. In the yellow coloured regions the DM annihilation cross-section at present time exceeds the limits determined by FERMI from searches of DM annihilations in DSPh.}}
\label{fig:FSiAportal}
\end{figure}

The comparison between DM relic density and (present and future) experimental constraints, is performed, in the usual fashion, in fig.~(\ref{fig:FSiAportal}). Here we have chosen the DM mass $m_\psi$ and the scalar mass $m_S$ as free parameters while $m_a$ has been set to 5 GeV, in order to avoid dangerous constraints from low energy physics. We have then considered three assignations for $(g_\psi, c_S)$, i.e., (1,0.25), (1,1) and (0.25,1) while the coupling $\lambda$ of the scalar potential has been set to 1. 

As evident the presence of annihilation channels, like $aa$ and $Sa$, involving non-SM light states allows to achieve the correct relic density, compatibly with constraints from DM direct detection, for relatively low values of the DM mass without necessarily rely on s-channel resonances. The presence of s-wave unsuppressed annihilation channels like $\ovl ff$ (contribution from s-channel exchange of the pseudoscalar) and $Sa$~\footnote{For simplicity, we have considered in our analysis only limits from searches in DSph. As pointed in~\cite{Arcadi:2016dbl,Ibarra:2013eda,Ibarra:2015tya} the annihilation into $Sa$ can lead to box shaped gamma-ray signals which could be probed in the next future by CTA. However, our choice for $m_a$, implies a too suppressed branching ratio of decay into gamma-rays for this field. For this reason we have not considered explicitly this possible signals in our analysis.} makes mandatory to consider, besides direct detection, also limits from Indirect Detection. However, for the chosen parameter assignation, these last constraints have no impact in the region corresponding to the correct DM relic density.





\section{Portals to secluded sectors}

In this section we will consider the case in which the mediator of the DM interactions has no direct coupling with the SM fermions. Dark portals can be nevertheless realized, at the renormalizable level. Indeed the SM features to Lorentz and gauge invariant bilinears, i.e. $H^\dagger H$ and $B^{\mu \nu}$. The first can be coupled to another scalar bilinear. In case also this second scalar field has non zero vacuum expectation value, a mass mixing with the SM Higgs is generate so that portal interactions with the SM fermions, as well as the gauge and the Higgs boson itself are generated. The field strength $B^{\mu \nu}$ can be coupled with the field strenght of another $U(1)$ gauge bosons. This kinetic mixing term is at the origin, after EW symmetry breaking, of a mixing between the $Z$ and the new $Z'$ boson.

In both these two scenarios, the DM interacts with the SM fields (both fermions and boson) through a double s-channel mediator. The relevant processes for phenomenology are substantially the same already investigated in the cases of the SM and BSM s-channel dark portals. Contrary to these scenarios we cannot consider order one couplings between the SM states and the BSM mediators, as well as between the DM and the SM ones, since the mixing between the Higgs and an additional scalar or the Z and the Z' are constrained to be small by several experimental and theoretical constraints.

\subsection{Higgs + Spin-0 portal}

In this subsection we will revisit the phenomenology of a real spin-0 mediator $\Phi$ considering the more realistic case in which it also features interaction with the SM Higgs doublet $H$.
We will thus assume the following scalar potential:
\begin{equation}
V=V_{H,\rm SM}+V(H,\Phi),
\end{equation}
where $V_{H,\rm SM}$ is the SM scalar potenti(${\lambda_h}{(H^\dagger H)}^2+\mu^2_h H^\dagger H$) while:
\begin{equation}
V(H,\Phi)={\lambda_{hS}} H^\dagger H {\Phi}^2+{\lambda_\Phi}{\Phi}^4+\mu^2_\Phi {\Phi}^2,
\end{equation}
here $\mu_h,\,\mu_\Phi$ are parameters with the dimension of
mass and $\mu^2_h,\,\mu^2_\Phi  <0$ for spontaneous symmetry breaking. 
%
One should note that in $V_{H,\rm SM}+V(H,\Phi)$, the condition
for getting a positive definite mass spectrum requires $4\lambda_h \lambda_\Phi > \lambda^2_{hS}$ while $\lambda_h,\,\lambda_\Phi >0$
are necessary to get $V_{H,\rm SM}+V(H,\Phi)$ bounded from below.
Combining these two conditions we see that $\lambda_{hS}$ can take
both the positive and negative values.
We denote the non-zero VEV of scalar field $\Phi$ as $v_\Phi$, so that it can be expanded as $\Phi=\left(v_\Phi+\phi\right)/\sqrt{2}$. The coupling with the SM Higgs doublet $H$ induces mass mixing so that, after EWSB (assuming unitary gauge), we define the following two mass eigenstates~\cite{Falkowski:2015iwa}:
\begin{equation}
\label{eq:BSMsmphimixing}
\left(
\begin{array}{c}
h \\
S
\end{array}
\right)=
\left(
\begin{array}{cc}
\cos\theta & \sin\theta \\
-\sin\theta & \cos\theta
\end{array}
\right)
\left(
\begin{array}{c}
\Re{H(0)} \\
\phi
\end{array}
\right),
\end{equation}
where $\Re{H(0)}$ represents the electrically neutral scalar part of the SM Higgs doublet $H$ and the mixing angle $\theta$ is defined by:
\begin{equation}
\label{eq:mixinghS}
\tan 2\theta=\frac{\lambda_{hS}v_h v_\Phi}{\lambda_\Phi v_\Phi^2-\lambda_h v_h^2}.
\end{equation}

The phenomenology of the mediator sector, thus, can be 
expressed as functions of the five parameters
$\lambda_h,\,\lambda_\Phi,v_h,v_\Phi,\,\lambda_{hS}$
or equivalently in terms of 
$m^2_h,\,m^2_S,v_h,\sin\theta,\,\lambda_{hS}$
using the following relations~\cite{Falkowski:2015iwa}: 
\bea
\label{eq:hSrelation}
\lambda_h &&= \frac{m^2_h}{2 v^2_h}
+ \frac{(m^2_S-m^2_h)\sin^2\theta}{2v^2_h},
\,\, \lambda_\Phi = \frac{2\lambda^2_{hS} v^2_h}{\sin^22\theta (m^2_S-m^2_h)}
\left(\frac{m^2_S}{(m^2_S-m^2_h)}-\sin^2\theta\right),\nonumber\\
{~~~~\rm and ~~~~} && v_\Phi = \frac{(m^2_h-m^2_S)\sin2\theta}{2\lambda_{hS}v_h}.
\eea

The mixing between $\Re{H(0)}$ and $\phi$
(see eq.~(\ref{eq:BSMsmphimixing})) indicates 
that both the mass eigenstates $(h,\,S)$ will couple to the SM states as well as to the DM and thus, represents a two-portal scenario.
The couplings of the two s-channel mediators with the SM $W^\pm,\,Z$-bosons and fermions are described by:
\begin{equation}
\mathcal{L}^{hS}_{\rm SM}=\frac{h \cos\theta - S \sin\theta}{v_h}\left[2 m_W^2 W^{+}_\mu W^{\mu -}+ m_Z^2 Z^\mu Z_\mu-\sum_f m_f \ovl f f \right],
\end{equation}
while their cubic self-couplings are given by:
\begin{equation}
\mathcal{L}_{hS} = -\frac{\kappa_{hhh}v_h}{2}~h^3-\frac{\kappa_{hhS}v_h}{2}\sin\theta~ h^2 S- \frac{\kappa_{hSS}v_h}{2}\cos\theta ~h S^2-\frac{\kappa_{SSS}v_h}{2}~ S^3,
\end{equation}
with:
\bea
 \kappa_{hhh}~&&= \frac{m^2_h}{v^2_h\cos\theta}
 \left(\cos^4\theta + \sin^2\theta \frac{\lambda_{hS}v^2_h}{(m^2_h-m^2_S)}\right),\,\,
  \kappa_{SSS}~= \frac{m^2_S}{v^2_h\sin\theta}
 \left(\sin^4\theta + \cos^2\theta \frac{\lambda_{hS}v^2_h}{(m^2_S-m^2_h)}\right),\nonumber\\
\kappa_{hhS} &&=\frac{2 m_h^2+m_S^2}{v_h^2}\left(\cos^2 \theta+\frac{\lambda_{hS}v_h^2}{(m_S^2-m_h^2)}\right),\,\,
\kappa_{hSS}=\frac{2 m_S^2+m_h^2}{v_h^2}\left(\sin^2 \theta+\frac{\lambda_{hS} v_h^2}{(m_h^2-m_S^2)}\right).
\eea

The parameters $\lambda_{hS}$ and $\sin\theta$ are subject of several experimental and theoretical constraints (see e.g.~\cite{Falkowski:2015iwa} for an extensive discussion). For example non null $\theta$ angle modify the couplings of the Higgs with SM particles and his then constrained by the measurement of the Higgs signal strength. The coupling $\lambda_{hS}$ is instead constrained by the stability of the scalar potential. Most of these constraints become increasingly stringent as $m_S$ decreases; for this reason we will focus in our analysis on the case $m_S > m_h$.

Analogously to the other spin-0 mediator scenarios, this extended Higgs sector will be coupled to a scalar DM $\chi$, a fermionic (we will restrict to the Dirac case) DM $\psi$ and a spin-1 DM $V_\mu$. We will consider the following Lagrangians for the corresponding interactions. 

In the case of scalar DM we consider the following interaction:
\begin{equation}
\mathcal{L}_{\chi}=\lambda_H^\chi {|\chi|}^2 H^\dagger H+\lambda_\Phi^\chi {|\chi|}^2 {\Phi}^2,
\end{equation}
which leads, after symmetry breaking, to the following effective lagrangian:
\begin{equation}
\mathcal{L}_\chi=g_{\chi \chi h}|\chi|^2 h+g_{\chi \chi S}|\chi|^2 S+g_{\chi \chi hh}|\chi|^2 h^2+g_{\chi \chi hS}|\chi|^2 h S+g_{\chi \chi SS}|\chi|^2 S^2,
\end{equation}
where:
\begin{align}
\label{eq:bsmhSDMS}
& g_{\chi \chi h}=\left(\lambda_H^\chi v_h \cos\theta
+ \lambda_\Phi^\chi \sin^2\theta\cos\theta \frac{(m^2_h-m^2_S)}{\lambda_{hS} v_h}\right),\nonumber\\ 
& g_{\chi \chi S}=\left(-\lambda_H^\chi v_h \sin\theta + \lambda_\Phi^\chi \cos^2\theta\sin\theta \frac{(m^2_h-m^2_S)}{\lambda_{hS}v_h} \right),\nonumber\\
& g_{\chi \chi h h}=\left(\lambda_H^\chi \cos^2\theta
+ \lambda_\Phi^\chi \sin^2\theta \right),\nonumber\\ 
& g_{\chi \chi h S}=2 \left(-\lambda_H^\chi \sin\theta \cos\theta
+ \lambda_\Phi^\chi \sin\theta\cos\theta \right),\nonumber\\
& g_{\chi \chi S S}=\left(\lambda_H^\chi \sin^2\theta
+ \lambda_\Phi^\chi \cos^2\theta \right)
\end{align}
here we have used eq.~(\ref{eq:hSrelation}). From eq.~(\ref{eq:bsmhSDMS}) it can be seen that without loss of generality one between the couplings $\lambda_H^\chi$ and $\lambda_\Phi^\chi$ can be set to zero. We will then pursuing this minimal choice setting $\lambda_H^\chi=0$. With this choice we end up with five free-parameters $\lambda^\chi_\Phi,\,\lambda_{hS},\,\sin\theta,m_\chi,m_S$.

In the case of fermionic DM it is natural, in this setup, to assume a Yukawa-type coupling with the field $\Phi$ of the form:
\begin{equation}
\mathcal{L}_\psi=y_\psi \ovl \psi \psi \Phi;
\end{equation}
the DM mass is dynamically generated by the VEV of $\Phi$ so that mass and DM coupling are not independent but are related as $y_\psi \propto {m_\psi}/{v_\Phi}$. This choice allows to reduce the number of free parameters compared to the scalar DM scenario. The effective couplings of the DM with the $S$ and $h$ field can be straightforwardly derived by using eqs.~(\ref{eq:BSMsmphimixing}) and (\ref{eq:hSrelation}).    

A dynamical generation of the DM mass has been considered also in the embedding of the interactions of vectorial DM. We indeed identify the DM as a stable gauge boson of a $U(1)$ dark gauge group~\footnote{Dark matter a gauge boson of dark sector connected to the SM through the Higgs sector has been also proposed, for different DM production mechanisms, in~\cite{Hambye:2008bq,Bernal:2015ova}.} spontaneously broken by the vev of a complex field $\Phi$. The interaction between the latter and the DM are then embedded in the covariant derivative $(D_\mu \Phi)^{*}D^\mu \Phi$ with $D_\mu=\partial_\mu -i \frac{\eta_V^S}{2}V_\mu$~\footnote{We have adopted the same definition of the covariant derivative as~\cite{Arcadi:2016qoz}.}. After symmetry breaking the DM lagrangian reads: 
\begin{equation}
\mathcal{L}_{V}=\frac{1}{2}{\eta_V}m_V V^\mu V_\mu S+ \frac{1}{8} S^2 V^\mu V_\mu + \frac{1}{2}m^2_V V^\mu V_\mu
\end{equation}
where we have used the expression of $\Phi$ in the unitary gauge $\Phi=\frac{1}{\sqrt{2}} \left(v_\Phi+S\right)$ and $m_V=\frac{1}{2} \eta_V v_\Phi$. The coupling $\eta_V$ here then represents a gauge coupling.

\begin{figure}[t]
\includegraphics[width=4.9cm]{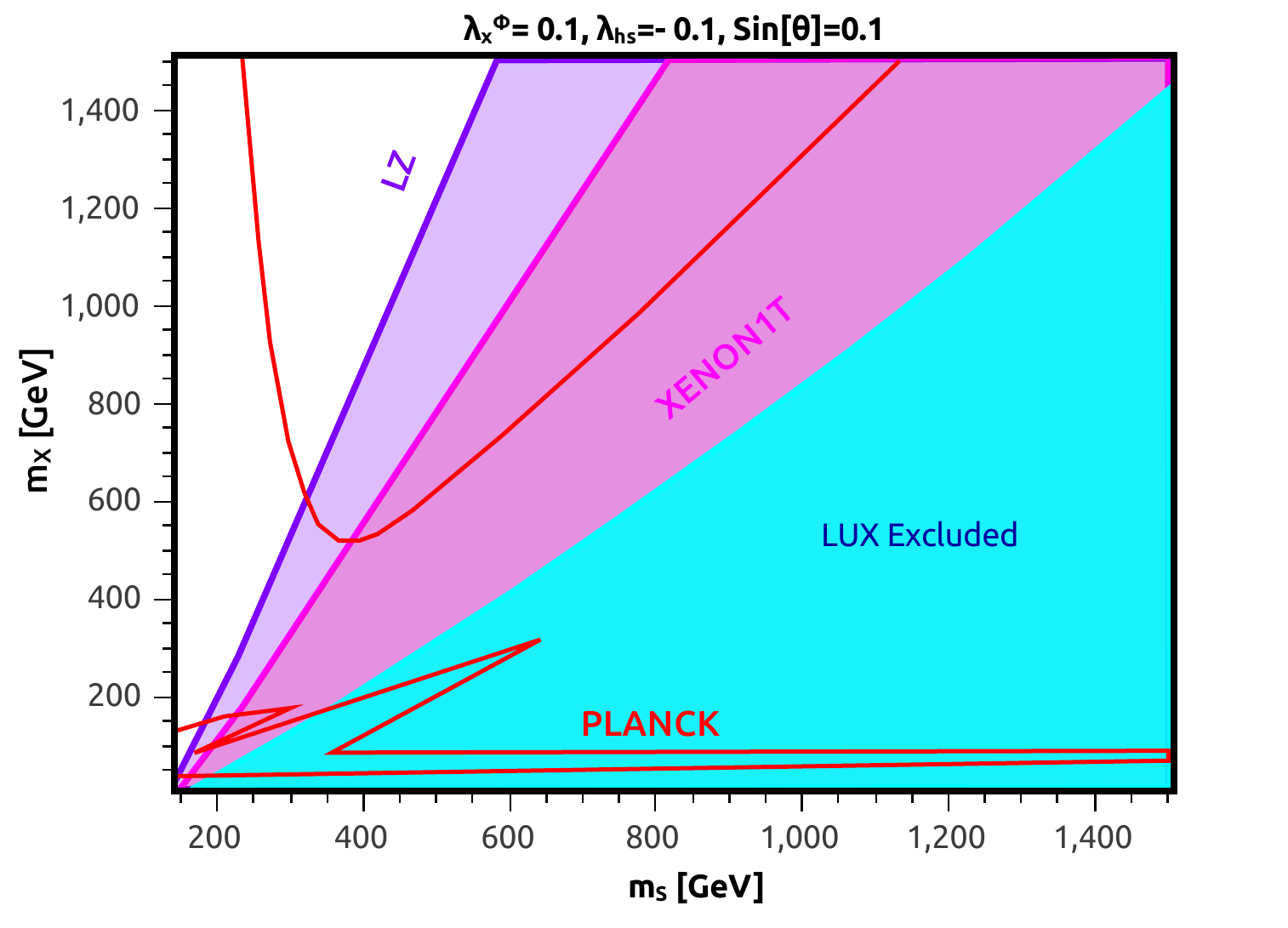}
\includegraphics[width=4.9cm]{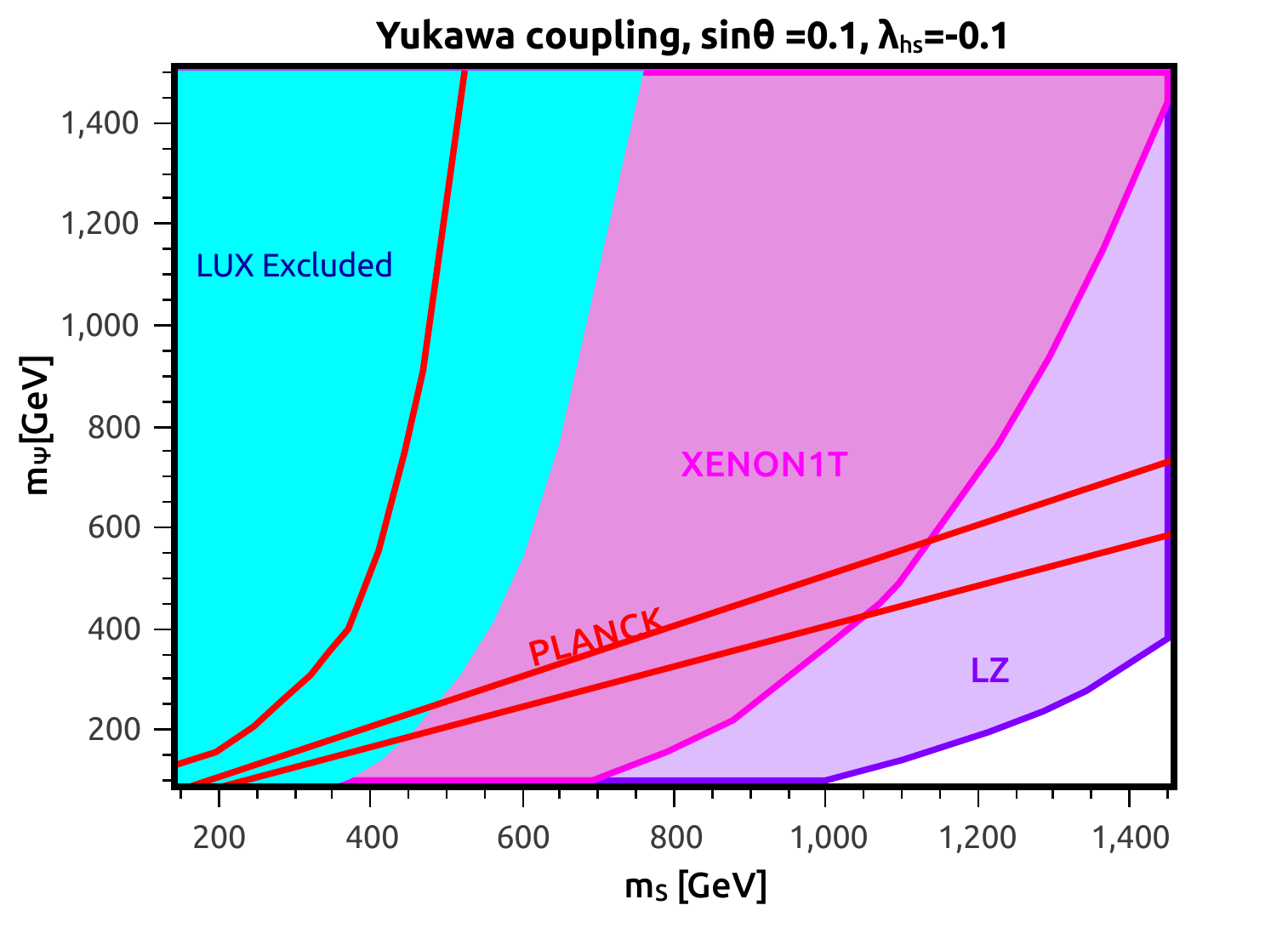}
\includegraphics[width=4.9cm]{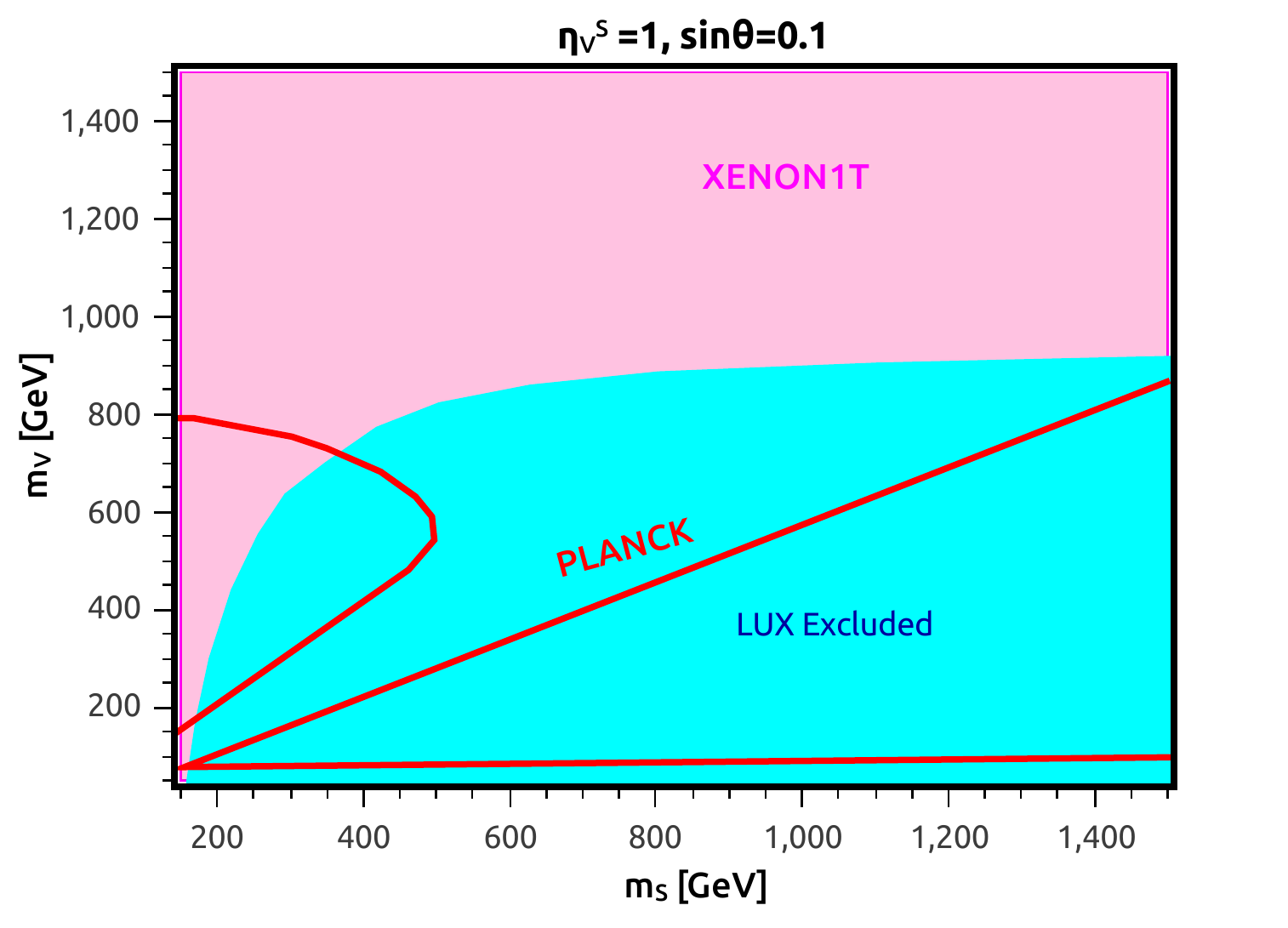}
\caption{\footnotesize{Summary of constraints for the SM-Higgs + BSM Scalar portal for a scalar (left), Dirac fermion (middle) and Vector (right) DM. The red coloured contours represent the correct DM relic density while the blue, magenta and purple coloured regions represent the current exclusion by LUX and the projected sensitivities of the XENON1T and LZ, respectively. The details concerning the assignations of the model parameters are discussed in the main text.}}
\label{fig:hSportal}
\end{figure}

The processes responsible for DM relic density and detection have been already discussed in detail for the cases of Higgs and Scalar portal individually; we then just illustrate the results of our analysis, as reported in fig.~(\ref{fig:hSportal}). We just remind that all the considered types of DM candidates feature SI interactions with nuclei.

The results are reported in the usual plane ($m_S,m_{\rm DM}$). The angle $\theta$ has been conservatively set to $\sin\theta=0.1$ in order to comply with the constraints from the Higgs signal strengths. Similarly the coupling $\lambda_{hS}$ has been set to values $-0.1$ in order to comply various constraints on the Higgs sector (see e.g., ref.~\cite{Falkowski:2015iwa} for an extensive discussion). The couplings $\lambda_H^\Phi$ and $\eta_V$ have been set to 1 while the DM coupling for fermionic DM is, in our construction, not a free parameter.

As evident the outcome of our analysis presents some sensitive differences with respect to the case of the spin-0 mediator discussed in the previous sections. In the case of fermionic and vector DM the limits from Direct Detection are rather effective, as due to the presence of an additional light mediator. In the case of vector DM the only viable region, for masses of the DM and the mediator below the TeV scale, corresponds to the case $m_V > m_S$ thanks to the enhancement of the DM annihilation cross-section due to the $VV \rightarrow SS$ process. In the case of fermion DM the region surviving current constraints corresponds to the ``pole'' $m_\psi \sim m_S/2$. In both cases, next generation of DD detectors will probe the WIMP paradigm for masses of the DM and the mediator up to few TeVs. More particular is the case of scalar DM. The shape of the DD contours is rather different with respect to the other spin-0 mediators. This is due to the different choice of the energy scale, the vev $v_\Phi$ rather than the mass of the mediator, which implies a larger cross-section at higher values of $m_S$. On the contrary, as can be noticed by eq.~(\ref{eq:bsmhSDMS}), for $\lambda_\chi^H=0$ the DM couplings become smaller as $m_h$ and $m_S$ get close in value~\footnote{In the case of scalar DM would be even possible to generate a ``blind spot'' in the scattering cross-section by a rather specific choice of the couplings $\lambda_\chi^{H,\Phi}$~\cite{Esch:2014jpa,Arcadi:2016kmk} which would be induce a destructive interference between the amplitudes associated to the $h$ and $S$ exchange. A different solution for relaxing Direct Detection constraints in presence of multiple scalars singlet coupled to the Higgs has been recently proposed in~\cite{Casas:2017jjg}}. For the chosen parameter assignations a large region of viable thermal DM is present in the regime $m_\chi > m_S$ for rather light $m_S$. A sizable part of this region will be excluded by absence of signals at XENON1T and LZ.

\subsection{Kinetic mixing}

We will reconsider in this section the scenario in which the DM is coupled to the gauge boson of a new $U(1)$ group. In this case, however, SM fermions won't be charged under the new gauge group so that no direct couplings with the new gauge boson are induced. The ``dark'' and visible sectors are nevertheless connected by a kinetic mixing operator $B_{\mu \nu} X^{\mu \nu}$ which can already exist at the tree level, being both Lorentz and gauge invariant, or be, alternatively, radiatively generated (for example if the new gauge sector features new fermions with non-trivial quantum numbers under the SM gauge group~\cite{Baumgart:2009tn}).

We will then consider the following Lagrangian with $\delta$ as the kinetic mixing parameter:
\begin{equation}
\label{eq:Klagrangian}
\mathcal{L}=-\frac{1}{4}B^{\mu \nu}B_{\mu \nu}-\frac{1}{4}X^{\mu \nu}X_{\mu \nu}-\frac{1}{2}\sin \delta B^{\mu \nu}X_{\mu \nu}+\frac{1}{2}m^2_{X} X^\mu X_\mu+\mathcal{L'}_{\rm SM}+\mathcal{L}_{\rm DM},
\end{equation}
where $B^{\mu \nu}$ and $X^{\mu \nu}$ are, respectively, the fields strength of the hypercharge and of new U(1) charge. $\mathcal{L'}_{\rm SM}$ is the SM Lagrangian besides the already written kinetic term of $B_{\mu \nu}$, while $\mathcal{L}_{\rm DM}$ is the DM Lagrangian including the kinetic and mass term as well as a coupling with the $X_\mu$ boson. These terms will depend on the spin assignation of the DM. We will consider the following cases:
\begin{equation}
\label{eq:KDMlagrangians}
\mathcal{L}_{\rm DM}=
\left \{
\begin{array}{cc}
\mathcal{L}_{\rm DM}=(D^\mu \chi)^*D_\mu \chi-m_\chi^2 \chi^{*}\chi & \mbox{(complex scalar)}\\  
\mathcal{L}_{\rm DM}=\ovl \psi \gamma^\mu D_\mu \psi-m_\psi \ovl \psi \psi & \mbox{(Dirac fermion)}\\
\mathcal{L}_{\rm DM}=\eta_V^X [[VVX]]+m_V^2 V_\mu^{\dagger}V^{\mu} & \mbox{(non-Abelian vector)}\\
\end{array}
\right.
\end{equation}
As already discussed, a very natural option to couple a scalar and or a fermionic DM to a new gauge boson is to assume it to be charged under the new symmetry group so that its interactions originate from the covariant derivative $D_\mu=\partial_\mu-i g_X X_\mu$~\footnote{Notice that in the definition of $g_X$ we have also encoded the value of the DM charge.}. Slightly more complicated is the case of vectorial DM. Here we have assumed a analogous coupling as the one, in the SM, between the Z and two W bosons~\footnote{The case of an Abelian vector DM will be object of a dedicated publication.}. The coupling $\eta_V^X$ encodes the gauge coupling $g_X$ and eventual extra factors entering in the definition of the couplings. $m_{\chi,\, \psi,\, V}$ denote the respective DM masses of different spin assignments.

The kinetic term in eq.~(\ref{eq:Klagrangian}) should be diagonalized and canonically normalized. After EWSB, it is possible to define three mass eigenstates for the neutral gauge bosons through the following two consecutive transformations~\cite{Babu:1997st,Chun:2010ve,Frandsen:2011cg,Mambrini:2011dw}: 
\begin{equation}
\label{eq:dtransformation}
\left(
\begin{array}{c}
B_\mu \\
W^3_\mu \\
X_\mu
\end{array}
\right)
=
\left(
\begin{array}{ccc}
1 & 0 & -\tan\delta \\
0 & 1 & 0 \\
0 & 0 & 1/\cos\delta
\end{array}
\right)
\left(
\begin{array}{ccc}
c_{\hat{W}} & -s_{\hat{W}} \cos\xi & s_{\hat{W}} \sin\xi \\
s_{\hat{W}} & c_{\hat{W}} \cos\xi & -c_{\hat{W}} \sin\xi \\
0 & \sin\xi & \cos\xi 
\end{array}
\right)
\left(
\begin{array}{c}
A_\mu \\
Z_\mu \\
Z'_\mu
\end{array}
\right),
\end{equation}
where, $c_{\hat W},\,s_{\hat W}=\cos\theta_{\hat W},\,\sin\theta_{\hat W}$ and the angle $\xi$ is defined by:
\begin{equation}
\tan 2\xi=\frac{-2 m_{\hat{Z}}^2 s_{\hat{W}} \cos\delta \sin\delta}{m_{X}^2-m_{\hat{Z}}^2 \cos\delta^2 +m_{\hat{Z}}^2 s_{\hat{W}}^2 \sin\delta^2}.
\end{equation}

Notice that the transformation given in eq.~(\ref{eq:dtransformation}) leads to physical solutions only if one of these conditions is met~\cite{Chun:2010ve}:
\bea
r_X &&\geq 1+2 s_{\hat{W}} \tan^2 \delta +2 \sqrt{s_{\hat{W}}^2 \tan^2 \delta \left(1+s_{\hat{W}}^2 \tan^2 \delta\right)}, \nonumber\\
r_X &&\leq 1+2 s_{\hat{W}} \tan^2 \delta -2 \sqrt{s_{\hat{W}}^2 \tan^2 \delta \left(1+s_{\hat{W}}^2 \tan^2 \delta\right)},\,{\rm where~} r_X=\frac{m_X^2}{m_{\hat{Z}}^2}.
\eea

In the expressions above $s_{\hat{W}},\,m_{\hat{Z}}$ do not represent the experimental measures of the Weinberg angle and of the $Z$-boson mass but are related to the latter, i.e., $s_W\equiv\sin\theta_W,\,m_Z$, as function of $\delta$.
Indeed, since the photon coupling does not change once passing to ``physical'' basis, one can write:
\begin{equation}
c_W^2 s_W^2=\frac{c_{\hat{W}}^2 s_{\hat{W}}^2}{1+s_{\hat{W}}\tan\xi \tan\delta},
\, {\rm ~with~~} c_W\equiv \cos\theta_W.
\end{equation}

In analogous fashion, the preservation of the mass of the $W$-boson by the transformation allows to relate the kinetic mixing parameter to the $\rho$ parameter as:
\begin{equation}
\rho=\frac{c_{\hat{W}}^2}{\left(1+s_{\hat{W}}\tan\delta \tan\xi\right)c_W^2}
\end{equation} 

The previous relation can be reformulated as:
\begin{equation}
\omega=s_W \tan\delta \tan\xi \simeq -\left(1-t_W^2\right) \Delta,
~~{\rm where~~}\Delta=\rho-1~~{\rm and~~} t^2_W=\tan^2\theta_W.
\end{equation}

So that $\rho-1=4^{+8}_{-4} \times 10^{-4}$ measurement~\cite{Amsler:2008zzb} can be used to constrain the parameter $\delta$. 

The kinetic mixing parameter is further constrained by
EW Precision Tests (EWPT)~\cite{Kumar:2006gm,Chang:2006fp} which reads as:
\begin{equation}
\label{eq:deltaEWPT}
\tan\delta \leq \frac{m_{Z'}}{2.5\,\mbox{TeV}}.
\end{equation}

The spectrum of the neutral gauge bosons features one massless eigenstate, coinciding with the SM photon, and two massive states. Their masses, in the (experimentally favored) $\sin\delta,\,\sin\xi \sim \delta,\,\xi \ll 1$ limit,
are given as:
\bea
m_Z^2 &&\simeq m_{\hat{Z}}^2+\left(m_{\hat{Z}}^2-m_X^2\right)\xi^2 ,\nonumber\\
m_{Z'}^2 &&\simeq m_X^2+m_X^2 \xi \left(\xi-s_{\hat{W}} \delta\right)-m_{\hat{Z}}^2\left(\xi-s_{\hat{W}}\delta\right)^2,
\eea
where $m_Z$ must coincide with the experimental value of the mass of the $Z$ boson.

The interactions (relevant for DM phenomenology) of the $Z,Z'$ with the SM states are described by~\cite{Chun:2010ve}:
\bea
 \mathcal{L}_{Z/Z',SM}&&=\ovl f \gamma^\mu \left(g_{f_L}^{Z}P_L+g_{f_R}^{Z}P_R\right) f Z_\mu+\ovl f \gamma^\mu \left(g_{f_L}^{Z'}P_L+g_{f_R}^{Z'}P_R\right) f Z'_\mu +g_W^Z [[W^+ W^-Z]] \nonumber\\
&& +g_W^{Z'} [[W^+ W^-Z']] +g_{hZZ}h Z^\mu Z_\mu + g_{hZZ'} h Z'_\mu Z^\mu+g_{hZ'Z'} h Z'_\mu Z'^{\mu},
\eea
where:
\bea
g_{f_L}^Z&&=-\frac{g}{c_W}\cos\xi \left \{T_3 \left(1+\frac{\omega}{2}\right)-Q \left[s_W^2+\omega \left(\frac{2-t_W^2}{2(1-t_W^2)}\right)\right] \right \}, \nonumber\\
g_{f_R}^Z&&=\frac{g}{c_W}\cos\xi \left \{Q \left[s_W^2+\omega \left(\frac{2-t_W^2}{2(1-t_W^2)}\right)\right] \right \}
\eea
\bea
 g_{f_L}^{Z'}&&=-\frac{g}{c_W}\cos\xi \left \{T_3 \left[s_W \tan\delta-\tan\xi+\frac{\omega}{2}\left(\tan\xi+\frac{s_W t_W^2 \tan\delta}{1-t_W^2}\right)\right]\right.\nonumber\\
&&\left.\hspace*{2.0cm}+ Q \left[s_W^2 \tan\xi-s_W \tan\delta +\frac{1}{2}t_W^2 \omega \left(\frac{\tan\xi-s_W \tan\delta}{1-t_W^2}\right)\right] \right \}, \nonumber\\
 g_{f_R}^{Z'}&&=-\frac{g}{c_W}\cos\xi \left \{Q \left[s_W^2 \tan\xi-s_W \tan\delta +\frac{1}{2}t_W^2 \omega \left(\frac{\tan\xi-s_W \tan\delta}{1-t_W^2}\right)\right]. \right\}
\eea
\begin{equation}
g_W^Z=g c_W \cos\xi \left(1-\frac{\omega}{2 (c_W^2-s_W^2)}\right), \,\,\,
g_W^{Z'}=-g c_W \sin\xi \left(1-\frac{\omega}{2 (c_W^2-s_W^2)}\right).
\end{equation}
\bea
 g_{hZZ}&&=\frac{m_Z^2}{v_h}\cos\xi^2 (1+\omega),\nonumber\\
 g_{hZZ'}&&=2\frac{m_Z^2}{v_h}\cos\xi^2\left[2 s_W \tan\delta-\tan\xi+\omega \left(\tan\xi+\frac{s_W t_W^2 \tan\delta}{1-t_W^2}\right)\right],\nonumber\\
 g_{hZ'Z'}&&=\frac{m_Z^2}{v_h}\cos\xi^2\left[\tan^2\xi+s_W^2 \tan\xi-\omega\left(2+\tan^2 \xi-\frac{s_W^2 t_W^2 \tan^2 \delta}{1-t_W^2}\right)\right].
\eea

Here $T_3,\,Q$ are the iso-spin quantum number and electric charges
of the associated SM fermions.
The coupling of the DM with the two mass eigenstates are given by:
\begin{align}
& \mathcal{L}_\chi=g_X \left(\chi^{*}\partial_\mu \chi-\chi \partial_\mu \chi^{*}\right) \left(g_{\rm DM}^X Z^\mu+g_{\rm DM}^{Z'} Z'^\mu\right)\nonumber\\
& \mathcal{L}_\psi=g_X \bar \psi \gamma_\mu \psi \left(g_{\rm DM}^X Z^\mu+g_{\rm DM}^{Z'} Z'^\mu\right)\nonumber\\
& \mathcal{L}_V=\eta_V^X \left( g_{\rm DM}^{Z'}  [[VVZ']]+g_{\rm DM}^{Z}  [[VVZ]]\right)
\end{align}
with $g_{\rm DM}^{Z'}=\frac{\cos\xi}{\cos\delta}$ and $g_{\rm DM}^{Z}=-\frac{\sin\xi}{\cos\delta}$ 
As evident, in the physical basis, the DM is connected to the SM sector by two s-channel mediators, the $Z$ and the $Z'$.
The DM relic density is determined by annihilation processes into fermion pair final states, $WW$ and $Z(Z')h$, induced by s-channel exchange of the mediators, and $ZZ$, $Z'Z$ and $Z'Z'$, induced by t-channel exchange of a DM state. The corresponding rates can be straightforwardly derived from the cases of $Z/Z'$ portals so won't be rediscussed in detail here. Similarly to scenario already described, DM direct detection relies, in all cases but the Abelian vector DM, on SI interactions, which induces a scattering cross-section written, for the case of the proton, as:
\bea
 \sigma_{\chi p/\psi p/V p}^{\rm SI}&&=\frac{\mu_{\chi p/\psi p/V p}^2 c_X^2}{\pi}{\left[b_p \frac{Z}{A}+b_n \left(1-\frac{Z}{A}\right)\right]}^2,\nonumber\\
 {\rm where}~~ b_p=&&2 b_u+b_d,\,\,\,\,b_n=b_u+2 b_d,
 ~{\rm and}~b_f=\frac{g_{DM}^Z\left(g_{f_L}^Z+g_{f_R}^{Z}\right)}{2 m_Z^2}+\frac{g_{DM}^{Z'}\left(g_{f_L}^{Z'}+g_{f_R}^{Z'}\right)}{2 m_{Z'}^2}.
\eea
here $\mu_{\chi p/\psi p/V p}$, as already illustrated, denotes the concerned reduced masses. $c_X$ encodes the DM couplings $g_X^2$ (for complex scalar and fermion) or $\eta_V^X$ (for vector) and eventual overall factors, depending on the kind of DM candidate, in the expression of the cross-section. 

\begin{figure}[t]
\includegraphics[width=4.5 cm]{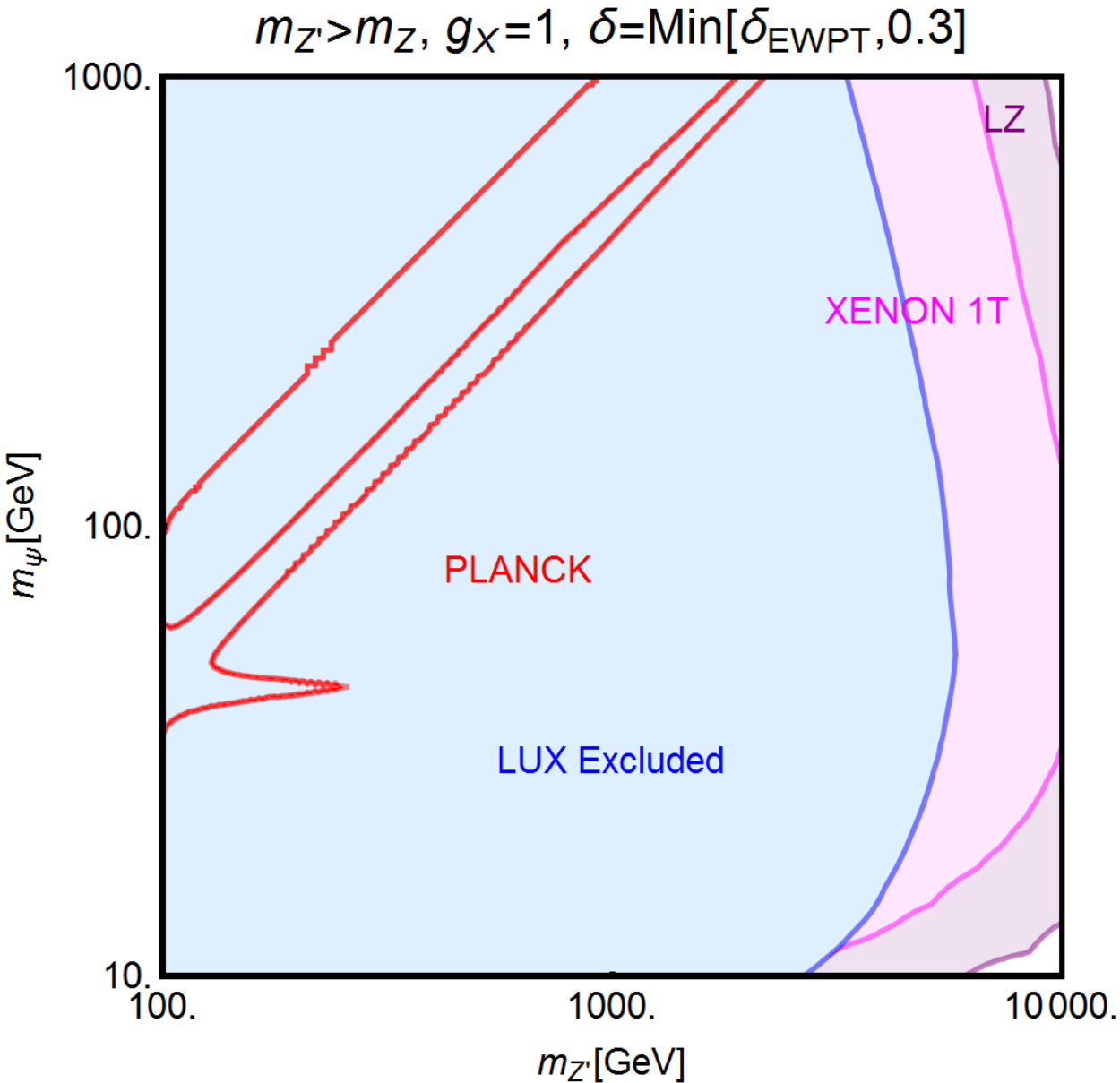}
\includegraphics[width=4.5 cm]{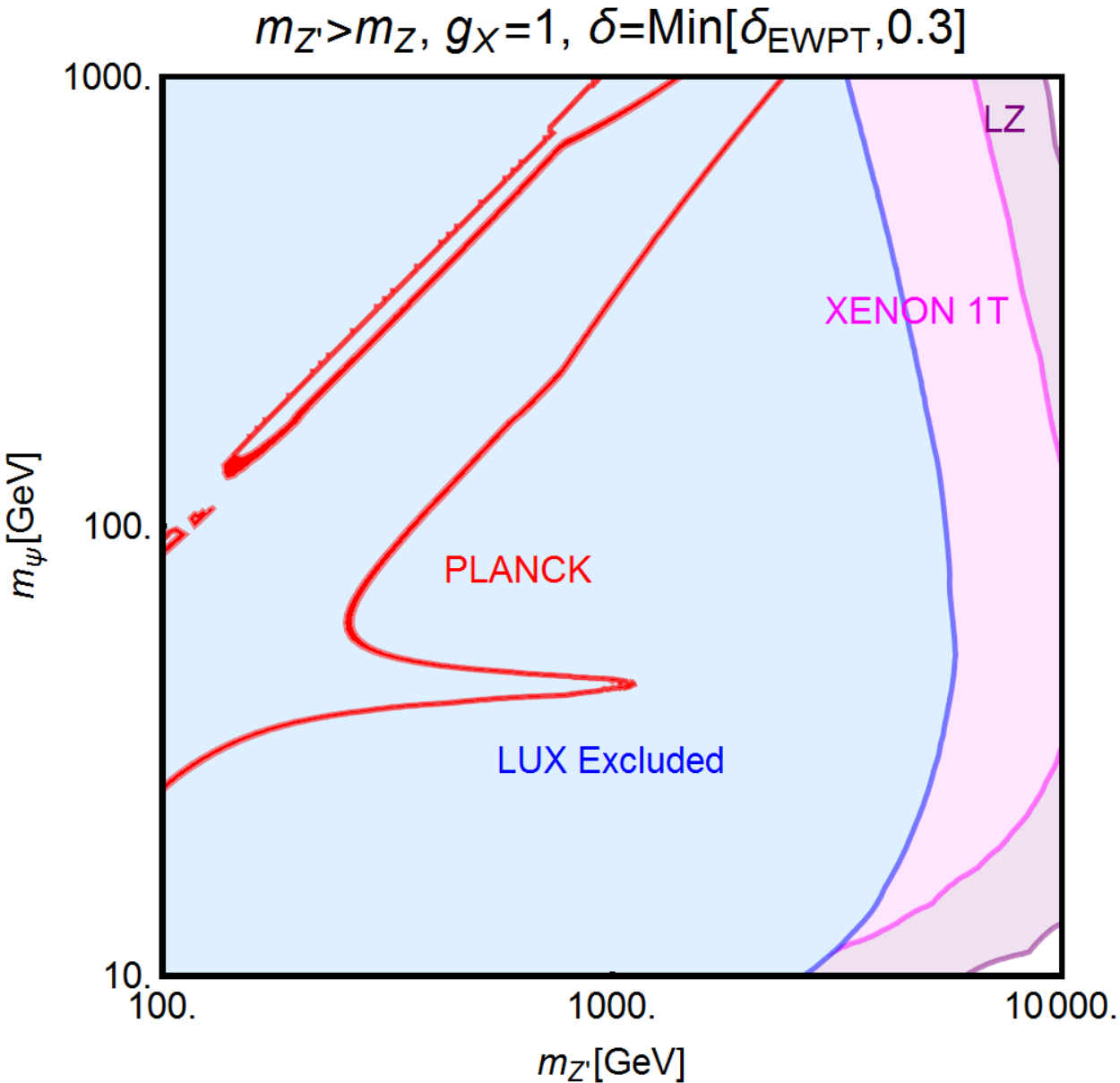}
\includegraphics[width=4.5 cm]{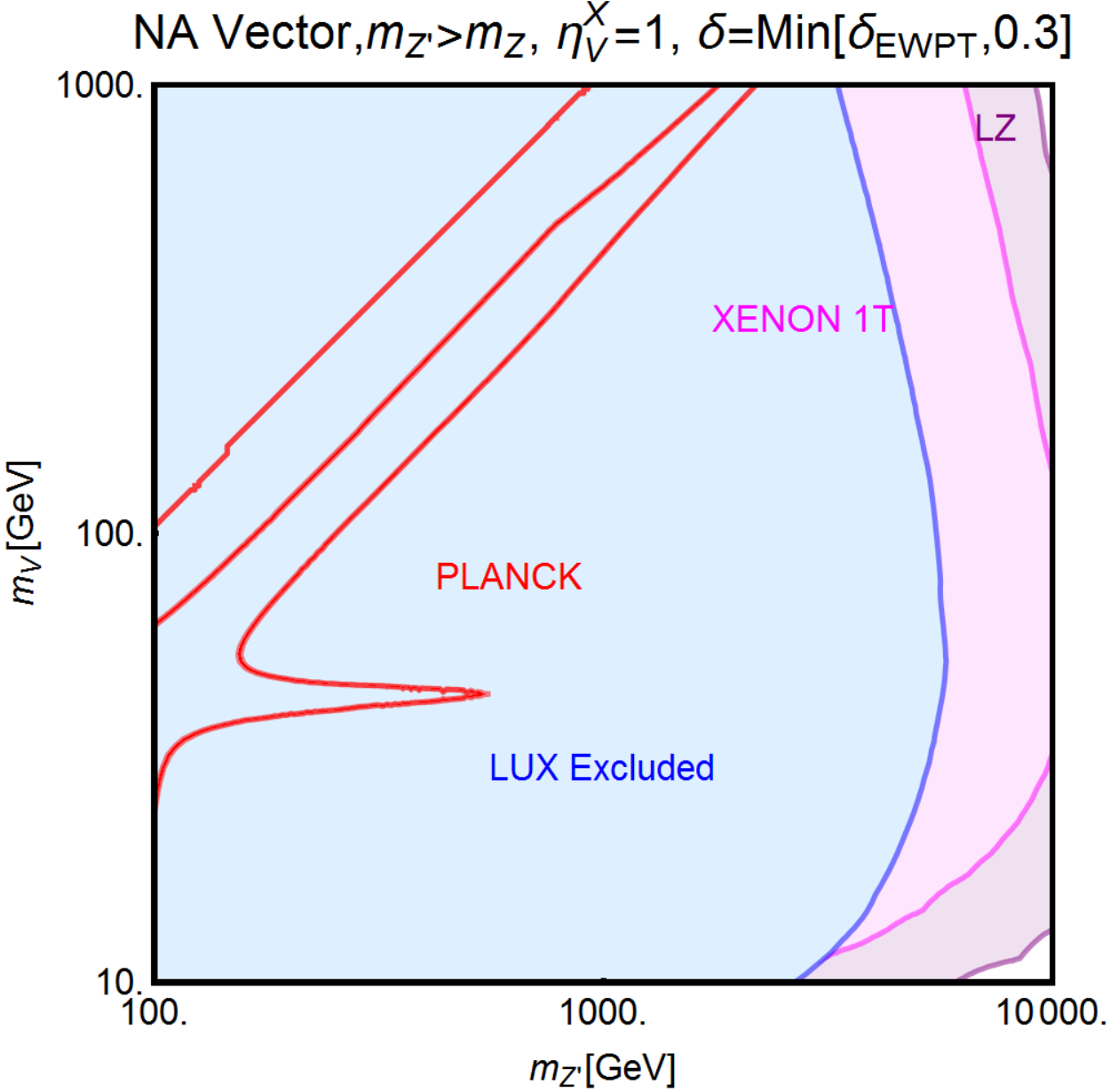}\\
\includegraphics[width=4.5 cm]{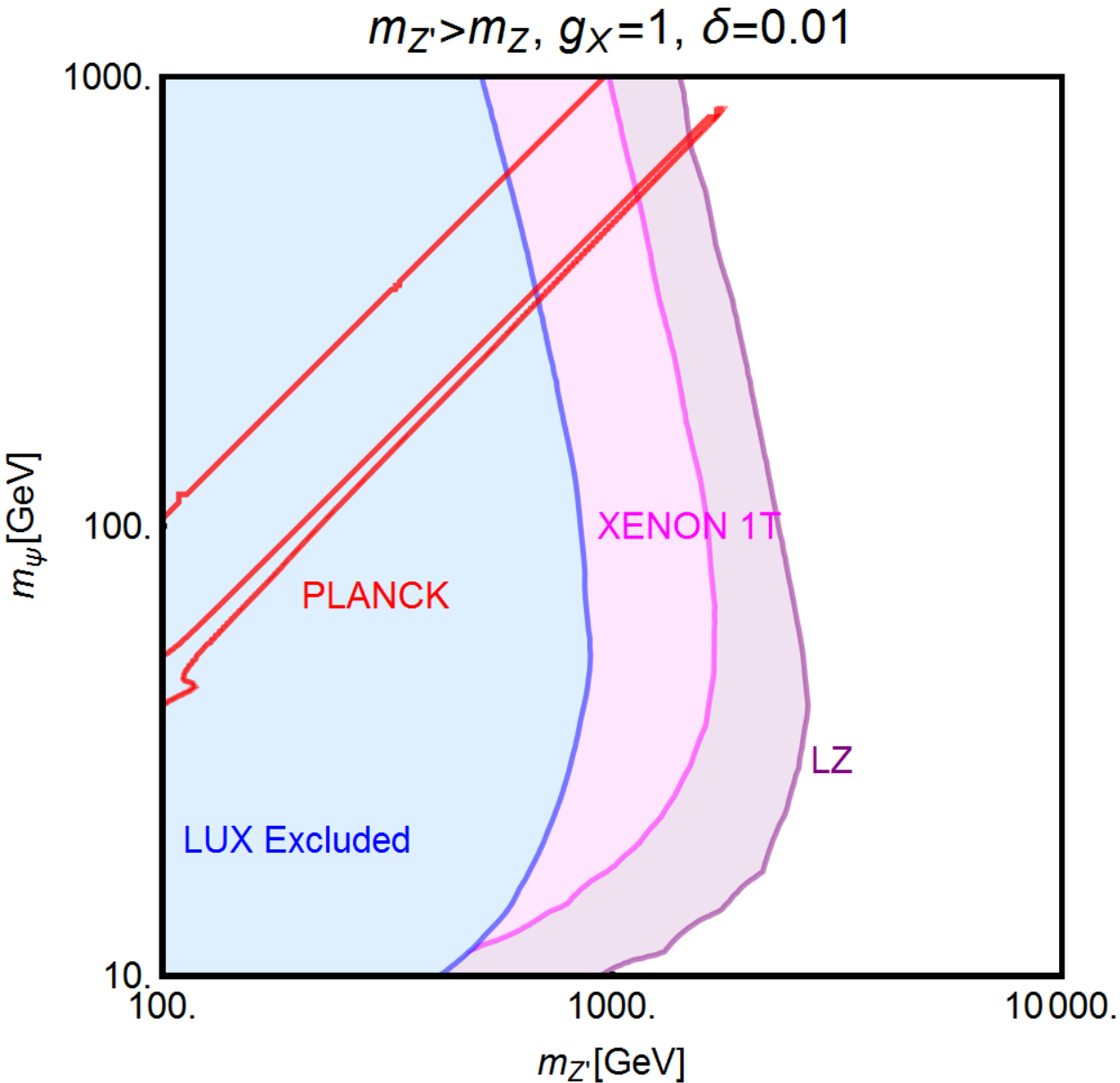}
\includegraphics[width=4.5 cm]{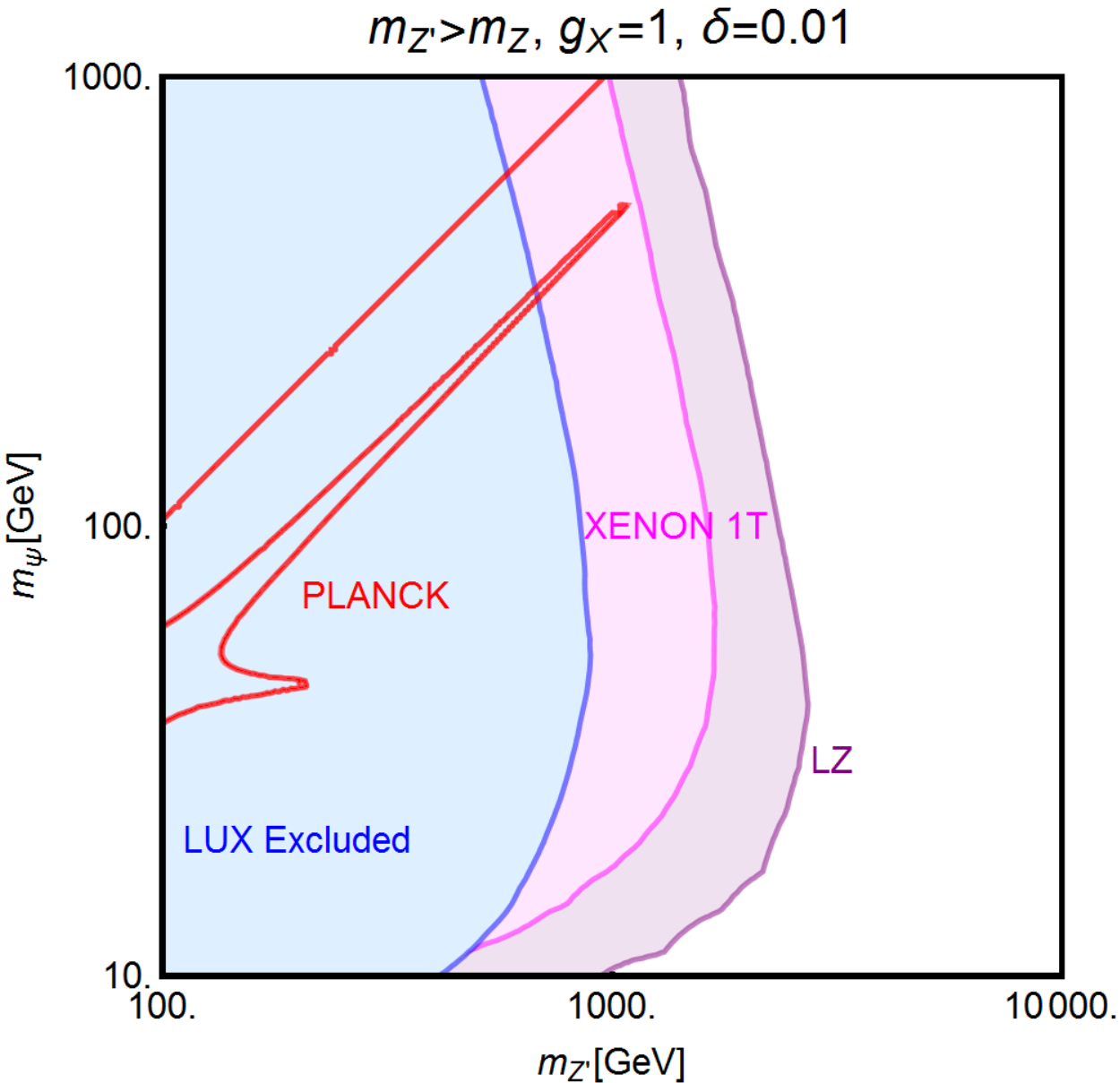}
\includegraphics[width=4.5 cm]{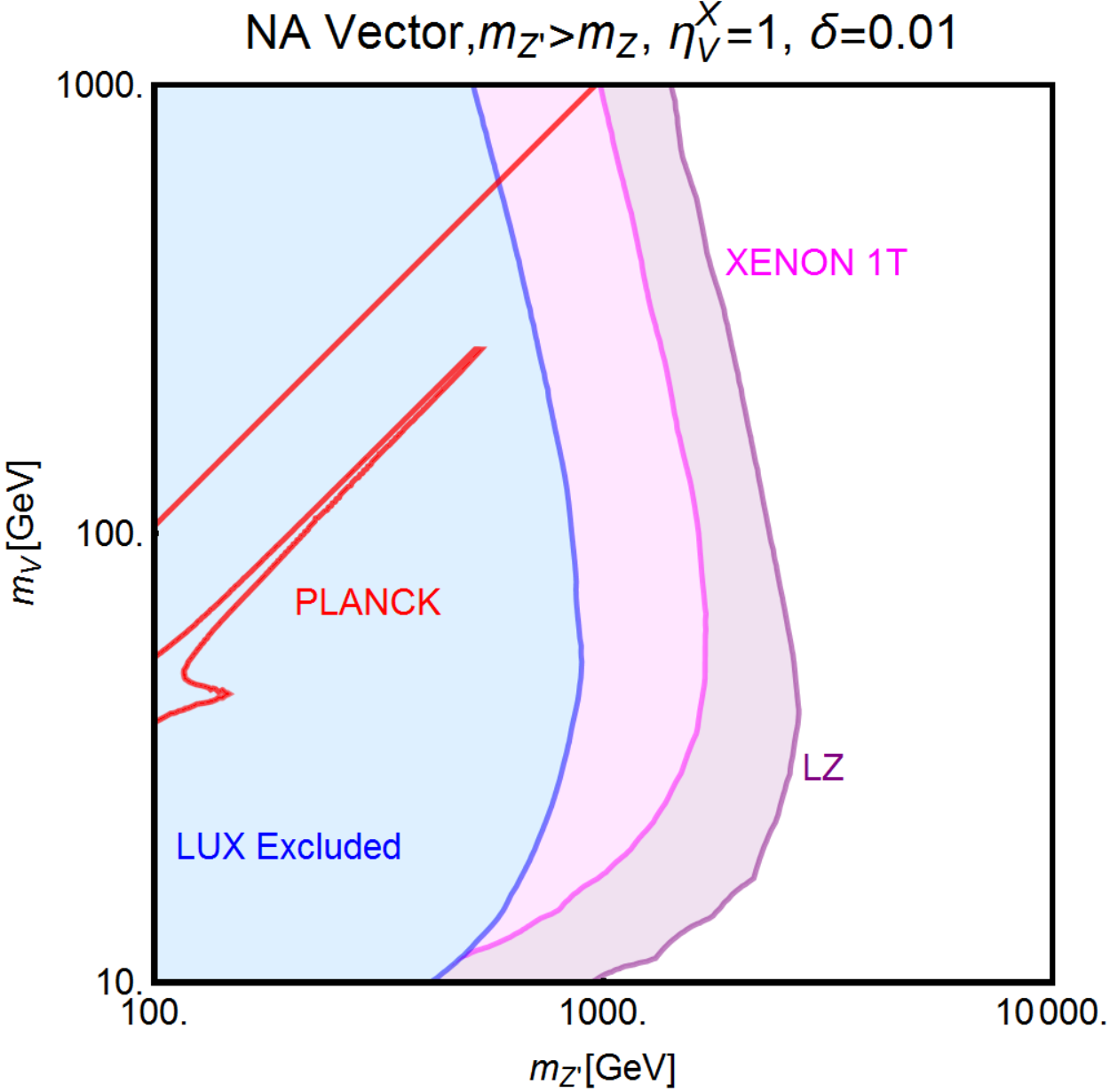}
\caption{\footnotesize{Combined constraints for the scalar (left column), fermion (middle column) and non-Abelian vector (right column) DM interacting with a $Z'$, kinetically coupled with the SM $Z$ boson. In top-row plots the kinetic mixing parameter $\delta$ has been set to the maximal value, as a function of $m_{Z'}$, consistent with the EWPT constraints while for the bottom-row plots $\delta$ has been set to a constant value of $0.01$. We set $g_X=1$ for all these plots. In this figure the red coloured curve represents the contour of correct DM relic density. The blue coloured region is excluded by the current constraints from LUX while the magenta and purple coloured regions would appear excluded in the absence of signals from XENON1T (after two years of exposure) and LZ, respectively.}}
\label{fig:SKinetic_high}
\end{figure}

The interplay between DM relic density and direct detection is shown, for the various spin assignations of the DM, in fig.~(\ref{fig:SKinetic_high}). 
We have considered two assignations of the kinetic mixing parameter, the first corresponding to the present limit from EWPT, i.e., taking the equal sign in eq.~(\ref{eq:deltaEWPT}), the second to the constant value $\delta=0.01$. This last choice is inspired by models in which the kinetic mixing parameter is radiatively generated upon integrating out heavy degrees of freedom charged under both $U(1)_V$ and $U(1)_X$~\cite{Carone:1995pu,Baumgart:2009tn}.


%

By comparing the outcome of fig~(\ref{fig:SKinetic_high}) with the scenarios of direct coupling of the $Z'$ with the SM fermions we notice that Direct Detection probes a more limited region of the parameters space. This because the scattering cross-section depends on coupling suppressed by the small parameter $\delta$. However, this kind of suppression affects also DM annihilation processes, ad exception of the $Z'Z'$ final state so that a strong tension with experimental constraints, analogously, for example, to the SSM, still persists. We notice indeed that the thermal DM is excluded for masses below the TeV scale unless small values, $\mathcal{O}\sim(0.01)$, of the kinetic mixing parameter are taken. Even in such a case, the correct relic density is achieved only at the $m_{Z'}/2$ pole or for $m_{\chi,\psi,V} > m_{Z'}$ where it is accounted annihilation into $Z'Z'$ without relying on the kinetic mixing parameter. These setups will be nevertheless excluded in absence of signals in next generation multi-TON experiments. 

As explicitly indicated in the panels of fig.~(\ref{fig:SKinetic_high}) we have considered only the case $m_{Z'}>m_Z$. The opposite regime (often dubbed as dark photon) would be also feasible, although the constraints on $\delta$ would be even stronger because of eq.~(\ref{eq:deltaEWPT}) and additional constraints from $g-2$ and parity violation effects in atomic physics~\cite{Chun:2010ve} as well as from low energy colliders \cite{Alexander:2016aln}. We have checked that the low mass $Z'$ regime is substantially excluded by current direct detection limits, unless considering DM masses below the sensitivity of DD experiments, and we have then not explicitly reported it.  

%

%

\section{t-channel portals}

We finally consider, in this section, the case in which the DM has not pair coupling but it is instead coupled with one mediator state and a SM quarks. Keeping the assumption that the DM is a SM singlet the simplest option is to consider the coupling with right-handed quarks through a color triplet (with suitable assignation of the hypercharge) mediator.

We will then consider the cases of a complex scalar DM $\chi$, coupled with a Dirac fermionic mediator $\Psi_q$, and a fermionic (Dirac or Majorana) DM $\psi$ coupled with a scalar field $\Sigma_q$ according the following Lagrangians~\footnote{Given the non-trivial gauge charges the t-channel mediators will be coupled with the gluon, the photon and the $Z$-boson. We will omit for simplicity of explicitly writing these interactions. In the case of the scalar mediator a 4-field coupling with the Higgs doublet can arise at tree-level, being renormalizable. We will assume that the corresponding coupling is negligible.}:

\be
\label{eq:tchannel_lagrangian}
 \mathcal{L}=\lambda_{\Psi_q} \ovl \Psi_q \chi q_R+h.c.
 ~~~~{\rm or}~~~~
\mathcal{L}=\lambda_{\Sigma_q} \ovl \psi \Sigma_q q_R+h.c.,\,\,\,\,q=u\,\mbox{~or~}\,d.
\ee

Under the assumption of the introduction of one mediator field, the DM can be coupled only either to up-type or down-type quark~\footnote{In eq.~(\ref{eq:tchannel_lagrangian}) the labels $u$ and $d$ refer globally to up- and down-type quarks. The couplings and the masses of the mediator fields carry also a generation index which is not explicitly reported (see main text for further clarification)}. Since there are not substantial differences we will focus just on the first possibility.

Given the coupling~(\ref{eq:tchannel_lagrangian}), cosmological stability of DM is achieved only if $m_\psi < m_{\Sigma_u} (m_\chi<m_{\Psi_u})$ and if both the DM and the mediator are charged under some new quantum number so that couplings of the mediator with only SM states are forbidden~\footnote{By relaxing this hypothesis it is possible to have a viable decaying DM candidate model with characteristic phenomenology~\cite{Arcadi:2013aba,Arcadi:2014tsa,Arcadi:2014dca}. In this case very low values of the couplings should be assumed.}. In order to avoid possible occurrence of flavor violation effects, with consequent strong constraints on the coupling $\lambda_{\Psi_\mu} (\lambda_{\Sigma_\mu})$ we assume that the mediator carries also a flavor quantum number (a ``flavored DM''~\cite{Agrawal:2011ze} would be equally feasible), i.e., $\Sigma_{u}\equiv(\sigma_u,\sigma_c,\sigma_t) (\Psi_{u}\equiv(\psi_u,\psi_c,\psi_t))$, and that the interactions~(\ref{eq:tchannel_lagrangian}) are flavor conserving. This is achieved by assuming the component of the mediator fields to be degenerate in mass, we call $m_{\Sigma_u}$ and $m_{\Psi_u}$ these masses, and the couplings $\lambda_{\Psi_u},\lambda_{\Sigma_u}$ (being actually matrices) to be diagonal in the flavor space. For further simplification we will assume all the couplings to be equal and the drop the flavor index.

Concerning DM phenomenology the relic density is determined, for all the three type of candidate, by annihilation processes into fermion pairs induced by t-channel exchange of the mediator field. For close values of the DM and mediator masses coannihilation processes, like $\psi \Sigma_u (\Psi_u \chi) \rightarrow q g$ and mediator pair annihilation processes like, $\Sigma_u \Sigma_u (\Psi_u \Psi_u) \rightarrow \ovl q q, gg$ induced by gauge interactions, might become also important. In analogous fashion as the other models reviewed in this work we have focused on the assignations $\lambda_{\Psi_u},\lambda_{\Sigma_u}=1$. In such a case the dominant contribution to the DM relic density comes from pair annihilations into SM fermions. We can then achieve an analytical description through some simple approximations of the corresponding cross-sections, based on the velocity expansion (more complete expressions are presented in the appendix): 
\bea
&& \langle \sigma v \rangle^{\rm Complex\, Scalar}~=\frac{3{(\lambda_{\Psi_u})}^4 m_t^2}{16 \pi {\left(m_\chi^2+m_{\Psi_u}^2-m_t^2\right)}^2}{\left(1-\frac{m_t^2}{m_\chi^2}\right)}^{3/2}+\frac{3{(\lambda_{\Psi_\mu})}^4 m_\chi^2 v^2}{8 \pi m_{\Psi_u}^4}{\left(1+\frac{m_\chi^2}{m_{\Psi_u}^2}\right)}^{-2}, \nonumber\\
 &&\langle \sigma v \rangle^{\rm Majorana}~=\frac{3{(\lambda_{\Sigma_u})}^4 m_t^2}{32 \pi {\left(m_\psi^2+m_{\Sigma_u}^2-m_t^2\right)}^2}\sqrt{\left(1-\frac{m_t^2}{m_\psi^2}\right)}+\frac{3{(\lambda_{\Sigma_\mu})}^4 m_\psi^2 v^2}{8 \pi m_{\Sigma_u}^4}{\left(1+\frac{m_\psi^2}{m_{\Sigma_u}^2}\right)}^{-2}, \nonumber\\
&&\langle \sigma v \rangle^{\rm Dirac}~=\sum_{f=u,c,t}~~ \frac{3{(\lambda_{\Sigma_u})}^4 m_\psi^2}{32 \pi {\left(m_\psi^2+m_{\Sigma_u}^2-m_f^2\right)}^2}\sqrt{1-\frac{m_f^2}{m_\psi^2}}.
\eea

In the cases of complex scalar and Majorana DM the s-wave term of the annihilation cross-section is helicity suppressed so that, ad exception of values of the DM mass close to the one of the top, the dominant contributions comes from the p-wave term, leading to the following estimate:
\begin{equation}
\langle \sigma v \rangle \approx 1.7\times 10^{-26} {\mbox{cm}}^3 {\mbox{s}}^{-1} {\left(\frac{m_\psi}{200\,\mbox{GeV}}\right)}^2 {\left(\frac{1\,\mbox{TeV}}{m_{\Sigma_u}}\right)}^4 {(\lambda_{\Sigma_u})}^4.
\end{equation}

On the contrary the annihilation cross section of Dirac DM is s-wave dominated and can be estimated as:
\begin{equation}
\langle \sigma v \rangle=4.2\times 10^{-26} {\mbox{cm}}^3 {\mbox{s}}^{-1} {\left(\frac{m_\psi}{200\,\mbox{GeV}}\right)}^2 {\left(\frac{1\,\mbox{TeV}}{m_{\Sigma_u}}\right)}^4 {(\lambda_{\Sigma_u})}^4.
\end{equation}

For what regards Direct Detection, it relies on scattering of the DM on up-quarks through s-channel exchange of the mediator. In the cases of complex scalar and Dirac fermionic DM these interactions lead to SI cross-sections (in the case of Dirac fermion also SD scattering is present but its impact is negligible given the much weaker experimental limits) whose expressions are: 
\bea
\sigma_{\chi p}^{\rm SI}~&&=\frac{{(\lambda_{\Psi_u})}^2 m_p^2}{32 \pi {\left(m_{\Psi_u}^2-m_\chi^2\right)}^2}{\left[f_{p}\frac{Z}{A}+f_{n}\left(1-\frac{Z}{A}\right)\right]}^2 \,\,\,\,\mbox{(Complex scalar)}, \nonumber\\
\sigma_{\psi p}^{\rm SI}~&&=\frac{{(\lambda_{\Sigma_u})}^2 \mu_{\psi p}^2}{64 \pi {\left(m_{\Sigma_u}^2-m_\psi^2\right)}^2}{\left[f_{p}\frac{Z}{A}+f_{n}\left(1-\frac{Z}{A}\right)\right]}^2 \,\,\,\,\mbox{(Dirac Fermion)},
\eea
with $f_{p}=2,f_{n}=1$~\cite{Jungman:1995df,Gondolo:2004sc}.

On the contrary, in the case of Majorana DM, one should consider SD interactions described by the following cross-section:
\begin{equation}
\sigma_{\psi n}^{\rm SD}=\frac{3 {(\lambda_{\Sigma_u})}^2 \mu_{\psi p}^2 \Delta_{nu}^2}{16 \pi {\left(m_{\Sigma_u}^2-m_\psi^2\right)}^2}.
\end{equation}

\begin{figure}[t]
\includegraphics[width=5.5 cm]{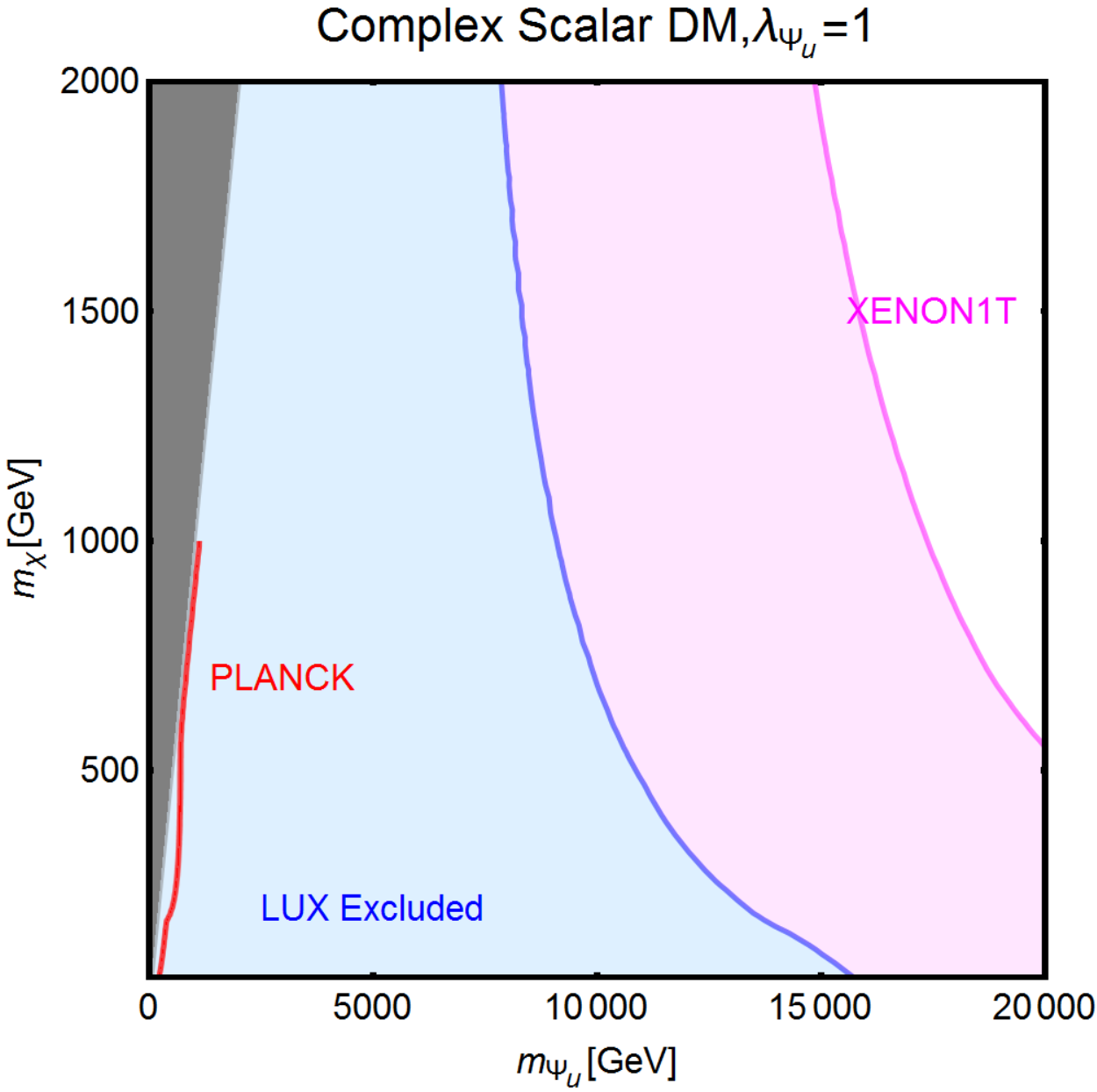}
\caption{\footnotesize{Combined constraints for a complex scalar DM $\chi$ coupled with the right-chiral up-type quarks through a Dirac fermionic t-channel mediator $\Psi_u$. The results are in the bi-dimensional plane $(m_{\Psi_u},m_\chi)$ and the coupling $\lambda_{\Psi_u}$ has been set to 1. The red colored curve corresponds to the contour of correct DM relic density. The blue colored region is excluded by the current constraints from LUX while the magenta colored region will be excluded in the absence of signals from XENON1T after two years of exposure. The gray colored region corresponds to $m_\chi > m_{\Psi_u}$ for which the DM would not be cosmologically stable.}}
\label{fig:ftchannelS}
\end{figure}

\begin{figure}[t]
\includegraphics[width=5.5 cm]{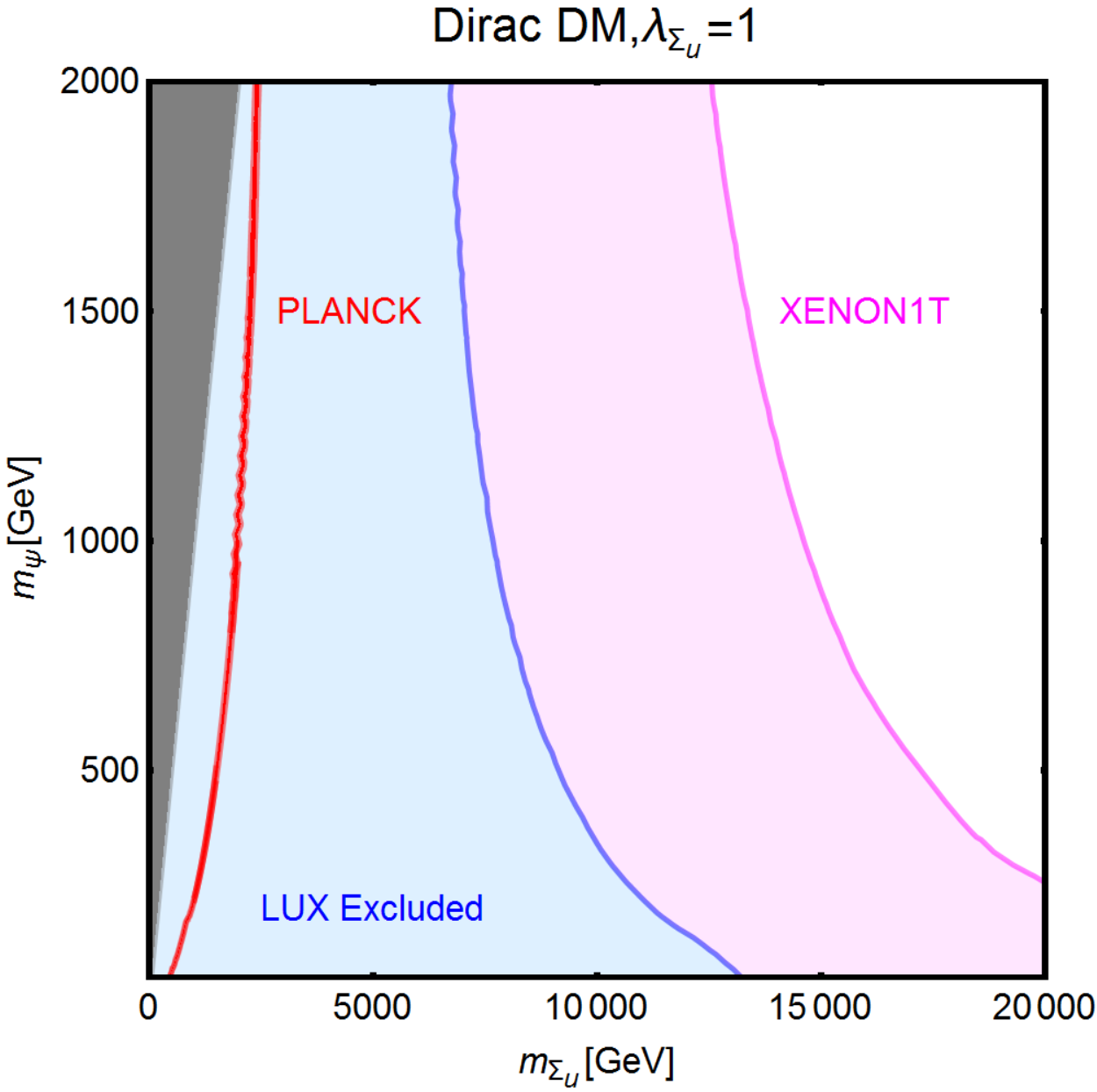}
\includegraphics[width=5.2 cm]{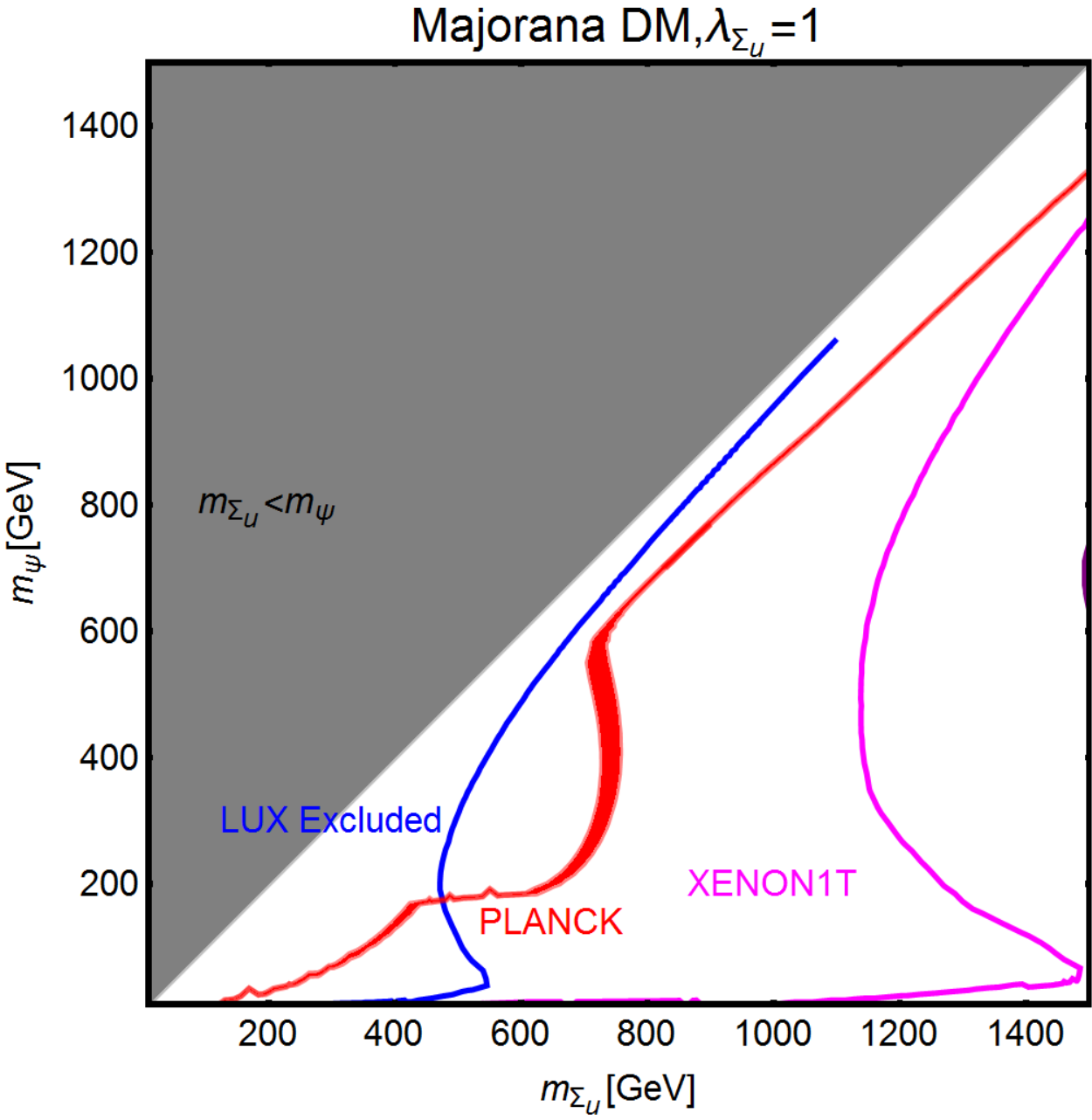}
\caption{\footnotesize{The same as fig.~(\ref{fig:ftchannelS}) but for the case of a Dirac (left panel) or Majorana (right panel) fermion DM and scalar t-channel mediator $\Sigma_u$.}}
\label{fig:ftchannelF}
\end{figure}

The results of our analysis are presented in the planes $(m_\chi,m_{\Psi_u})$ and $(m_\psi,m_{\Sigma_u})$, with couplings set to 1, in fig.~(\ref{fig:ftchannelS}) and~(\ref{fig:ftchannelF}) for fermionic DM.

As evident, the cases of scalar and Dirac fermion DM, in which unsuppressed SI interactions are present, are already substantially ruled out~\footnote{One could weaken the constraints by taking smaller values of the couplings. However relic density would be achieved only in the very fine-tuned coannihilation regime.}. Only the case of Majorana DM survives to present constraint and will be probed, for masses of both the DM and the mediator up to few TeV, by next future DD experiments.



\section{Summary and discussion}

We have discussed impact of current, and possible future, direct detection limits, possibly complemented by ones from collider searches, in several simplified realizations of WIMP DM.

The first and simplest classes of models considered are the ones in which the interactions of pairs of SM singlet scalar, fermionic and vectorial DM and pairs of SM fermions, are mediated by electrically neutral s-channel (portal) mediators. In the most minimal case the particle spectrum of the SM should be complemented by just a new state, i.e., the DM candidate, since portal interactions can be mediated either by the Higgs or by the Z-boson, although in the last case a theoretically consistent construction is more contrived. In the case of Higgs portal, for all the DM spin assignations, SI interactions with nucleons are induced. The consequent very strong limits, due to the light mediator, are incompatible with thermal relic density ad exception of masses above the TeV scale or the ``pole'', i.e., $m_{\rm DM} \simeq m_h/2$, region. This last scenario would be nevertheless ruled out in case of absence of signals at XENON1T and LZ. In the $Z$-portal scenario current limits on the SI cross-section already exclude the pole region. These strong limits can be nevertheless partially overcome in two setups: fermionic DM with only axial couplings with the $Z$, as naturally realized in the case of Majorana DM, and vector DM coupled through Chern-Simons term. In these two cases the DM features SD interactions with nuclei, whose constrains are sensitively weaker. In particular, in the case of Majorana fermion, thermal DM with mass of few hundreds of GeV would remain viable even in absence of signals at next generation detectors. 

The Higgs and Z-portal setups are easily extended to the case of BSM spin-0 and spin-1 mediators. In the case of scalar mediators we have imposed, in order to preserve $SU(2)$ invariance, a yukawa structure for the couplings of the mediator with SM fermions. This, on one side, implies a suppression of the DM annihilation cross-section for masses below the one of the top (unless the $SS$ final state is kinematically accessible). At the same time also possible collider signals are strongly suppressed so that corresponding limits are not competitive with respect to the ones from Direct Detection and have been neglected for simplicity. Despite the different velocity dependence of the annihilation cross section, the regions of the correct relic density are then mostly determined by the Yukawa structure of the couplings of the mediator with the SM fermions. The correct relic density is indeed obtained, far from resonance regions, only when the $\ovl t t$ and/or $SS$ annihilation channels are kinematically open. Regarding Direct Detection, the limits are associated to the SI component of the DM scattering cross-section for all the different assignations of the DM spin. The shape of the DD isocontours are, however, different in the various DM scenarios. This is due to the different assignation of the dimension-1 couplings of the scalar and vectorial DM. Theoretical considerations suggested, indeed, to parametrize these couplings in terms of a fundamental mass scale, the mass of the mediator and the DM mass in the cases, respectively, of scalar and vectorial DM and an unknown dimensionless coupling. Current limits still allow masses of few hundreds GeV for both the DM and the mediator while XENON1T, in absence of signals after two years of exposure, will exclude mediator masses up to approximately 1 TeV and DM masses of up to few TeVs. Given the several free parameters, for clarity of the picture, we have focused our investigation on the masses of the new particle states and fixed the couplings to be close to order one (see in alternative e.g., ref.~\cite{Arina:2016cqj}). We notice on the other hand that lowering the couplings would contemporary suppress both the direct detection rate and the DM annihilation cross-section, in particular the $SS$ channel becomes negligible as soon as the DM couplings $\lambda_{\chi,\psi,V}^S$ deviates sensitively from order 1 values. As a consequence, in this setup, thermal DM is achieved only in the pole region which results particularly fine tuned because of the typically small width of the scalar mediator.

The scenario of spin-1 BSM s-channel mediator is even more constrained than the spin-0 case. Indeed the constraints from SI cross-section are typically much stronger, because of an effective enhancement of the cross section due to the isospin violation interactions of the $Z'$ with nucleons, as the scalar case, so that masses of the DM and the mediator below approximately 5 TeV are already excluded. In case of no signals at next generation DD experiments the exclude regions will extend up to masses of the order of 10 TeV, beyond the reach of LHC. In addition, the (reasonable) assumption of a $Z'$ coupled with both quark and leptons implies a strong complementarity with LHC searches of dilepton resonances. The corresponding limits, exclude, for the models considered here, masses of the $Z'$ between 2 and 3 TeV (the exclusion can be even above 4 TeV in other realizations~\cite{Alves:2016fqe}), even in setups in which the SI component of the DD cross section is suppressed or absent. We remark again that, despite in our analysis we have limited to some fixed assignations of the couplings, our results have general validity because of the strong correlation between the DM relic density and the scattering rate of nucleons. For example reducing the size of the couplings would actually reduce the viable parameter regions since the correct relic density would be achieved only in correspondence of s-channel resonances.

Despite our work is focused on scenarios probed by current and next future Direct Detection experiments, we have nevertheless also discussed the a setup in which DD is, in general evaded: the pseudoscalar portal. Under the assumption of CP conservation only fermionic DM is considered in this case. Most of the parameter space is substantially insensitive to Direct Detection (we remind that we have, conservatively, considered values of the pseudoscalar mass above 1 GeV in order to avoid flavour constraints) since tree level interactions with nucleons are momentum suppressed and, furthermore, are not subject to coherent enhancement. A rather limited region of the parameter space might be still probed by 1- and multi-TON detectors because of a 1-loop induced SI cross-section. Thermal DM is nevertheless sensitively constrained from Indirect Detection. In addition, there is again a strong complementarity from collider constraints, dominated, for this scenario, by monojets. A light pseudoscalar mediator could be interpreted as the pseudo-goldstone boson of a spontaneously broken global U(1) symmetry. We have then considered the case of complex scalar mediator which can be decomposed into a scalar and light pseudoscalar component. Although in this case sizable direct detection limits are reintroduced, thermal DM is viable in large portions of the parameter space due to the presence of efficient annihilation processes in the $aa$ and $Sa$ final states.

We have then performed some steps towards more theoretically motivated realizations of Dark portals. As well known, the bilinears $H^{\dagger} H$ and $B^{\mu \nu}$ are Lorentz and gauge invariant, so naturally lead to portal interactions with a dark sector (even if this is completely secluded). We have thus considered the cases of scalar mediator is coupled both to the DM and Higgs boson and mixes with the latter because of a non zero vev and of a $Z'$ coupled to the $Z$ boson and, in turn, with the other SM states through a kinetic mixing term, also responsible of a mixing between the two spin-1 states. Also for what regards the coupling of the DM with the mediators we have considered less generic assignations with respect to the cases considered in the previous sections. In the case of Higgs+Scalar portal we have explicitly considered a dynamical origin for the DM mass. Indeed fermion DM has been assumed to have Yukawa coupling so that its mass is originated by the vev of the new scalar field. Similarly, vector DM has been assumed to be the vector boson of a spontaneously broken dark U(1) gauge symmetry and its mass is again related to vev of the new field.

As last case of study we have relaxed the hypothesis that of a SM singlet BSM mediator and rather assigned it non trivial quantum numbers under color and electromagnetism. In this case the a single DM state is coupled with the mediator and a SM fermion, according the coupling assignation of the mediator (for simplicity we have restricted our analysis to couplings with right-handed up-type quarks). Contrary to the other scenarios considered in this work DM pair annihilation occurs through t-channel exchange of the mediator while DD scattering is induced by its s-channel exchange. Restricting for simplicity to the case in which the mediator field has the same quantum numbers, with to respect the SM gauge group, as the right-handed up quarks, the scenarios, i.e. Complex Scalar and Dirac fermion, in which SI interactions are present are excluded, for order 1 values of the couplings, for masses up to order of 10 TeV. On the contrary, thermal Majorana DM is still viable for scales below the TeV and will be extensively probed by next generation of DD experiments.    

\section{Conclusions}

We have reviewed the theoretical foundations of the WIMP paradigm and discussed the detection methods and a multitude of models and constraints encompassing scalar, vector and fermionic dark matter setups. In light of the extensive search for dark matter combing complementary probes namely, direct indirect and collider, we assessed the status of simplified models accounting for current and projected limits.

In particular, we have reviewed well known portals such as the Higgs portal and the Z-portal. We have also addressed the so popular dark $Z^{\prime}$ portal and many others models that possess in their spectrum more than one mediator. Moreover, we have also investigated new models dictated by the kinetic mixing, often used in dark photon models.

We concluded that the simplest constructions, i.e., the SM dark portals will be substantially ruled out, ad exception of the case of fermionic DM with only axial couplings with the Z-boson (e.g., Majorana DM), in absence of signals in next generation of Direct Detection experiments.

The most straightforward extension of SM dark portals, represented by the introduction of BSM s-channel mediators, are, similarly, strongly constrained in presence of Spin-independent interactions of the DM off of nuclei. In particular, the case of spin-1 mediator is strongly disfavored because of the presence of complementary constraints from searches of resonances at the LHC, pushing the dark matter mass to the multi-TeV scale.

The tension with direct detection constraints can be relaxed in somehow next-to-minimal scenarios, featuring multiple mediators or new states lighter than the DM (we have reviewed the example of a light pseudo-scalar).

In summary, we combined a plethora of experimental data set and theoretical models, computing the relic density, direct, indirect and collider observables to have a clear picture of where the WIMP paradigm stands and the prospects. It is clear that most of the WIMP models will be scrutinized in the next decades, highlighting the paramount role of the next generation of experiments.

\section*{Acknowledgements}

The authors thank Werner Rodejohann, Miguel Campos, Alexandre Alves, Carlos Yaguna and Chris Kelso for discussions. The authors also thank Francesco D'Eramo for his valuable comments. P.~G. acknowledges the support from P2IO Excellence Laboratory (LABEX). S.~P.'s work was partly supported by the U.S. Department of Energy grant number DE-SC0010107. 
This work is also supported by the Spanish MICINN's Consolider-Ingenio 2010 Programme under grant Multi-Dark {\bf CSD2009-00064}, the contract 
{\bf FPA2010-17747}, the France-US PICS no. 06482 and the LIA-TCAP of CNRS. Y.~M. acknowledges partial support the ERC advanced grants Higgs@LHC and MassTeV. 
This research was also supported in part by the Research
Executive Agency (REA) of the European Union under
the Grant Agreement {\bf PITN-GA2012-316704} (``HiggsTools'').

\appendix


\section{Annihilation cross-sections}

The thermal average is defined as:
\begin{equation}
\langle \sigma v \rangle=\frac{1}{8 m_\chi^4 T {K_2\left(\frac{m_\chi}{T}\right)}^2}\int_{4 m_\chi^2}^{\infty} ds \sigma(s) \sqrt{s} \left(s-4 m_\chi^2\right)K_1\left(\frac{\sqrt{s}}{T}\right)
\end{equation}


Away from resonances, a manageable analytical expression for the thermally averaged pair annihilation cross section is obtained by performing the formal velocity expansion, in the non-relativistic limit, as defined in \cite{Gondolo:1990dk}. The thermally averaged cross section can be computed as:
\begin{equation}
\label{eq:Gondolo_int}
\langle \sigma v \rangle = \frac{2 x^{3/2}}{\pi^{1/2}}\int_0^{\infty} \sigma v_{\rm lab}\epsilon^{1/2}e^{-\epsilon x}
\end{equation}
where:
\begin{align}
& v_{\rm lab}=\frac{2 \epsilon^{1/2}{\left(1+\epsilon\right)}^{1/2}}{\left(1+2\epsilon\right)} \nonumber\\
& \epsilon=\frac{s-4 m_\chi^2}{4 m_\chi^2}\,\,\,\,\,\,x=\frac{m_\chi}{T}
\end{align}
This kind of integral can be analytically computed by considering an expansion in series of $\epsilon$ of $\sigma v_{\rm lab}$, namely:
\begin{equation}
\sigma v_{\rm lab}=a_0 +a_1 \epsilon +a_2 \epsilon^2 \cdots
\end{equation}

We will report in the following the full analytical expressions for the thermally averaged cross-section in the dark portal scenarios investigated in this work, in the velocity expansion. We will adopt a general notation of the couplings so that the various expression can be applied to the different models discussed in the main text as well as being of more general utility. Notice that the coupling of scalar mediators with scalar and vector field are dimensional, rather than be decomposed, as the main text, into a physical scale and an dimensionless parameter. In the case of spin-0 and spin-1 portals, the expressions for, respectively, $hh$ and $ZZ$ and $Zh$ final states can be derived directly, by suitable substitution, from the ones corresponding to the $SS$, $Z'Z'$ and $Z'h$ final states. The Higgs+scalar portal and the kinetic mixing models feature, for the DM, also annihilation processes into, respectively, $hS$ and $ZZ'$ final states. We won't explicitly report the corresponding expressions since particularly complicated.
We remind that the velocity expansion is not valid in vicinity of s-channel resonances and kinematic threshold of annihilation channels~\cite{Griest:1990kh,Gondolo:1990dk}.

\subsection{Scalar portal}

\footnotesize{

\subsubsection{Scalar Dark Matter}

$\chi^{*} \chi \rightarrow \ovl f f$:

\begin{equation}
\langle \sigma v \rangle_{\rm ff} =  \sum_f N^c_f \frac{|g_{\chi \chi S}|^2 c_S^2 m_f^2 \left(m_{\chi }^2-m_f^2\right){}^{3/2}}{8 \pi v_h^2  m_{\chi }^3 \left(m_S^2-4 m_{\chi }^2\right)^2}
\end{equation}

$\chi^{*} \chi \rightarrow W^+ W^-$:

\begin{equation}
\langle \sigma v \rangle_{WW} =  \frac{|g_{\chi \chi S}|^2 |g_{WWS}|^2 \sqrt{m_{\chi }^2-m_W^2} \left(-4 m_{\chi }^2 m_W^2+4 m_{\chi }^4+3 m_W^4\right)}{64 \pi  m_{\chi }^3 m_W^4 \left(m_S^2-4 m_{\chi }^2\right)^2}
\end{equation}


$\chi^{*} \chi \rightarrow ZZ$

\begin{equation}
\langle \sigma v \rangle_{ZZ} =  \frac{|g_{\chi \chi S}|^2 |g_{Z Z S}|^2 \sqrt{m_{\chi }^2-m_Z^2} \left(-4 m_{\chi }^2 m_Z^2+4 m_{\chi }^4+3 m_Z^4\right)}{128 \pi  m_{\chi }^3 m_Z^4 \left(m_S^2-4 m_{\chi }^2\right)^2}
\end{equation}


$\chi^{*} \chi \rightarrow SS$

\begin{align}
& \langle \sigma v \rangle_{SS} = \frac{1}{64 \pi m_\chi^2} \sqrt{1-\frac{m_S^2}{m_\chi^2}} \left(2 |g_{\chi \chi SS}|^2-\frac{2 g_{\chi \chi S} g_{SSS} g_{\chi \chi SS}}{m_S^2-4 m_\chi^2}+\frac{|g_{\chi \chi S}|^2 |g_{SSS}|^2}{{\left(m_S^2-4 m_\chi^2\right)}^2}+\frac{4|g_{\chi \chi S}|^4}{{\left(m_S^2-2 m_\chi^2\right)}^2} \right.\nonumber\\
& \left. -\frac{4 g_{\chi \chi S} g_{SSS} g_{\chi \chi SS}}{\left(m_S^2-4 m_\chi^2\right) \left(m_S^2-2 m_\chi^2\right)}+\frac{4 g_{\chi \chi S S} |g_{\chi \chi S}|^2}{m_S^2-2 m_\chi^2}\right)
\end{align}

\subsubsection{Fermionic Dark Matter}

$\ovl \psi \psi \rightarrow \ovl f f$:

\begin{equation}
\langle \sigma v \rangle_{ff} =  |\lambda_\psi^S|^2 \sum_f n_c^f \frac{c_S^2 m_f^2 \left(m_{\psi }^2-m_f^2\right){}^{3/2}}{4 \pi  m_{\psi } v_h^2 \left(m_S^2-4 m_{\psi }^2\right)^2}v^2
\end{equation}

$\ovl \chi \chi \rightarrow W^+ W^-$:

\begin{equation}
\langle \sigma v \rangle_{WW} = |\lambda_\psi^S|^2 |g_{SWW}|^2  \frac{\sqrt{m_{\psi }^2-m_W^2} \left(-4 m_{\psi }^2 m_W^2+4 m_{\psi }^4+3 m_W^4\right)}{64 \pi   m_{\psi } v_h^2 \left(m_S^2-4 m_{\psi }^2\right)^2}v^2
\end{equation}

$\ovl \psi \psi \rightarrow ZZ$:

\begin{equation}
\langle \sigma v \rangle_{ZZ} = |\lambda_\psi^S|^2 |g_{SZZ}|^2  \frac{\sqrt{m_{\psi }^2-m_Z^2} \left(-4 m_{\psi }^2 m_Z^2+4 m_{\psi }^4+3 m_Z^4\right)}{128 \pi   m_{\psi } v_h^2 \left(m_S^2-4 m_{\psi }^2\right)^2}v^2
\end{equation}

$\ovl \psi \psi \rightarrow SS$:
\begin{eqnarray}
&&\langle \sigma v \rangle_{SS}=\frac{v^2}{192 \pi m_\psi^2} \sqrt{1-\frac{m_S^2}{m_\chi^2}} \left(\frac{3 |g_{SSS}|^2 |\lambda_\psi^S|^2 m_\psi^2}{(m_S^2-4 m_\psi^2)^2}+\frac{8 g_{SSS} (\lambda_\psi^{S})^{3} m_\psi^3 (2 m_S^2-5 m_\psi^2)}{(m_S^2-4 m_\psi^2) (m_S^2-2 m_\psi^2)^2}\right.\nonumber\\
&&\left.+\frac{16 |\lambda_\psi^S|^4 \left(9 m_\psi^8-8 m_\psi^6 m_S^2+2 m_S^8\right)}{(m_S^2-2 m_\psi^2)^4}\right)
\end{eqnarray}

\subsubsection{Vectorial Dark Matter}

$VV \rightarrow \ovl f f$
\begin{eqnarray}
&&\langle \sigma v \rangle_{ff}=\sum_f n_c^f |g_{VVS}|^2 \frac{c_S^2 m_f^2}{v_h^2} \left(\frac{\sqrt{4-\frac{4 m_f^2}{m_V^2}} \left(4 m_V^2-4 m_f^2\right)}{96 \pi  m_V^2 \left(4 m_V^2-m_S^2\right)^2}\right.\nonumber\\
&&\left.\frac{v^2 \sqrt{1-\frac{m_f^2}{m_V^2}} \left(m_f^2 \left(7 m_S^2-76 m_V^2\right)+2 m_V^2 \left(m_S^2+20 m_V^2\right)\right)}{144 \pi 
   m_V^2 \left(m_S^2-4 m_V^2\right)^3}\right)
\end{eqnarray}


$VV \rightarrow W^+ W^-$:

\begin{eqnarray}
& \langle \sigma v \rangle_{WW} = \left(\frac{ |g_{WWS}|^2 |g_{VVS}|^2 \sqrt{4-\frac{4 m_W^2}{m_V^2}} \left(16 m_V^4-16 m_V^2 m_W^2+12 m_W^4\right)}{768 \pi  m_V^2 m_W^4 \left(4
   m_V^2-m_S^2\right)^2}\right.\nonumber\\
&\left. -\frac{|g_{WWS}|^2 |g_{VV S}|^2 v^2 \left(3 m_W^6 \left(76 m_V^2-7 m_S^2\right)+16 m_V^2 m_W^4 \left(m_S^2-25
   m_V^2\right)-32 m_V^6 \left(m_S^2+2 m_V^2\right)+4 m_V^4 m_W^2 \left(7 m_S^2+68 m_V^2\right)\right)}{1152 \pi  m_V^4 m_W^4\left(m_S^2-4
   m_V^2\right)^3 \sqrt{1-\frac{m_W^2}{m_V^2}}}\right)\nonumber\\
\end{eqnarray}

$VV \rightarrow ZZ$:

\begin{eqnarray}
& \langle \sigma v \rangle_{ZZ} = \left(\frac{ |g_{VVS}|^2 |g_{ZZS}|^2 \sqrt{4-\frac{4 m_Z^2}{m_V^2}} \left(16 m_V^4-16 m_V^2 m_Z^2+12 m_Z^4\right)}{1536 \pi  m_V^2 m_Z^4 \left(4
   m_V^2-m_S^2\right)^2}\right.\nonumber\\
&\left. -\frac{|g_{ZZS}|^2 |g_{VV S}|^2 v^2 \left(3 m_Z^6 \left(76 m_V^2-7 m_S^2\right)+16 m_V^2 m_Z^4 \left(m_S^2-25
   m_V^2\right)-32 m_V^6 \left(m_S^2+2 m_V^2\right)+4 m_V^4 m_Z^2 \left(7 m_S^2+68 m_V^2\right)\right)}{2304 \pi  m_V^4 m_Z^4 \left(m_S^2-4
   m_V^2\right)^3 \sqrt{1-\frac{m_Z^2}{m_V^2}}}\right)\nonumber\\
\end{eqnarray}

$VV \rightarrow SS$

\begin{eqnarray}
& \langle \sigma v \rangle_{SS}=\frac{\sqrt{1-\frac{m_S^2}{m_V^2}}}{288 \pi  m_V^2} \left[\frac{3 g_{SSS}^2 g_{VVS}^2}{\left(m_S^2-4 m_V^2\right)^2}+\frac{4 g_{SSS} g_{VVS}^3}{m_S^2 m_V^2-2
   m_V^4}-\frac{3 g_{SSS} g_{VVS} g_{VVSS}}{m_S^2-4 m_V^2}\right.\nonumber\\
&\left. +4 g_{VVS}^4 \left(\frac{2}{\left(m_S^2-2 m_V^2\right)^2}+\frac{1}{m_V^4}\right)-\frac{2
   g_{VVS}^2 g_{VVSS} \left(m_S^2-4 m_V^2\right)}{m_V^2 \left(m_S^2-2 m_V^2\right)}+\frac{3 g_{VVSS}^2}{4}\right]\nonumber\\
&+\frac{v^2}{27648 \pi  m_V^7 \left(m_S^2-4 m_V^2\right)^3 \left(m_S^2-2
   m_V^2\right)^4 \sqrt{m_V^2-m_S^2}} \left[-4
   g_{VVS}^2 m_V^2 \left(m_S^2-2 m_V^2\right)\right.\nonumber\\
& \left. \left(2 g_{VVSS} \left(m_S^2-4 m_V^2\right)^3 \left(21 m_S^8-192 m_S^6 m_V^2+602 m_S^4
   m_V^4-784 m_S^2 m_V^6+344 m_V^8\right)\right.\right.\nonumber\\
&\left. \left.-3 g_{SSS}^2 m_V^2 \left(m_S^2-2 m_V^2\right)^3 \left(7 m_S^4-80 m_S^2 m_V^2+64
   m_V^4\right)\right)\right.\nonumber\\
&\left. +16 g_{SSS} g_{VVS}^3 m_V^2 \left(m_S^2-4 m_V^2\right)^2 \left(m_S^2-2 m_V^2\right) \left(21 m_S^8-216 m_S^6
   m_V^2+722 m_S^4 m_V^4-976 m_S^2 m_V^6+440 m_V^8\right)\right.\nonumber\\
&\left. -12 g_{SSS} g_{VVS} g_{VVSS} m_V^4 (m_S-2 m_V) (m_S+2 m_V)
   \left(m_S^2-2 m_V^2\right)^4 \left(7 m_S^4-56 m_S^2 m_V^2+40 m_V^4\right)\right.\nonumber\\
&\left. +16 g_{VVS}^4 \left(m_S^2-4 m_V^2\right)^3 \left(19 m_S^{10}-180
   m_S^8 m_V^2+706 m_S^6 m_V^4-1408 m_S^4 m_V^6+1416 m_S^2 m_V^8-544 m_V^{10}\right)\right.\nonumber\\
&\left. +3 g_{VVSS}^2 m_V^4 \left(m_S^2-4
   m_V^2\right)^3 \left(7 m_S^2-4 m_V^2\right) \left(m_S^2-2 m_V^2\right)^4\right]
\end{eqnarray}

\subsection{Spin-1 portal}

\subsubsection{Scalar Dark Matter}

$\chi \chi^{*} \rightarrow \ovl f f$

\begin{equation}
\langle \sigma v \rangle_{\rm ff}=g^2 |g_{\chi \chi Z}|^2 v^2 \sum_{f} n_c^f\sqrt{m_\chi^2-m_f^2}
\frac{ \left(2|A_f^{Z'}|^2  \left(m_{\chi}^2- m_f^2\right)+ |V_f^{Z'}|^2 (2 m_\chi^2+ m_f^2)\right)}{3 \pi m_\chi  \left(m_{Z'}^2-4 m_\chi^2\right)^2}
\end{equation}

$\chi \chi^{*} \rightarrow W^+ W^-$:

\begin{align}
\langle \sigma v \rangle_{W^{+}W^{-}}=\frac{ 
   {\left(1-\frac{m_W^2}{m_\chi^2}\right)}^{3/2} m_\chi^2 |g_{\chi \chi Z}|^2 |g_{WWZ}|^2 v^2 \left(3 m_W^4+20 m_W^2 m_\chi^2 +4 m_\chi^4\right)}{6 \pi  m_W^4
   \left(m_Z^2-4 m_\chi^2\right)^2}
\end{align}

$\chi \chi^{*} \rightarrow Z^\prime Z^\prime$:

\begin{eqnarray}
& \langle \sigma v \rangle_{Z'Z'}=\frac{\sqrt{m_\chi^2-m_{Z'}^2}}{16 \pi  m_{Z'}^4 m_\chi^3 \left(m_{Z'}^2-2 m_\chi^2\right)^2} \left(8 |g_{\chi \chi Z'}|^2 |g_{\chi \chi Z'Z'}|^2 m_\chi^2 (m_{Z'}-m_\chi) (m_{Z'}+m_\chi)
   \left(m_{Z'}^2-2 m_\chi^2\right)^2\right.\nonumber\\
& \left.+16 |g_{\chi \chi Z'}|^4 m_\chi^4 \left(m_{Z'}^2-m_\chi^2\right)^2+|g_{\chi \chi Z'Z'}|^4 \left(m_{Z'}^2-2
   m_\chi^2\right)^2 \left(3 m_{Z'}^4-4 m_{Z'}^2 m_\chi^2+4 m_\chi^4\right)\right)\nonumber\\
& +\frac{v^2}{192 \pi  m_{Z'}^2 m_\chi^3 \left(m_{Z'}^2-2 m_\chi^2\right)^4 \sqrt{m_\chi^2-m_{Z'}^2}} \left(16 |g_{\chi \chi Z'}|^4 m_\chi^4 \left(m_{Z'}^2-m_\chi^2\right)^2 \left(3 m_{Z'}^4-20 m_{Z'}^2 m_\chi^2+36 m_\phi^4\right)\right.\nonumber\\
& \left.+8 |g_{\chi \chi Z'}|^2 |g_{\chi \chi Z'Z'}|^2 m_\chi^2 (m_{Z'}-m_\chi) (m_{Z'}+m_\chi) \left(m_{Z'}^2-2 m_\chi^2\right)^2
   \left(5 m_{Z'}^4-26 m_{Z'}^2 m_\chi^2+36 m_\chi^4\right)\right.\nonumber\\
&\left. +9 |g_{\chi \chi Z'Z'}|^4 \left(m_{Z'}^2-2 m_\chi^2\right)^4 \left(5 m_{Z'}^4-8 m_{Z'}^2
   m_\chi^2+4 m_\chi^4\right)\right)
\end{eqnarray}


$\chi \chi^{*} \rightarrow Zh$:

\begin{eqnarray}
& \langle \sigma v \rangle_{Zh}=|g_{\chi \chi Z}|^2 |g_{ZZh}|^2 v^2\sqrt{1-\frac{(m_h-m_Z)^2}{4 m_\phi^2}}\sqrt{1-\frac{(m_h+m_Z)^2}{4 m_\chi^2}}\frac{(m_h^2-m_Z^2)^2-8 (m_h^2-5 m_Z^2) m_\chi^2+16 m_\chi^4}{384 \pi m_Z^2 m_\chi^2 \left(m_Z^2-4 m_\chi^2\right)^2}\nonumber\\
\end{eqnarray}

\subsubsection{Fermionic Dark Matter}

For simplicity we explicitly report the case of Dirac DM. The expressions for Majorana DM are straightforwardly from these.

$\bar{\psi} \psi \rightarrow \ovl f f$

\begin{align}
& \langle \sigma v \rangle_{\rm ff}=g^4 \sum_{f} n_c^f\sqrt{m_\psi^2-m_f^2}\nonumber\\
&\frac{2  \left[|A^{Z'}_f|^2 |A_\psi^{Z'}|^2 m_f^2 \left(m_{Z}^2-4 m_\psi^2\right)^2+m_{Z'}^4 2 |V_\psi^{Z'}|^2\left(2 |A^{Z'}_f|^2 \left(m_\psi^2-m_f^2\right)+|V_f|^2 \left(m_f^2+2 m_\psi^2\right)\right)\right]}{4 \pi m_\psi m_{Z'}^4 \left(m_{Z'}^2-4 m_\psi^2\right)^2}\nonumber\\
&-\frac{1}{24 \pi  m_\psi m_{Z'}^4 \sqrt{m_\psi^2-m_f^2} \left(m_{Z'}^2-4 m_\psi^2\right)^3}v^2 \left(|A^{Z'}_f|^2 \left(2
   m_{Z'}^4 |V_\psi^{Z'}|^2  (m_f-m_\psi) (m_f+m_\psi)\right. \right.\nonumber\\
& \left. \left.\left(-2 m_\psi^2 \left(46 m_f^2+m_{Z'}^2\right)+11 m_f^2 m_{Z'}^2+56 m_\psi^4\right)-|A_\psi^{Z'}|^2 \left(m_{Z'}^2-4 m_\psi^2\right)\right.\right.\nonumber\\
& \left.\left. \left(23 m_f^4 m_{Z'}^4-192 m_f^2 m_\psi^6-4 m_f^2 m_\psi^2 m_{Z'}^2 \left(30 m_f^2+7
   m_{Z'}^2\right)+8 m_\psi^4 \left(30 m_f^4+12 m_f^2 m_{Z'}^2+m_{Z'}^4\right)\right)\right)\right.\nonumber\\
&\left.+m_{Z'}^4 |V^{Z'}_f|^2 \left(4 |A_\psi^{Z'}|^2 \left(m_f^4+m_f^2 m_\psi^2-2 m_\psi^4\right) \left(m_{Z'}^2-4 m_\psi^2\right)\right.\right.\nonumber\\
&\left.\left.+ |V_\psi^{Z'}|^2\left(-11 m_f^4 m_{Z'}^2+4 m_\psi^4 \left(14 m_f^2+m_{Z'}^2\right)-2 m_f^2 m_\psi^2 \left(m_{Z'}^2-46 m_f^2\right)-112 m_\psi^6\right)\right)\right)
\label{Eq:sigvff2}
\end{align}

$\bar{\psi} \psi \rightarrow W^+ W^-$:

\begin{eqnarray}
& \langle \sigma v \rangle_{W^{+}W^{-}}=g^2 |g_{WWZ'}|^2 \left[ \frac{ 
   |V_\psi^{Z'}|^2\sqrt{m_\psi^2-m_W^2} \left(-3 m_W^6-17
   m_W^4 m_\psi^2+16 m_W^2 m_\psi^4+4
   m_\psi^6\right)}{4 \pi  m_W^4 m_\psi
   \left(m_{Z'}^2-4 m_\psi^2\right)^2} \right. \nonumber\\
& +\left. \frac{v^2 \sqrt{m_\psi^2-m_{W}^2}}{48 \pi  m_W^4 m_\psi
   \left(m_{Z'}^2-4 m_\psi^2\right)^3}
    \left(4 |A_\psi^{Z'}|^2
   \left(-3 m_W^6-17 m_W^4 m_\psi^2+16 m_W^2
   m_\psi^4+4 m_\psi^6\right) \left(m_{Z'}^2-4
   m_\psi^2\right)\right. \right.\nonumber\\
 &\left. \left.+  |V_\psi^{Z'}|^2 \left(33 m_W^6
   m_{Z'}^2+8 m_\psi^6 \left(58 m_W^2+5
   m_{Z'}^2\right)+4 m_W^2 m_\psi^4 \left(19
   m_{Z'}^2-298 m_W^2\right)+2 m_W^4 m_\psi^2
   \left(47 m_{Z'}^2-138 m_W^2\right)\right.\right. \right.\nonumber\\
& \left. \left. \left. +32 m_\psi^8\right)\right) \right]\nonumber\\
 \end{eqnarray}

$\bar{\psi} \psi \rightarrow Z' Z'$:

\begin{eqnarray}
&\langle \sigma v \rangle_{Z'Z'}=
g^4 |g_{\psi \psi Z'}|^4 \left[ \frac{\left(m_\psi^2-m_{Z'}^2\right)^{3/2} \left(|A_\psi^{Z'}|^4 m_{Z'}^2+2 |A_\psi^{Z'}|^2 |V_\psi^{Z'}|^2 \left(4 m_\psi^2-3 m_{Z'}^2\right)+m_{Z'}^2 |V_\psi^{Z'}|^4\right)}{\pi 
   m_\psi \left(m_{Z'}^3-2 m_\psi^2 m_{Z'}\right)^2}\right.\nonumber\\
&\left.+\frac{v^2\sqrt{m_\psi^2-m_{Z'}^2}}{4 \pi  m_\psi  \left(m_{Z'}^3-2 m_\psi^2 m_{Z'}\right)^4} \left(|A_\psi^{Z'}|^4 \left(128 m_\psi^{10}+23 m_{Z'}^{10}-118 m_\psi^2
   m_{Z'}^8+172 m_\psi^4 m_{Z'}^6+32 m_\psi^6 m_{Z'}^4-192 m_\psi^8 m_{Z'}^2\right)\right.\right.\nonumber\\
&\left. \left.-2 |A_\psi^{Z'}|^2 m_{Z'}^2 |V_\psi^{Z'}|^2 \left(160 m_\psi^8+21 m_{Z'}^8-182
   m_\psi^2 m_{Z'}^6+508 m_\psi^4 m_{Z'}^4-528 m_\psi^6 m_{Z'}^2\right)\right.\right.\nonumber\\
&\left.\left. +m_{Z'}^6 |V_\psi^{Z'}|^4 \left(76 m_\psi^4+23 m_{Z'}^4-66 m_\psi^2
   m_{Z'}^2\right)\right)\right]\nonumber\\
 \end{eqnarray}

$\bar{\psi} \psi \rightarrow Zh$:

\begin{eqnarray}
& \langle \sigma v \rangle_{Zh}=g^4 |g_{ZZh}|^2 \frac{\sqrt{m_h^4-2 m_h^2 \left(4 m_\psi^2+m_Z^2\right)+\left(m_Z^2-4 m_\psi^2\right)^2}}{3072 \pi  m_\psi^4 m_Z^6} \left[ \left(3 |A_\psi^{Z}|^2 \left(m_h^4-2 m_h^2 \left(4 m_\psi^2
+m_Z^2\right)+\left(m_Z^2-4 m_\chi^2\right)^2\right)\right. \right.\nonumber\\
& \left.  +|V_\psi^{Z}|^2\frac{3 m_Z^4  \left(-8 m_\psi^2 \left(m_h^2-5
   m_Z^2\right)+\left(m_h^2-m_Z^2\right)^2+16 m_\psi^4\right)}{\left(m_Z^2-4 m_\psi^2\right)^2}\right) \nonumber\\
& -\frac{v^2 }{ \left(m_Z^2-4 m_\psi^2\right)^3
   \left((m_h-m_Z)^2-4 m_\psi^2\right) \left((m_h+m_Z)^2-4 m_\psi^2\right)} \nonumber\\
&	\left(|A_\psi^{Z}|^2 \left(m_Z^2-4 m_\psi^2\right) \left(-96 m_\psi^6 \left(5
   m_h^2+7 m_Z^2\right)+5 m_Z^4 \left(m_h^2-m_Z^2\right)^2+8 m_\psi^4 \left(12 m_h^4+6 m_h^2 m_Z^2+43 m_Z^4\right)\right.\right.\nonumber\\
& \left.\left. -2 m_\chi^2 m_Z^2 \left(24
   m_h^4-37 m_h^2 m_Z^2+59 m_Z^4\right)+384 m_\psi^8\right) \left(m_h^4-2 m_h^2 \left(4 m_\psi^2+m_Z^2\right)+\left(m_Z^2-4 m_\psi^2\right)^2\right)\right.\nonumber\\
& +\left. m_Z^4 |V_\psi^{Z}  \left(128 m_\psi^8 \left(37 m_h^2-82 m_Z^2\right)+5 m_Z^2 \left(m_h^2-m_Z^2\right)^4+32 m_\psi^6 \left(-69
   m_h^4+217 m_h^2 m_Z^2+242 m_Z^4\right)\right.\right.\nonumber\\
& \left.\left.  +2 m_\psi^2 \left(m_h^2-m_Z^2\right)^2 \left(-16 m_h^4+m_h^2 m_Z^2+37 m_Z^4\right)+8 m_\chi^4 \left(55
   m_h^6-178 m_h^4 m_Z^2+147 m_h^2 m_Z^4-200 m_Z^6\right)\right.\right.\nonumber\\
& \left.\left.\left.   -3584 m_\psi^{10}\right)\right) \right]
 \end{eqnarray}

\subsubsection{Vectorial Dark Matter}

$VV \rightarrow \ovl f f$:

\begin{align}
& \langle \sigma v \rangle_{\rm ff}^{\rm A}= {g'}^4 |\eta_V^{Z'}|^2\sum_f n_c^f \left[\frac{|A_f^{Z'}|^2 m_f^2 v^2 \sqrt{m_V^2-m_f^2}}{9 \pi  m_V m_{Z'}^4}+\frac{5 v^4}{324 \pi  m_V m_{Z'}^4 \sqrt{m_V^2-m_f^2} \left(m_{Z'}^2-4 m_V^2\right)^2}\right. \nonumber\\
&\left.  \left(|A_f^{Z'}|^2 \left(m_f^4 \left(240 m_V^4-120 m_V^2 m_{Z'}^2+23
   m_{Z'}^4\right)-4 m_f^2 \left(48 m_V^6-24 m_V^4 m_{Z'}^2+7 m_V^2 m_{Z'}^4\right)+8 m_V^4 m_{Z'}^4\right)\right.\right.\nonumber\\
&\left.\left.-4 m_{Z'}^4 |V_f^{Z'}|^2
   \left(m_f^4+m_f^2 m_V^2-2 m_V^4\right)\right)\right]
\end{align}

\begin{align}
& \langle \sigma v \rangle_{\rm ff}^{\rm NA}={g'}^4 |\eta_V^{Z'}|^2 v^2 \sum_{f} n_c^f\sqrt{m_V^2-m_f^2}\nonumber\\
&\frac{ \left(2|A_f|^2  \left(m_V^2- m_f^2\right)^2+ |V_f|^2 (2 m_V^2+ m_f^2)\right)}{\pi m_V  \left(m_{Z}^2-4 m_V^2\right)^2}
\end{align}

$VV \rightarrow W^+ W^-$:

\begin{align}
\langle \sigma v \rangle_{W^{+}W^{-}}^{\rm A}=g^2 |\eta_V^Z|^2 |g_{WWZ'}|^2 v^4 \frac{5}{108 \pi m_W^4 (m_W^2-4 m_V^2)^2} \sqrt{1-\frac{m_W^2}{m_V^2}}\left(4 m_V^6-8 m_V^4 m_W^2-9 m_V^2 m_W^4-3 m_W^6\right)
\end{align}

\begin{align}
& \langle \sigma v \rangle_{W^{+}W^{-}}^{\rm NA}=g^2 |\eta_V^Z|^2 |g_{WWZ'}|^2 v^2 {\left(1-\frac{m_W^2}{m_V^2}\right)}^{3/2} \frac{m_V^2
    \left(3 m_W^4+20 m_W^2 m_V^2 +4 m_V^4\right)}{2 \pi  m_W^4
   \left(m_Z^2-4 m_V^2\right)^2}
\end{align}

$VV \rightarrow Z' Z'$:

\begin{eqnarray}
& \langle \sigma v \rangle_{Z' Z'}^{\rm A}=g^4 |\eta_V^{Z'}|^4\left(\frac{\sqrt{(m_V-m_{Z'}) (m_V+m_{Z'})} \left(32 m_V^8-56 m_V^6 m_{Z'}^2+69 m_V^4 m_{Z'}^4-50 m_V^2 m_{Z'}^6+14
   m_{Z'}^8\right)}{144 \pi  m_V^3 m_{Z'}^4 \left(m_{Z'}^2-2 m_V^2\right)^2}\right.\nonumber\\
& \left.+\frac{v^2 \left(512 m_V^{14}-832 m_V^{12} m_{Z'}^2-952
   m_V^{10} m_{Z'}^4+4292 m_V^8 m_{Z'}^6-5500 m_V^6 m_{Z'}^8+3391 m_V^4 m_{Z'}^{10}-994 m_V^2 m_{Z'}^{12}+110 m_{Z'}^{14}\right)}{1728 \pi 
   m_V^3 \sqrt{(m_V-m_{Z'}) (m_V+m_{Z'})} \left(m_{Z'}^3-2 m_V^2 m_{Z'}\right)^4}\right)\nonumber\\
\end{eqnarray}

\begin{eqnarray}
& \langle \sigma v \rangle_{Z'Z'}^{\rm NA}=g^4 |\eta_V^{Z'}|^4\left( \frac{\sqrt{(m_V-m_{Z'})
   (m_V+m_{Z'})}}{144 \pi 
   m_V^7 m_{Z'}^4 \left(m_{Z'}^2-2
   m_V^2\right)^2} \left(144 m_V^{12}-272
   m_V^{10} m_{Z'}^2+316 m_V^8
   m_{Z'}^4-264 m_V^6 m_{Z'}^6\right. \right.\nonumber\\
&\left. \left.	+168
   m_V^4 m_{Z'}^8-14 m_V^2
   m_{Z'}^{10}+3 m_{Z'}^{12}\right)+\frac{ v^2}{1728 \pi  m_V^7
   \sqrt{(m_V-m_{Z'}) (m_V+m_{Z'})}
   \left(m_{Z'}^3-2 m_V^2 m_{Z'}\right)^4}\right.\nonumber\\
&	\left.  \left(3584 m_V^{18}-8256 m_V^{16}
   m_{Z'}^2+6208 m_V^{14} m_{Z'}^4+3104
   m_V^{12} m_{Z'}^6-17456 m_V^{10}
   m_{Z'}^8+21372 m_V^8 m_{Z'}^{10}-10488
   m_V^6 m_{Z'}^{12}\right.\right. \nonumber\\
& \left. \left.+2480 m_V^4
   m_{Z'}^{14}-346 m_V^2 m_{Z'}^{16}+41
   m_{Z'}^{18}\right)\right)\nonumber\\
\end{eqnarray}

$VV \rightarrow Zh$:

\begin{align}
& \langle \sigma v \rangle_{\rm Zh}^{\rm NA}=g^2 |g_{ZZh}|^2 v^2 \sqrt{(m_h+2 m_V-m_Z)
   (-m_h+2 m_V+m_Z)}
   \sqrt{(-m_h+2 m_V-m_Z) (m_h+2
   m_V+m_Z)}\nonumber\\
&	\frac{\left(-2 m_Z^2
   \left(m_h^2-20
   m_V^2\right)+\left(m_h^2-4
   m_V^2\right)^2+m_Z^4\right)}{512 \pi 
   m_V^4 \left(m_Z^3-4 m_V^2
   m_Z\right)^2}
\end{align}

\subsection{Pseudoscalar portal}

\begin{align}
& \langle \sigma v \rangle_{ff}=\frac{c_a^2 \frac{m_f^2}{v_h^2} |\lambda_\psi^a|^2 m_\psi \sqrt{m_\psi^2-m_f^2}}{2 \pi  \left(m_a^2-4 m_\psi^2\right)^2}
\end{align}

\begin{equation}
\langle \sigma v \rangle_{aa}=\frac{1}{12 \pi}|\lambda_\psi^{a}|^4 \frac{m_\psi^6}{{\left(m_a^2-2 m_\psi^2\right)}^4}{\left(1-\frac{m_a^2}{m_\psi^2}\right)}^{5/2}
\end{equation}

\subsection{Scalar+Pseudoscalar portal}

The expression for the cross-section into $\ovl f f$ and $SS$ final state can be straightforwardly derived from the previous one. Will then just report the cross section relative to the $aa$ and $Sa$ final states.

The process $\psi \psi \rightarrow aa$ receives, with respect to the case of the pure pseudoscalar portal, and additional contribution from s-channel exchange of the scalar field $S$, so that:
\begin{equation}
\langle \sigma v \rangle_{aa}=\frac{1}{128 \pi m_\psi^2}\sqrt{1-\frac{m_a^2}{m_\psi^2}}\left(\frac{32}{3}\frac{g_\psi^2 m_\psi^4 {\left(m_a^2-m_\psi^2\right)}^2}{{\left(m_a^2-2 m_\psi^2\right)}^4}+\frac{4 \lambda g_\psi^2 m_\psi^2 m_S^2}{{\left(m_S^2-4 m_\psi^2\right)}^2}\right)v^2
\end{equation}

\begin{align}
& \langle \sigma v \rangle_{Sa}=\frac{g_\psi^2 \sqrt{m_a^4-2 m_a^2 \left(4 m_\psi^2+m_S^2\right)+\left(m_S^2-4 m_\psi^2\right)^2}}{64 \pi  m_\psi^4}\nonumber\\
& \left[\frac{g_\psi^2
   \left(m_a^2+4 m_\psi^2-m_S^2\right)^2}{\left(m_a^2-4 m_\psi^2+m_S^2\right)^2}+\frac{2 \sqrt{2} g_\psi \sqrt{\lambda } m_\psi
   m_S \left(m_a^2+4 m_\psi^2-m_S^2\right)}{\left(m_a^2-4 m_\psi^2\right) \left(m_a^2-4 m_\psi^2+m_S^2\right)}+\frac{2
   \lambda  m_\psi^2 m_S^2}{\left(m_a^2-4 m_\psi^2\right)^2}\right]+\nonumber\\
& \frac{v^2}{2} \left[\frac{g_\psi^2 \sqrt{m_a^4-2 m_a^2 \left(4 m_\psi^2+m_S^2\right)+\left(m_S^2-4 m_\psi^2\right)^2}}{12 m_\psi^4}\right. \nonumber\\
&\left. \left(\frac{g_\psi^2 \left(m_a^8-8 m_a^6 \left(m_S^2-2 m_\psi^2\right)+14 m_a^4 \left(m_S^2-4 m_\psi^2\right)^2-8 m_a^2 \left(m_S^2-4
   m_\psi^2\right)^2 \left(10 m_\psi^2+m_S^2\right)+\left(m_S^2-4 m_\psi^2\right)^4\right)}{\left(m_a^2-4 m_\psi^2+m_S^2\right)^4}\right.\right.\nonumber\\
& \left. \left. +\frac{2 \sqrt{2} g_\psi \sqrt{\lambda } m_\psi m_S}{\left(m_a^2-4 m_\psi^2\right)^2 \left(m_a^2-4
   m_\psi^2+m_S^2\right)^3}\right. \right.\nonumber\\
& \left. \left. \left(m_a^8+m_a^6 \left(28 m_\psi^2-3 m_S^2\right)+m_a^4
   \left(-240 m_\psi^4+3 m_S^4+40 m_\psi^2 m_S^2\right)-m_a^2 \left(-320 m_\psi^6+m_S^6+12 m_\psi^2 m_S^4+16
   m_\psi^4 m_S^2\right)\right.\right.\right.\nonumber\\
& \left. \left. \left. -8 m_\psi^2 \left(m_S^2-4 m_\psi^2\right)^3\right)+\frac{6 \lambda  m_\psi^2 m_S^2 \left(m_a^2+4 m_\psi^2\right)}{\left(m_a^2-4 m_\psi^2\right)^3}\right)\right. \nonumber\\
& \left. -\frac{g_\psi^2 \left(-20 m_\psi^2 \left(m_a^2+m_S^2\right)+3 \left(m_a^2-m_S^2\right)^2+32
   m_\psi^4\right)}{4 m_\psi^4 \left(m_a^2-4 m_\psi^2\right)^2 \left(m_a^2-4 m_\psi^2+m_S^2\right)^2 \sqrt{m_a^4-2 m_a^2 \left(4 m_\psi^2+m_S^2\right)+\left(m_S^2-4 m_\psi^2\right)^2}}\right.\nonumber\\
	& \left. \left(g_\psi^2 \left(m_a^2-4 m_\psi^2\right)^2 \left(m_a^2+4 m_\psi^2-m_S^2\right)^2+2 \sqrt{2} g_\psi
   \sqrt{\lambda } m_\psi m_S \left(m_a^2-4 m_\psi^2\right) \left(m_a^4-\left(m_S^2-4 m_\psi^2\right)^2\right)\right.\right.\nonumber\\
	&\left. \left. +2 \lambda  m_\psi^2
   m_S^2 \left(m_a^2-4 m_\psi^2+m_S^2\right)^2\right)\right]	
\end{align}

\subsection{t-channel portals}

As already mentioned in the main text, in the case of $O(1)$ coupling the DM relic density is mainly accounted by DM pair annihilations into SM fermions. We will then explicitly report only the cross-sections corresponding to these processes.

\subsubsection{Complex Scalar Dark Matter}

\begin{align}
& \langle \sigma v \rangle_{ff}=\sum_f n^c_f \frac{\lambda_{\Psi_u}^4 m_f^2 \left(m_\chi^2-m_f^2\right)^{3/2}}{16 \pi  m_\chi^3 \left(-m_f^2+m_\chi^2+m_{\Psi_u}^2\right)^2} \nonumber\\
&+\sum_f n^c_f \frac{\lambda_{\Psi_u}^4 \sqrt{m_\chi^2-m_f^2} v^2}{192 \pi  m_\chi^3  \left(-m_f^2+m_\chi^2+m_{\Psi_u}^2\right)^4} \left(15 m_f^8-6 m_f^6 \left(9 m_\chi^2+5 m_{\Psi_u}^2\right)+m_f^4 \left(m_\chi^2+m_{\Psi_u}^2\right) \left(71 m_\chi^2+15 m_{\Psi_u}^2\right)\right.\nonumber\\
& \left. -8 m_f^2 m_\chi^2 \left(5 m_\chi^4+9 m_\chi^2 m_{\Psi_u}^2+m_{\Psi_u}^4\right)+8 m_\chi^4 \left(m_\chi^2+m_{\Psi_u}^2\right)^2\right)
\end{align}

\subsubsection{Dirac Fermion Dark Matter}
\begin{align}
& \langle \sigma v \rangle_{ff}=\sum_f n^c_f \frac{\lambda_{\Sigma_u}^4 m_\psi \sqrt{m_\psi^2-m_f^2}}{32 \pi  \left(-m_f^2+m_\Sigma^2+m_\psi^2\right)^2}+\sum_f n^c_f\frac{\lambda_{\Sigma_u}^4 v^2}{384 \pi  m_\psi \sqrt{m_\psi^2-m_f^2} \left(-m_f^2+m_{\Sigma_u}^2+m_\psi^2\right)^4}\nonumber\\
& \times  \left(3 m_\psi^2 \left(5 m_f^2-4 m_\psi^2\right) \left(-m_f^2+m_{\Sigma_u}^2+m_\psi^2\right)^2-2 \left(m_\psi^2-m_f^2\right) \left(m_f^2-m_{\Sigma_u}^2+m_\psi^2\right) \left(m_f^4-m_f^2 \left(m_{\Sigma_u}^2+2 m_\psi^2\right)\right. \right.\nonumber\\
& \left.\left.+7 m_{\Sigma_u}^2
   m_\psi^2+m_\psi^4\right)\right)
\end{align}

\subsubsection{Majorana Fermion Dark Matter}
\begin{eqnarray}
& \langle \sigma v \rangle_{ff}=\sum_f n^c_f \frac{\lambda_{\Sigma_u}^4 m_f^2 \sqrt{m_\psi^2-m_f^2}}{8 \pi m_\psi  \left(-m_f^2+m_{\Sigma_u}^2+m_\psi^2\right)^2} \nonumber\\
&+ \sum_f n^c_f \frac{\lambda_{\Sigma_u}^4 m_\psi v^2}{96 \pi \sqrt{m_\psi^2-m_f^2} \left(-m_f^2+m_{\Sigma_u}^2+m_\psi^2\right)^4} \left(-42 m_f^2 m_\psi^6+13 m_f^4 \left(m_f^2-m_{\Sigma_u}^2\right)^2+m_\psi^4 \left(49 m_f^4-44
   m_f^2 m_{\Sigma_u}^2+16 m_{\Sigma_u}^4\right)\right.\nonumber\\
& \left. -2 m_f^2 m_\psi^2 \left(18 m_f^4-35 m_f^2 m_{\Sigma_u}^2+13 m_{\Sigma_u}^4\right)+16 m_\psi^8\right)
\end{eqnarray}

}

	

\bibliographystyle{JHEPfixed}
\bibliography{bibfile}{}

\end{document}